\documentclass[a4paper,11pt]{article}
\pdfoutput=1
\usepackage{jcappub}
\usepackage[T1]{fontenc} 
\usepackage{hyperref}
\usepackage{graphics}
\usepackage{amssymb, amsmath}
\usepackage{bm}
\usepackage{xcolor,color,colortbl}
\usepackage{multirow,bigdelim}
\usepackage{bbold}
\usepackage{verbatim}
\usepackage{booktabs}
\usepackage[nameinlink,capitalise]{cleveref}

\author[a,b]{Alex Krolewski}
\author[b,c,a]{Will J.~Percival}
\author[d,e]{Simone Ferraro}
\author[f]{Edmond Chaussidon}
\author[g]{Mehdi Rezaie}
\author[d]{Jessica Nicole Aguilar}
\author[h]{Steven Ahlen}
\author[i]{David Brooks}
\author[j]{Kyle Dawson}
\author[k]{Axel de la Macorra}
\author[i]{Peter Doel}
\author[l,m]{Kevin Fanning}
\author[n]{Andreu Font-Ribera}
\author[d]{Satya Gontcho A Gontcho}
\author[d]{Julien Guy}
\author[l,m]{Klaus Honscheid}
\author[o]{Robert Kehoe}
\author[d]{Theodore Kisner}
\author[d]{Anthony Kremin}
\author[d]{Martin Landriau}
\author[d]{Michael E.~Levi}
\author[l,p]{Paul Martini}
\author[q]{Aaron M.~Meisner}
\author[r,l]{Ramon Miquel}
\author[s]{Jundan Nie}
\author[d,t]{Claire Poppett}
\author[l,p]{Ashley J.~Ross}
\author[u]{Graziano Rossi}
\author[v]{Michael Schubnell}
\author[d,w]{Hee-Jong Seo}
\author[v]{Gregory~Tarl\'{e}}
\author[k]{Mariana~Vargas-Maga\~na}
\author[q]{Benjamin Alan Weaver}
\author[f]{Christophe Y\`eche}
\author[d]{Rongpu Zhou}
\author[s]{Zhimin Zhou}

\affiliation[a]{Perimeter Institute for Theoretical Physics, 31 Caroline St. North, Waterloo, ON NL2 2Y5, Canada}
\affiliation[b]{Waterloo Centre for Astrophysics, University of Waterloo, Waterloo, ON N2L 3G1, Canada} 
\affiliation[c]{Department of Physics and Astronomy, University of Waterloo, Waterloo, ON N2L 3G1, Canada}
\affiliation[d]{Physics Division, Lawrence Berkeley National Laboratory, Berkeley, CA 94720}
\affiliation[e]{Department of Physics, University of California, Berkeley, CA 94720}
\affiliation[f]{IRFU, CEA, Universit\'{e} Paris-Saclay, F-91191 Gif-sur-Yvette, France}
\affiliation[g]{Department of Physics, Kansas State University,
116 Cardwell Hall, Manhattan, KS 66506, USA}
\affiliation[h]{Physics Department, Boston University, 590 Commonwealth Avenue, Boston, MA 02215, USA}
\affiliation[i]{Department of Physics \& Astronomy, University College London, Gower Street, London, WC1E 6BT, UK}
\affiliation[j]{Department of Physics and Astronomy, The University of Utah, 115 South 1400 East, Salt Lake City, UT 84112, USA}
\affiliation[k]{Instituto de F\'{\i}sica, Universidad Nacional Aut\'{o}noma de M\'{e}xico,  Cd. de M\'{e}xico  C.P. 04510,  M\'{e}xico}
\affiliation[l]{Center for Cosmology and AstroParticle Physics, The Ohio State University, 191 West Woodruff Avenue, Columbus, OH 43210, USA}
\affiliation[m]{Department of Physics, The Ohio State University, 191 West Woodruff Avenue, Columbus, OH 43210, USA}
\affiliation[n]{Institut de F\'{i}sica d'Altes Energies (IFAE), The Barcelona Institute of Science and Technology, Campus UAB, 08193 Bellaterra Barcelona, Spain}
\affiliation[o]{Department of Physics, Southern Methodist University, 3215 Daniel Avenue, Dallas, TX 75275, USA}
\affiliation[p]{Department of Astronomy, The Ohio State University, 4055 McPherson Laboratory, 140 W 18th Avenue, Columbus, OH 43210, USA}
\affiliation[q]{NSF's NOIRLab, 950 N. Cherry Ave., Tucson, AZ 85719, USA}
\affiliation[r]{Instituci\'{o} Catalana de Recerca i Estudis Avan\c{c}ats, Passeig de Llu\'{\i}s Companys, 23, 08010 Barcelona, Spain}
\affiliation[s]{National Astronomical Observatories, Chinese Academy of Sciences, A20 Datun Rd., Chaoyang District, Beijing, 100012, P.R. China}
\affiliation[t]{Space Sciences Laboratory, University of California, Berkeley, 7 Gauss Way, Berkeley, CA  94720, USA}
\affiliation[u]{Department of Physics and Astronomy, Sejong University, Seoul, 143-747, Korea}
\affiliation[v]{University of Michigan, Ann Arbor, MI 48109, USA}
\affiliation[w]{Department of Physics and Astronomy, Ohio University, Athens, OH 45701, USA}
\emailAdd{akrolewski@perimeterinstitute.ca}

\title{Constraining primordial non-Gaussianity from DESI quasar targets and Planck CMB lensing}


\abstract{We detect the cross-correlation between 2.7 million DESI quasar targets across 14,700 deg$^2$ (180 quasars deg$^{-2}$) and Planck 2018 CMB lensing at $\sim$30$\sigma$. We use the cross-correlation
on very large scales to constrain local primordial non-Gaussianity via the scale dependence of quasar bias. The DESI quasar targets lie at an effective redshift of 1.51 and are separated into four imaging regions of varying depth and image quality.
We select quasar targets from Legacy Survey DR9 imaging, apply additional flux and photometric redshift cuts to improve the purity and reduce the fraction of unclassified redshifts, and use early DESI spectroscopy of 194,000 quasar targets to determine their redshift distribution and stellar contamination fraction (2.6\%). Due to significant excess large-scale power in the quasar autocorrelation, we apply weights to mitigate contamination from imaging systematics such as depth, extinction, and stellar density.
We use realistic contaminated mocks to determine the greatest number of systematic modes that we can fit, before we are biased by overfitting and spuriously remove real power.
We find that linear regression with one to seven imaging templates removed per region accurately recovers the input cross-power, $f_{\textrm{NL}}$ and linear bias.
As in previous analyses, our $f_{\textrm{NL}}$ constraint depends on
the linear primordial non-Gaussianity bias parameter, $b_{\phi} = 2(b - p)\delta_c$ assuming universality of the halo mass function.
We measure $f_{\textrm{NL}} = -26^{+45}_{-40}$ with $p=1.6$ ($f_{\textrm{NL}} = -18^{+29}_{-27}$ with $p=1.0$), and find that this result is robust under several systematics tests. Future spectroscopic quasar cross-correlations with Planck lensing can tighten the $f_{\textrm{NL}}$ constraints by a factor of 2 if they can remove the excess power on large scales in the quasar auto power spectrum.
}

\begin{document}
\maketitle
\flushbottom

\section{Introduction}

Single-field slow roll inflation (SFSR) generates nearly Gaussian, nearly scale-invariant primordial fluctuations.
Deviations from Gaussianity are of order of the slow-roll parameter, $10^{-2}$. In fact, the consistency relations for SFSR inflation \citep{Maldacena03,Creminelli04,Acquaviva03} state that the squeezed-limit primordial bispectrum follows the ``local'' shape, with $f_{\textrm{NL}}^{\textrm{loc}} = 0.015$ for $n_s = 0.963$.
In contrast, multi-field inflation models can produce large amounts of local primordial non-Gaussianity, with $f_{\textrm{NL}}^{\textrm{loc}}$ generically around 1 \citep{Bartolo04}.
Hence, a detection of $f_{\textrm{NL}}^{\textrm{loc}} \gg 0.01$ would rule out single-field inflation, whereas constraining $f_{\textrm{NL}}^{\textrm{loc}} < 1$ would put significant pressure on multi-field models. In tandem with the tensor-to-scalar ratio measured from CMB B modes, local primordial non-Gaussianity measurements can disfavor significant portions of inflationary parameter space.

The current best constraints on $f_{\textrm{NL}}$ are from Planck's measurement of the bispectrum, $f_{\textrm{NL}} = -0.9 \pm 5.1$ \citep{PlanckFnl}.
In what follows, we will drop the superscript ``loc'' for clarity, but all constraints should be understood as referring to the local shape only.
The Planck measurements are close to the cosmic variance limit, and CMB-S4 is only expected to improve the constraint by a factor of 2 \citep{CMBS4-book}.
Thus, future CMB experiments will not be sufficient to reach $\mathcal{O}(1)$ constraints on $f_{\textrm{NL}}$.
Local primordial non-Gaussianity also induces a scale-dependent halo bias $\propto 1/k^2$, leading to an enhancement in clustering on very large (low $k$) scales \citep{Dalal08,Slosar08}. The scale dependence of the bias can also be used to probe interactions in multi-field inflation \cite{Ferraro:2014jba, Smith:2011ub} and to measure the mass of massive fields during inflation \cite{Baumann:2012bc}.

The current best constraints on $f_{\textrm{NL}}$ from large scale structure are $\sigma_{f\textrm{NL}} \sim 20$---30 \citep{Mueller,Castorina,Cabass22,DAmico22}.
The most robust current measurements are from the 3D power spectrum, either of eBOSS quasars ($\sigma_{f\textrm{NL}}$ = 21) \citep{Mueller,Castorina}, or from BOSS LRGs  ($\sigma_{f\textrm{NL}}$ = 29) \citep{Cabass22,DAmico22}.\footnote{Note that the constraining power is improved by 20\% from the inclusion of the bispectrum.} Analyses using photometric galaxy clustering have generally obtained similar precision measurements ($\sigma_{f\textrm{NL}} \sim 20$), although systematic errors can be more challenging than in the 3D case \citep{Slosar08,Xia11,Nikoloudakis12,Pullen13,Leistedt_Peiris,Giannantonio14a,Leistedt14,Ho15}. In particular, \cite{McCarthy22} and \cite{Giannantonio14b} have also constrained $f_{\textrm{NL}}$ using cross-correlations between CMB lensing and large-scale structure, finding $\sigma_{f\textrm{NL}} \sim 41$ \citep{McCarthy22} and $\sigma_{f\textrm{NL}} \sim 71$ \citep{Giannantonio14b}.

 Future LSS observations can potentially obtain $\sigma_{f\textrm{NL}} < 1$
 \citep{Seljak09,Dore14,Yamauchi14,Karagiannis18,SchmittfullSeljak,Ferraro20,Gualdi21,Schlegel22}.

Measuring galaxy clustering on very large scales is challenging. The fluctuations are small, and systematic effects can be quite significant, leading to significant added noise power on large scales. Indeed, all LSS $f_{\textrm{NL}}$ constraints to date have been systematics-limited. CMB lensing cross-correlations 
\citep{Pullen16,Doux18,Singh20,Hang21,Kitanidis21,White22,Chen22,Krolewski21,Darwish21,Omori2018b}
are therefore an attractive alternative to galaxy auto-correlations.
Most sources of noise are uncorrelated between galaxy and CMB surveys \citep[e.g.][]{Smith2007,Hirata2008,Chang10,Rhodes13}. Hence, galaxy survey systematics add a noise bias to the galaxy autocorrelation, but only add noise to the cross-correlation, since the cross-correlation covariance is proportional to the square root of the autocorrelation.
Therefore, cross-correlations can be quite useful for controlling large-scale systematics
\citep{Rhodes13,Giannantonio14b}.
CMB lensing is particularly sensitive to high redshifts, where the $f_{\textrm{NL}}$ signal is stronger. Last, the Planck CMB lensing map covers 70\% of the sky, critical for reducing cosmic variance on the largest scales. Ref.~\cite{SchmittfullSeljak} forecast that LSST-CMB lensing cross-correlations can reach $\sigma_{f\textrm{NL}} < 1$, taking into account sample-variance cancellation for future low-noise surveys.

In this paper, we cross-correlate DESI quasar targets with Planck 2018 CMB lensing to constrain $f_{\textrm{NL}}$ solely from the cross-correlation, neglecting information from the highly-contaminated auto-correlation. In Section~\ref{sec:theory}, we describe the theory necessary to compute angular correlation functions at low $\ell$. In Section~\ref{sec:data}, we describe the quasar and CMB lensing data, and in Section~\ref{sec:ps}, we describe our angular power spectrum pipeline. In Section~\ref{sec:mock_tests}, we validate our pipeline on mocks, including contaminated mocks to verify that we do not overcorrect when mitigating imaging systematics. Finally, in Section~\ref{sec:measurement}, we present the results, and in Section~\ref{sec:discussion} we compare them to previous $f_{\textrm{NL}}$ constraints.
Throughout this paper, we fix the cosmological parameters
to the Planck 2018  flat $\Lambda$CDM model \citep{PlanckLegacy18} with $h = 0.6766$, $A_s = 2.105 \times 10^{-9}$, $n_s = 0.9665$, $\Omega_m = 0.3096$, $\Omega_b = 0.049$, one neutrino with mass 0.06 eV, and $\sigma_8 = 0.8102$.

\section{Theory}
\label{sec:theory}

Local primordial non-Gaussianity can be parameterized by an additional
term in the primordial potential, proportional to $f_{\rm NL}$ \citep{KomatsuSpergel01}
\begin{equation}
    \Phi_{\rm NG} = \phi_{\rm G} + f_{\rm NL}(\phi_{\rm G}^2 - \langle \phi_{\rm G}^2\rangle)
\label{eqn:fnl_def}
\end{equation}
where $\phi_{\rm G}$ is the initial potential.
Correspondingly, non-Gaussianity also increases the density
\begin{equation}
    \delta_{\rm NG} = \delta_{\rm G} + 2 f_{\textrm{NL}} \phi_G \delta_G
\end{equation}
which comes from taking the Laplacian of Eq.~\ref{eqn:fnl_def} and setting $\nabla \phi_{\rm G} = 0$ since the first derivative is zero at a peak.

By changing the height of rare peaks, $f_{\rm NL}$
modifies the response of the halo abundance
to a background long-wavelength mode, i.e.\ the halo bias \citep{Dalal08,Slosar08,SeljakDesjacques}.
At the threshold for collapse $\delta_c$, the peak height is enhanced by $2 f_{\textrm{NL}} \phi_G \delta_c$. Therefore, defining the relationship between the late-time density field and the primordial potential with $\delta(k) = \alpha(k) \phi_G(k)$, the bias is enhanced by
\begin{equation}
    \Delta b(k) = b_\phi \frac{f_{\textrm{NL}}}{\alpha(k)} = 2 (b - p) f_{\rm NL} \frac{\delta_c}{\alpha(k)}
\label{eqn:delta_b}
\end{equation}
where the response $p=1$ for a mass-selected sample (hence $b-p$ is simply the Lagrangian bias).
For a sample dominated by recent mergers, $p=1.6$ is more appropriate.

The Laplacian in Fourier space becomes $k^2$. Hence $\alpha(k)$ is proportional to $k^2$ and $\Delta b(k)$ is proportional to $k^{-2}$, leading to a distinctive upturn in halo clustering on very large scales.
More specifically, $\alpha(k)$ is given by
\begin{equation}
\alpha(k) = \frac{2k^2 T(k) D(z)}{3 \Omega_m} \frac{c^2}{H_0^2} \frac{g(0)}{g({\infty})}
\end{equation}
where $T(k)$ is the transfer function, $D(z)$ is the linear growth factor normalized to 1 at $z = 0$, and $g(z) = (1+z)D(z)$. $\alpha(k)$ is evaluated using the CMB normalization \citep{Mueller19}, in
which $\Phi$ refers to the (constant) potential in the matter-dominated era, not the potential at $z = 0$, which is lower
by a factor of $\sim 1.3$ due to dark energy.

It has been argued the $p = 1$ is appropriate for luminous red galaxies and star-forming emission line galaxies,
and $p = 1.6$ is appropriate for quasars, which may result from mergers \citep{Castorina,Mueller}.
Following Ref.~\cite{Castorina,Mueller}, in Eq.~\ref{eqn:delta_b} we use $\delta_c= 1.686$, fix $p$ to a fiducial value of 1.6, and also test $p=1$. In deriving Eq.~\ref{eqn:delta_b}, we have assumed universality of the halo mass function to relate $b_\phi$ to the linear bias. Recent work finds that a significantly different form of the $b_\phi(b_1)$ relation may be possible for various galaxy samples selected from hydrodynamical simulations \citep{Barreira22,Biagetti16,Baldauf16,Barreira21,Barreira21b}. In light of these theoretical
uncertainties, this work may be regarded as placing constraints on $b_\phi f_{\textrm{NL}}$, and further theoretical work or additional data is needed to break the $b_\phi f_{\textrm{NL}}$ degeneracy for the tracer of interest.

We probe $\Delta b(k)$ using the matter-galaxy cross-power spectrum, $P_{gm}(k)$. Specifically, since CMB lensing
is a 2-dimensional projected field, our observable is the angular cross-power spectrum
\begin{equation}
C_{\ell}^{\kappa g} = \frac{2}{\pi} \int dz_1 \frac{dN}{dz} b(z_1) \int dz_2 W_\kappa(z_2)
\int \frac{dk}{k} k^3 P_{mm}(k, z_1, z_2) j_\ell(k\chi(z_1)) j_\ell(k\chi(z_2))
\label{eqn:gg_full}
\end{equation}
where $dN/dz$ is the quasar target redshift distribution, and $W_\kappa(z)$ is the CMB lensing kernel
\begin{equation}
   W_\kappa(z) =  \frac{3 \Omega_m H_0^2 (1+z)}{2 H(z)} \frac{\chi(z) (\chi_\star - \chi(z))}{\chi_\star}
\end{equation}
with $\chi_\star$ the distance to the surface of last scattering.

When adding scale-dependent bias due to $f_{\textrm{NL}}$, we substitute $b(z)$ with $b(z) + \Delta b(k,z)$ using
Eq.~\ref{eqn:delta_b}.
This assumes a linear galaxy bias, which is a valid assumption over the range of scales the we consider,
correponding to $k < 0.1$ $h$ Mpc$^{-1}.$
We fix the redshift evolution of the bias to the functional form of Ref.~\cite{Laurent17}, $b_L(z)$ (similar to other fits from \cite{Croom05,Chehade}), and scale it with a free amplitude $b_0$
\begin{equation}
    b(z) = b_0 b_L(z) = 0.278 b_0 \left((1 + z)^2 - 6.565\right) + 2.393
\label{eqn:laurent_b}
\end{equation}
We also investigate alternatives for the bias evolution in Section~\ref{sec:measurement} and Appendix~\ref{sec:highz}.

We also consider the contributions from lensing magnification and redshift-space distortions.
In lensing magnification, foreground structure
changes the background number density. This is correlated with CMB lensing, leading to an extra contribution to the cross-correlation:
\begin{equation}
    C_\ell^{\kappa \mu} = \frac{2}{\pi} \int dz_1 W_\mu(z_1) \int dz_2 W_\kappa(z_2)
\int \frac{dk}{k} k^3 P_{mm}(k, z_1, z_2) j_\ell(k\chi(z_1)) j_\ell(k\chi(z_2))
\label{eqn:magbias}
\end{equation}
where the magnification bias window function $W_\mu(z)$ is
\begin{equation}
   W_\mu(z) = (5s-2) \frac{3 \Omega_m H_0^2 (1+z)}{2 H(z)} \int_z^{z_\star} dz' \frac{dN}{dz} \frac{\chi(z') (\chi(z) - \chi(z'))}{\chi(z)}
\end{equation}
The inner integral runs over the distribution of galaxies, and $s \equiv \frac{d\log_{10}{n}}{dm}$ \citep{Scranton05} is the response
of the number density $n$ to achromatic changes in the brightness $dm$.

Redshift space distortions also contribute to the observed number counts
\begin{equation}
    C_\ell^{\kappa \textrm{RSD}} = \frac{2}{\pi} \int dz_1 f(z_1) \frac{dN}{dz} \int dz_2 W_\kappa(z_2)
\int \frac{dk}{k} k^3 P_{mm}(k, z_1, z_2) j_\ell(k\chi(z_1)) j_\ell''(k\chi(z_2))
\end{equation}
where $f(z)$ is the logarithmic derivative of the growth rate with respect to scale factor, $f \equiv \frac{d\ln{D}}{d\ln{a}} \approx \Omega_m(z)^{0.55}$, and the spherical Bessel function in Eqs.~\ref{eqn:gg_full} and~\ref{eqn:magbias} is replaced with its second derivative.
Then our final model is 
\begin{equation}
    C_{\ell} = C_{\ell}^{\kappa g} + C_{\ell}^{\kappa \mu} + C_{\ell}^{\kappa \textrm{RSD}}
\label{eqn:total_model}
\end{equation}
The only sensitivity to $f_{\textrm{NL}}$ is through $C_\ell^{\kappa g}$, as neither
$C_{\ell}^{\kappa \mu}$ nor $C_{\ell}^{\kappa \textrm{RSD}}$ depend on the galaxy bias.

\begin{figure}
    \centering
    \includegraphics[width=1.0\linewidth]{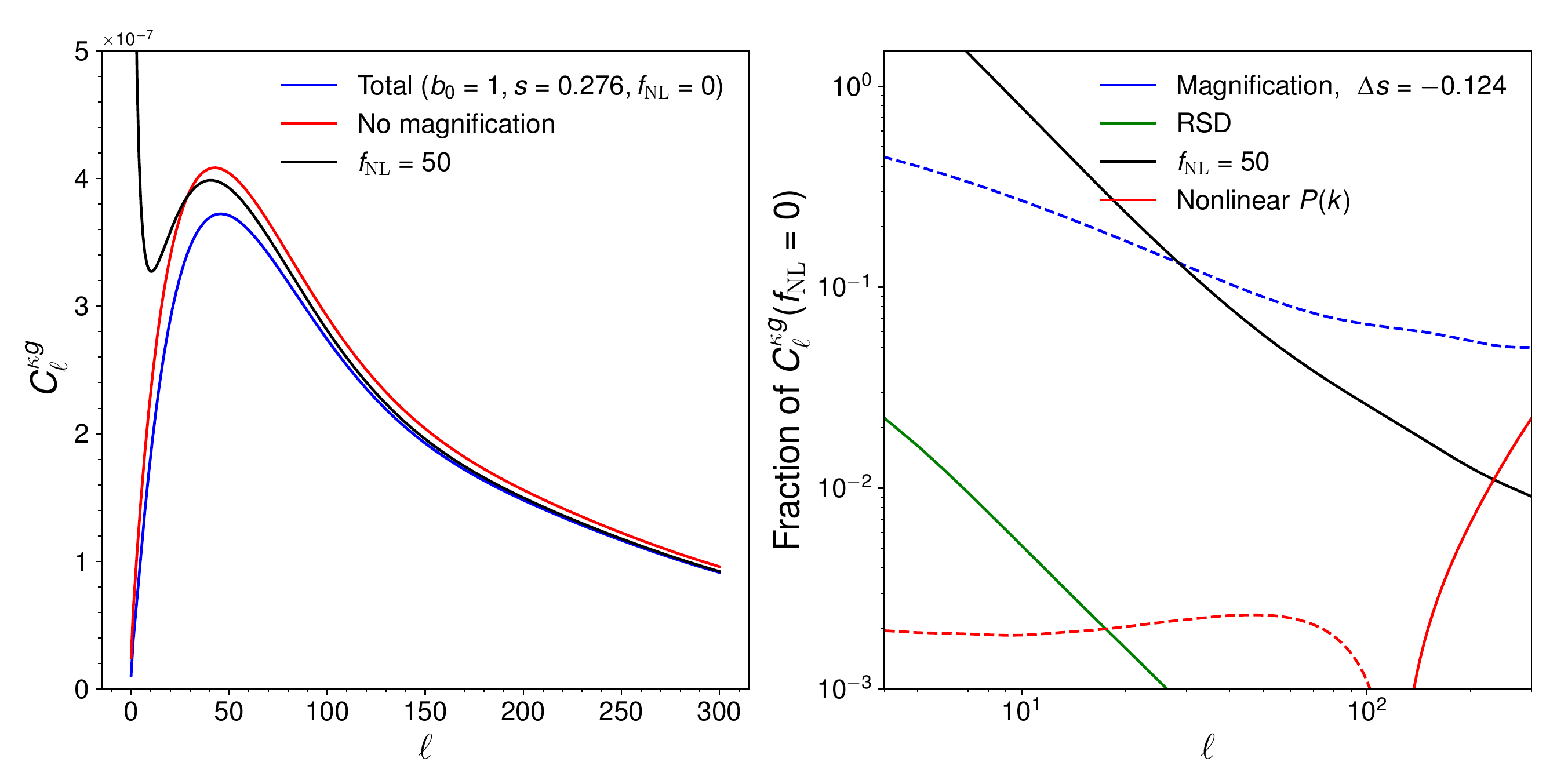}
    \caption{\textit{Left:} Total quasar-CMB lensing cross-correlation in the fiducial model (blue), without lensing magnification (red), and with $f_{\textrm{NL}} = 50$ (black). \textit{Right:} Contributions of each term to Eq.~\ref{eqn:total_model} as a fraction of the fiducial model with $f_{\textrm{NL}} = 0$.
    Negative terms are shown as dashed lines.}
    \label{fig:components_of_clkg}
\end{figure}

Fig.~\ref{fig:components_of_clkg} shows the fractional contributions of magnification and RSD (using the fiducial values of $b_0 = 1$ and $s = 0.276$), compared to $f_{\textrm{NL}} = 50$.
The magnification term is a considerable fraction ($\sim 10\%$) of the clustering term, and rises in a scale-dependent way that is approximately degenerate with $f_{\textrm{NL}}$ at $\ell > 30$, emphasizing the importance of low multipoles. 
Despite this degeneracy, the magnification bias slope is measured sufficiently accurately
that it does not worsen the $f_{\textrm{NL}}$ constraints (see discussion at end of Section~\ref{sec:measurement}).
The RSD term is very subdominant, only rising to 1\% of the fiducial model at $\ell < 6$. The nonlinear term is only non-negligible at $\ell > 200$, and it smallness shows that the measurement is well within the linear regime, using large scales at high redshift.

Equation~\ref{eqn:gg_full} is a numerically challenging three-dimensional integral over oscillatory spherical Bessel functions.
The Limber approximation \citep{Limber53} $j_\ell(k \chi) \rightarrow
\sqrt{\frac{\pi}{2 \ell + 1}} \delta_D(\ell + 1/2 - k\chi)$ reduces the three-dimensional integral to a one-dimensional integral in $k$ or $\chi$, but is not valid on the very large angular scales that we must
model to constrain $f_{\textrm{NL}}$.
We simplify the problem by splitting the power spectrum into linear and nonlinear regions, $P(k) = P_{\textrm{lin}} + (P_\textrm{NL}-P_\textrm{lin})$ \citep{Fang21}.
The redshift and $k$ dependence of the linear power spectrum are separable, $P(k,z) = P(k)D^2(z)$,
allowing us to perform the $z_1$ and $z_2$ integrals as 1D integrals, and subsequently the $k$ integral. We handle the spherical Bessel functions
using the FFTLog algorithm \citep{Fang21}. The nonlinear part is negligible until $\ell > 200$, where the Limber approximation works very well.

\section{Data}
\label{sec:data}


\subsection{Quasar target selection}
\label{sec:qso_target_selection}

DESI is a redshift survey targeting 40 million luminous red galaxies, emission line galaxies, and quasars, mainly at $0.5 < z < 2$ \citep{Aghamousa16,DESI_Instrument}. DESI measures redshifts using 5000 robotically-positioned fibers across a 7.5 deg$^2$ field of view \citep{DESI16b,Silber23,Miller23}. Three spectrographs with $R \sim 2000-4000$ span 360---980 nm.
DESI's goal is to determine the nature of dark energy
through the most precise measurement
of the Universe's expansion history ever made \citep{Levi13}.
We use data drawn from both the DESI Early Data Release (EDR) \citep{DESI23b} and from the first two months of first year release (DR1). We validate the redshifting and quasar classification pipeline
with a small ($\sim$1500) sample of visually inspected quasars from DESI Survey Validation \citep{DESI23a}, and check these results using the first two months of DR1 over a much wider area with a larger number of targets.
DR1 redshift catalogs beyond the first two months of observing are blinded; we therefore only use the first two months of observations.

\begin{figure}
    \centering
    \includegraphics[width=0.45\linewidth]{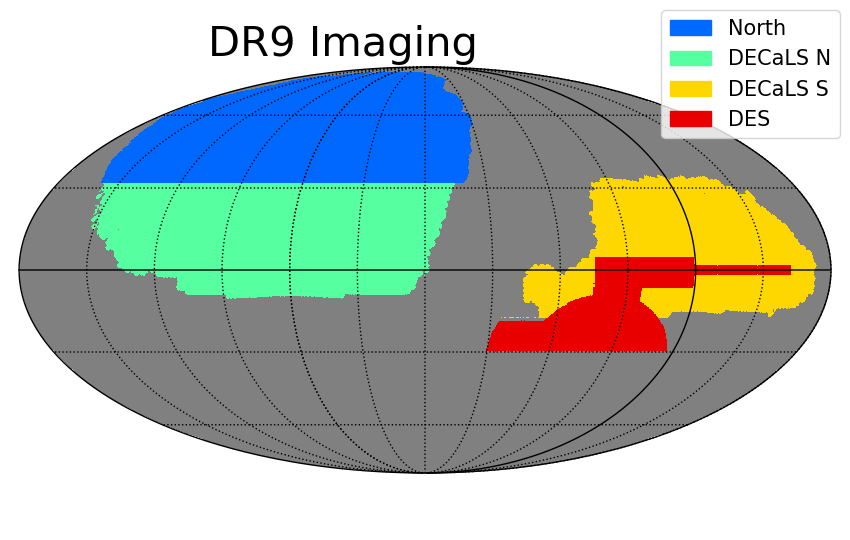}
     \includegraphics[width=0.45\linewidth]{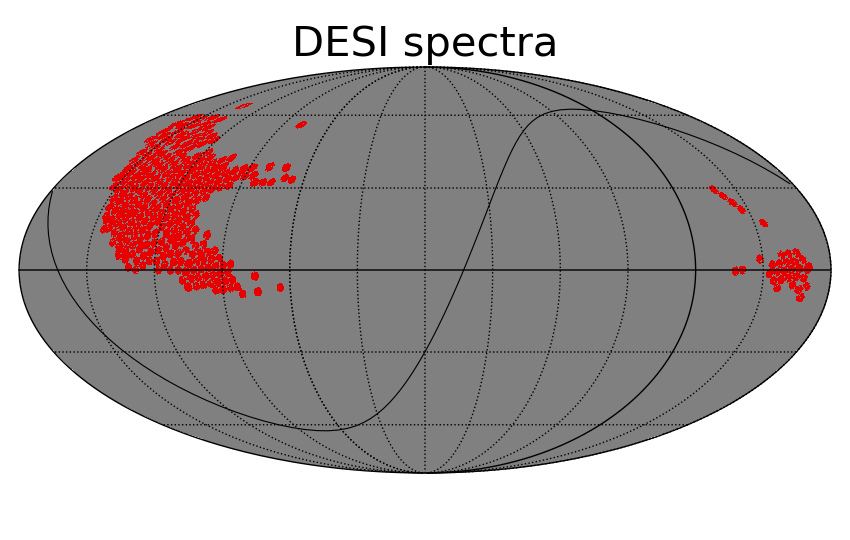}
    \caption{\textit{Left:} DR9 imaging regions. \textit{Right:} Sky coverage of early DESI spectroscopy used to measure the redshift distribution in Fig.~\ref{fig:dndz}.}
    \label{fig:imaging_plus_spec}
\end{figure}

DESI targets quasars as a direct tracer of large-scale structure, as well as backlights for the Ly$\alpha$ forest at $z > 2.1$. Quasars are selected from DR9 of the Legacy Imaging Surveys \citep{Zou17,Dey19,Schlegel22b} covering nearly 20,000 deg$^2$ in $g$, $r$, and $z$ bands. The Legacy Surveys consist of three independent programs, the Beijing-Arizona Sky Survey (BASS) and the Mayall z-Band Legacy survey (MzLS) in the north; and the DECam Legacy Survey (DECaLS) in the south.
Due to varying depths and the non-contiguity imposed by the Galactic plane, we divide the sky into four independent regions
(Fig.~\ref{fig:imaging_plus_spec}). In the north ($\delta > 32.375^{\circ}$), BASS observed $g$ and $r$ over 5100 deg$^2$ using the 2.3-meter
Bok telescope \citep{Zou17}.
$z$ band observations in this region come from MzLS \citep{Silva16}, using the 4-meter Mayall telescope. In the south, DECaLS include the publicly available Dark Energy Survey imaging across 5000 deg$^2$ \citep{Abbott_2021} using the Dark Energy Camera (DECam) \citep{Flaugher15}. DECaLS subsequently expanded the southern imaging to encompass the entire southern DESI footprint. Because DES was a dedicated weak lensing survey with separate goals, it is substantially deeper than the DECaLS imaging, with typically 4 or more passes of the sky rather than 3. In addition to the optical data, infrared data in W1 (3.4 $\mu$m) and W2 (4.6 $\mu$m) from the all-sky Wide-field Survey Explorer (WISE) \citep{Wright10} is critical for separating stars and quasars. The unWISE coadds \citep{Meisner17} use \textit{Tractor} \citep{Lang16} forced photometry to improve and deblend WISE flux measurements for optical sources. 

The differences in surveys and imaging quality lead us to consider the quasar sample as consisting of 4 disjoint areas: North (MzLS + BASS), DECaLS-North (the northern galactic cap observed by DECaLS), DECaLS-South (the southern galactic cap observed by DECaLS), and DES. As DECaLS-North and DECaLS-South are drawn from the same underlying imaging catalog, they are sometimes considered as one area. However, the stellar density and extinction are quite different between the two regions, and they are also calibrated separately. As a result, each region may show a different relationship between observed quasar density and observational properties such as depth or seeing. In order to test against the impact of these varying imaging systematics, we consider DECaLS-North and DECaLS-South separately.

We also find an unexplained excess power in quasar targets in the DES region, which is not mitigated by systematics weights. 
This excess power is driven by visually apparent large fluctuations in the quasar density field at $\delta < -30^{\circ}$. The $\delta < -30^{\circ}$ part of DES was found to have slightly stronger systematics trends than the entire DES footprint in Ref.~\cite{Chaussidon22}. 
This is due to a known shift of approximately 0.02 mag in the $z$-band, where the calibration method transitions from Pan-STARRS1 to Ubercal \citep{Padmanabhan08,Schlegel22b}.
We find that excluding the 60\% of the DES footprint at $\delta < -30^{\circ}$ reduces the excess low $\ell$ power (Fig.~\ref{fig:clgg}), and therefore define DES as only including the region above $\delta = -30^{\circ}$.

DESI uses quasars' near-infrared excess to separate them from stars, using 5 band photometry ($grzW1W2$) \citep{Myers15}. Quasar target selection is extensively described in Ref.~\cite{Chaussidon22b},
a preliminary version is described in Ref.~\cite{Yeche20}, and target selection for the entire survey is described in Ref.~\cite{Myers23}.
Angular clustering of quasar targets is measured in Ref.~\cite{Chaussidon22}.
To maximize the number of quasar targets, a machine-learning Random Forest (RF) algorithm is used to separate stars from quasars.
Before applying the random forest algorithm,
basic color cuts are applied ($16.5 < r < 23$, W1$<22.3$, W2$<22.3$), and targets are restricted to point-like sources ('PSF') in DR9 imaging.
The RF algorithm is trained on equal numbers of quasars and stars (332,650 each), using 11 input parameters: the 10 possible colors and $r$ band magnitude. The RF threshold $p$ is a function of $r$ magnitude and varies slightly within the 3 imaging regions North, DECaLS (non-DES) and DES: $p(r) = \alpha - \beta \tanh(r - 20.5)$, with $(\alpha, \beta)$ equal to (0.88, 0.04), (0.7, 0.05) and (0.84, 0.04), respectively. 
After selecting quasar targets, we additionally remove targets with any of maskbits\footnote{\url{https://www.legacysurvey.org/dr9/bitmasks/}} 1, 8, 9, 11, 12, or 13 set (near optical or WISE bright stars, SGA\footnote{\url{https://www.legacysurvey.org/sga/sga2020/}} large galaxies, or globular clusters).
The density of quasar targets is strongly dependent on imaging properties, and \cite{Chaussidon22} develops a method to correct for these systematic fluctuations and validates the angular correlation function of the quasar targets. The DESI target selection pipeline is described and documented in \cite{Myers22}.

\begin{figure}
    \centering
    \includegraphics[width=0.45\linewidth]{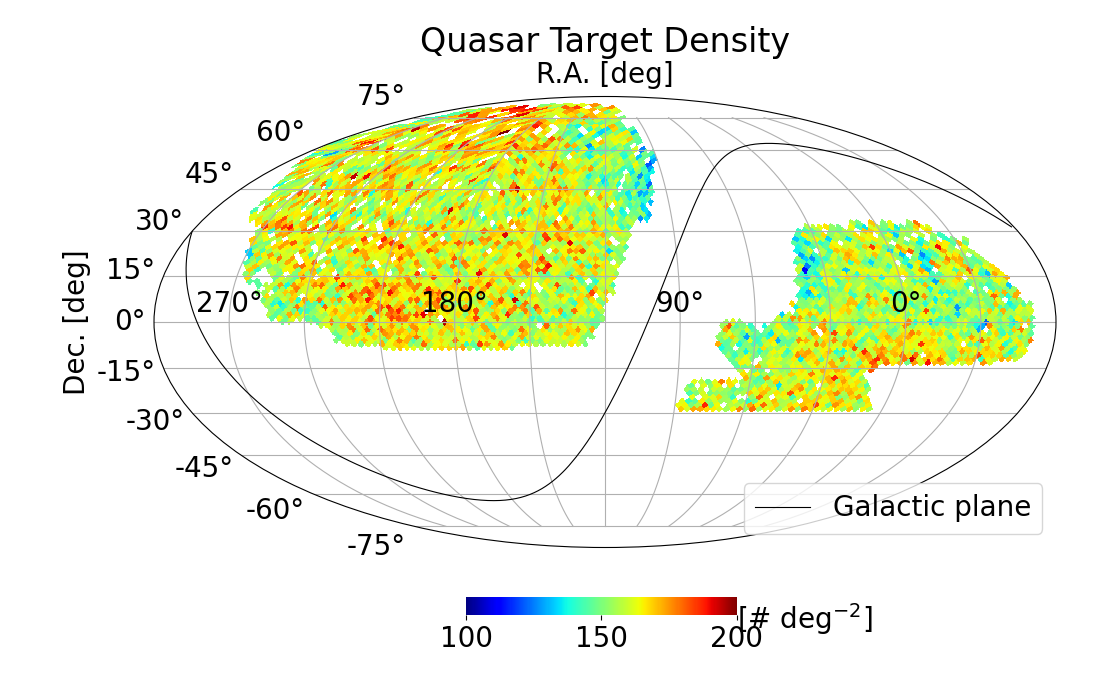}
    \includegraphics[width=0.45\linewidth]{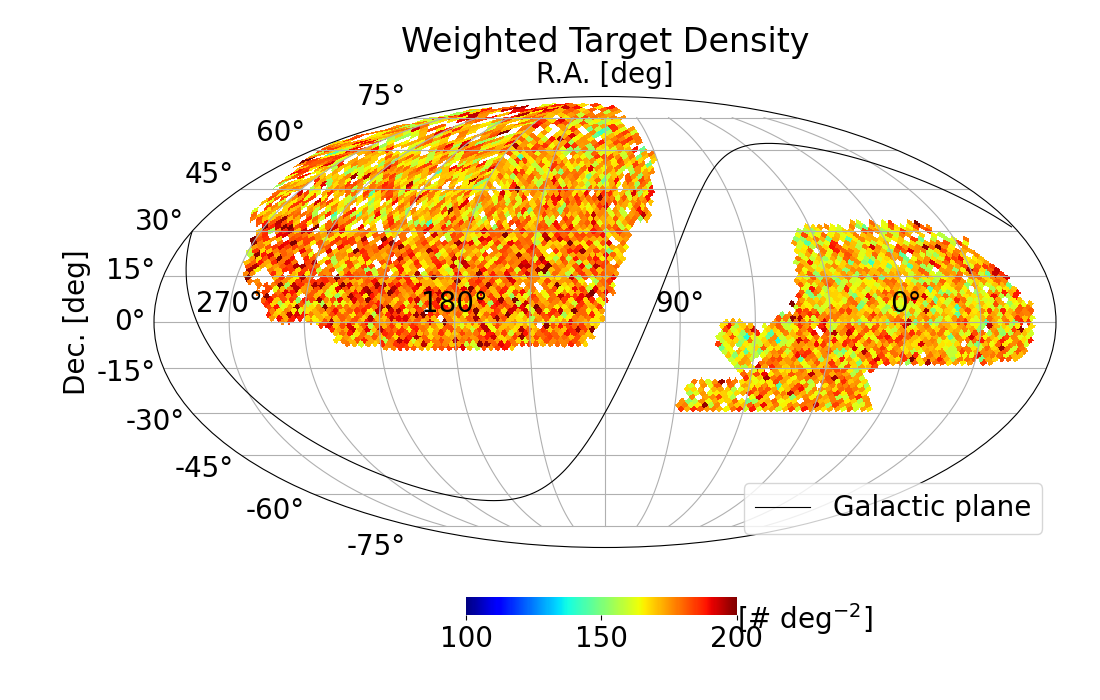}
   \includegraphics[width=0.45\linewidth]{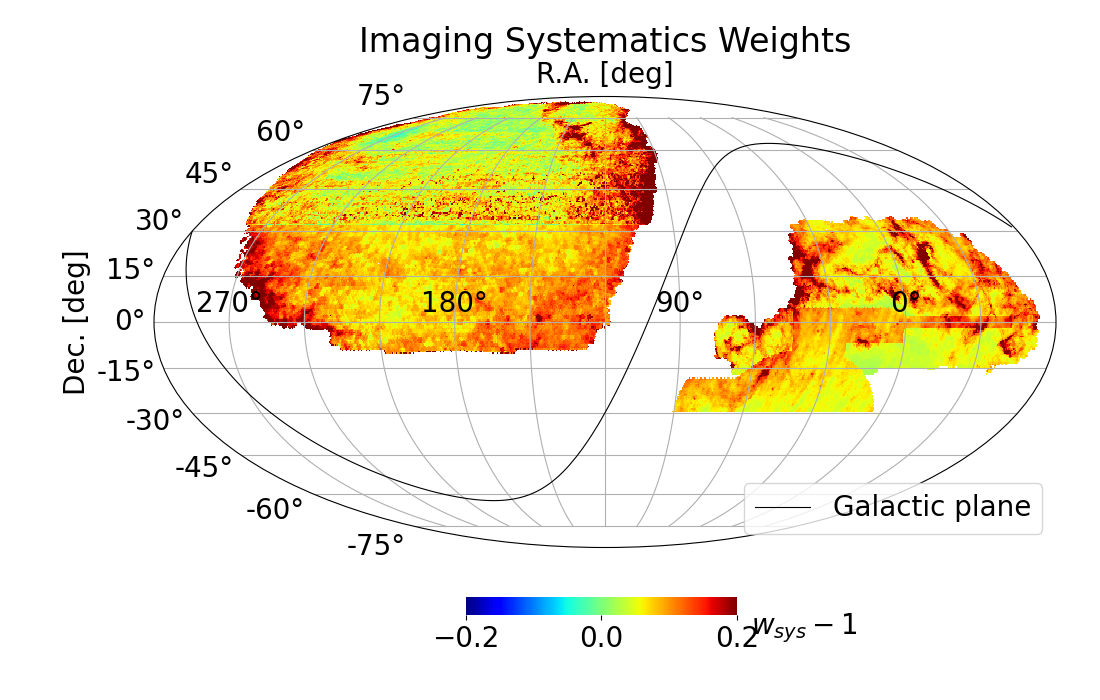}
    \includegraphics[width=0.45\linewidth]{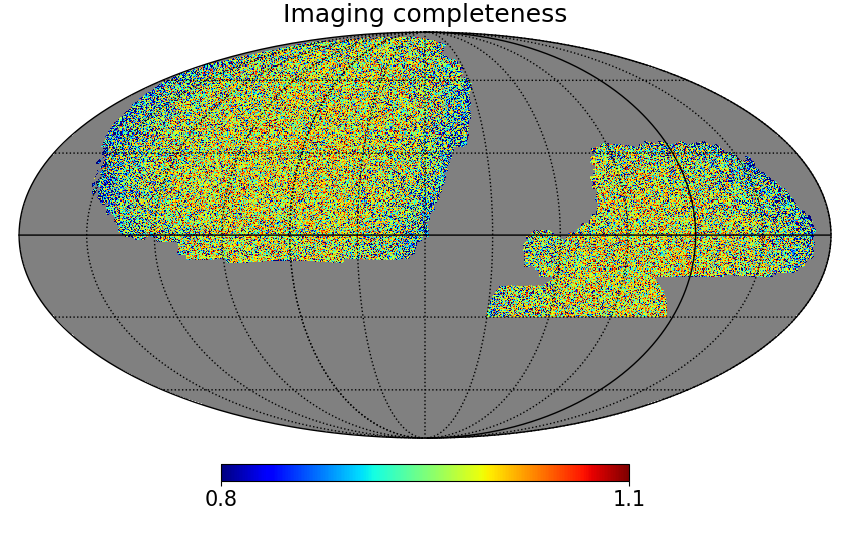}
    \caption{\textit{Top:} Quasar targets before (\textit{left}) and after (\textit{right}) applying imaging systematics weights (\textit{bottom left}), all in equatorial coordinates. \textit{Bottom right:} Fraction of each pixel within the DR9 imaging mask, as computed from the number density of random points.}
    \label{fig:skymaps}
\end{figure}

Sky maps of the quasar targets and imaging systematic weights are shown in Fig.~\ref{fig:skymaps}, and key statistics for the quasar sample are given in Table~\ref{tab:qso_summary}. The angular extent of the usable imaging regions is represented by a random catalog
with density 45000 deg$^{-2}$.
Since the random catalog has features on sub-pixel scales (e.g.\ excluded areas around moderately bright stars), we measure the random density in NSIDE=2048 HEALPixels \citep{Gorski05} to create an imaging completeness mask. We also create binary masks for each of the four imaging regions, and set any pixel with completeness $<80\%$ to zero in the binary masks. To compute the density in each pixel, we divide the quasar counts by the imaging completeness mask in each pixel. After removing low-completeness areas, the binary imaging masks cover 14,719 deg$^2$ (35.7\% of the sky). 

Imprints from the various feature maps are clearly visible in the map of systematic weights in Fig.~\ref{fig:skymaps}.
In DECaLS S, regions of high extinction are clearly visible as up-weighted regions; another high-extinction region is visible in the North, near RA 150$^{\circ}$, Dec 75$^{\circ}$.
Striping in ecliptic longitude is visible in the DES region, from the WISE scan strategy.
Close to the boundary between them, the systematics weights in DeCALS N appear to have much lower resolution than in the North. This is because the systematics weights are dominated by imaging depth and PSF size, which vary from field to field. The northern weights are dominated by PSFISZE\_Z, which comes from the relatively small field of view (0.4 deg$^2$) MzLS survey \citep{Dey16}.
In contrast, DECam has a 3.2 deg$^2$ field of view \citep{Dey19}, leading to much coarser structure in DECaLS N.

We use DESI spectroscopy from the first two months of DR1 to measure the redshift distribution of the quasar target sample (Fig.~\ref{fig:dndz}). The quasar redshift measurement is based on the automated RedRock pipeline \citep{Bailey23}
supplemented by MgII and QuasarNet \citep{Busca18} afterburners (Fig.\ 9 in \cite{Chaussidon22} describes the path to a successful quasar redshift). This pipeline has been validated for high purity (99\%) and completeness (95\%) with a visually inspected sample of quasars from 3 deep Survey Validation tiles with 7000--9000 s exposure time \citep{Alexander22}. These tiles are 7--9 times deeper than the Main Survey tiles, allowing both for true redshifts to be definitively identified, and for the creation of Main Survey like exposures to study the pipeline's performance.

\begin{table}[]
    \centering
    \begin{tabular}{c|cccc}
       Region  &  Density (deg$^{-2}$) & Shot Noise & $f_{\rm sky}$ & Area (deg$^2$) \\
       \hline
       North  &  161 & $1.89 \times 10^{-6}$ & 0.1101 & 4542 \\
       DECaLS N & 163 & $1.87 \times 10^{-6}$ & 0.1248 & 5148 \\
       DECaLS S & 156 & $1.95 \times 10^{-6}$ & 0.0787 & 3247 \\
     DES & 174 & $1.75 \times 10^{-6}$ & 0.0432 & 1782 \\
       \hline
       High Purity & 162 & $1.88 \times 10^{-6}$ & 0.3568 & 14719 \\
       \hline
       Main & 302 & $1.01 \times 10^{-6}$ & 0.4145 & 17099
    \end{tabular}
    \caption{Density of quasars and unmasked area in each imaging region. The bottom two lines give the average density and total area covered by the High Purity sample, and the same statistics for the Main sample used in Ref.~\cite{Chaussidon22}. The areas differ because high purity includes our cut on DES to exclude $\delta < -30^{\circ}$, whereas main includes the entire DES region.}
    \label{tab:qso_summary}
\end{table}

\subsection{Classifying quasar targets as quasars, stars, or galaxies}
\label{sec:classification}

The DESI quasar target selection maximizes the number density of quasars, but at the cost of a large fraction of interlopers. Within the deep visual inspection sample of main targets (1361 targets\footnote{This number differs from the 1455 main quasars quoted in Table 5 of \cite{Alexander22} because of the additional imaging bitmasks that we apply.}), 70.9\% are quasars, 16.1\% are galaxies, 6.3\% are stars, and 6.7\% are unclassified low-quality objects \citep{Alexander22}.
We also determine the quasar fraction for the automated redshifts from the first two months of DR1 (396,957 quasar targets).

The spectroscopic quasar classification pipeline is described in Section~\ref{sec:qso_target_selection} above.
Stars are defined as any object with a RedRock redshift $< 0.01$ that is not included in the quasar catalog. Galaxies are more complex to classify. Most of the quasar targets that are classified as galaxies show strong emission lines, and indeed there is substantial overlap between the quasar sample and the emission line galaxy sample. However, the galaxies show significant diversity, and a minority have strong 4000 \AA\ breaks indicative of an older stellar population. As a result, we blend components of the ELG and LRG selection criteria \citep{Raichoor22,Zhou22,Lan22}. We define a successful galaxy redshift as any object that is not a quasar or a star, which either:

1.\ meets the [OII] flux-DELTACHI2 criterion of \cite{Raichoor22}: $\rm{log}_{10} (\texttt{FOII\_SNR}) > 0.9 - 0.2 \times \rm{log}_{10} (\texttt{DELTACHI2})$;

2.\ or has $\texttt{DELTACHI2} > 30$;

3.\ or has [OIII] SNR $> 5$, for a small minority of objects that have strong [OIII] emission but not much [OII].

We validate this selection on the deep-field VI tiles,
using custom coadds reaching the exposure time of the DESI main survey  (1000 s).
We average our results across the five coadds that were created. Treating the VI classifications and redshifts as truth (and restricting to VI sources with quality $>=$ 2.5), we compute the completeness and purity of quasars, galaxies, and stars (Table~\ref{tab:vi_validation}).
We have on average 933 high-quality VI quasars, 212 galaxies, and 72 stars.
We remove targets which do not pass the spectroscopic quality flag COADD\_FIBERSTATUS.
Our results are slightly different from those of Ref.~\cite{Alexander22}, since we apply extra maskbits 8, 9, and 11 compared to the DESI main selection.

The completeness of the galaxy sample is comparable to the quasar sample, though the purity is 10\% lower. But the redshift accuracy of the 10\% mis-classified galaxies is sufficient for our purposes, and the good redshift purity is very high. It is key to note that since we are performing a 2D analysis, redshift errors of less than $\sim 0.1$ are completely negligible, while they are catastrophic to the DESI spectroscopic survey. Additionally, we do not care if the classifications of the spectra are wrong, only if their redshifts are catastrophically in error (or mis-classified as stars). Of the 116 mis-classified galaxies (summing over the five subsets), 113 are quasars and only 3 (2.6\%) are stars. The redshifts are nearly all correct; only 14 (12.1\%) are significant outliers with $|\Delta z| > 0.1$.
Therefore, the good redshift purity, or the fraction of pipeline-selected galaxies with redshifts accurate to within $|\Delta z| < 0.1$,
is 98.5\%. The pipeline redshifts match the VI redshifts extremely closely, with a median offset of -0.06 km s$^{-1}$ and median absolute deviation scaled by 1.4828 of 9 km s$^{-1}$. This is merely a statement that the pipeline and VI redshifts agree extremely well, not a quantification of the absolute redshift error. The absolute redshift error is measured in Ref.~\cite{Alexander22} by comparing DESI and SDSS redshifts, and is 60 km s$^{-1}$ across the entire quasar sample. Therefore, the agreement between the pipeline and VI redshifts for the quasar targets classified as galaxies implies that their redshift errors are typical for VI galaxies.

While we use the union of three separate galaxy cuts to create as inclusive a catalog as possible, most good redshift galaxies satisfy all three.
198 of the 211 VI galaxies have both 
$\rm{log}_{10} (\texttt{FOII\_SNR}) > 0.9 - 0.2 \times \rm{log}_{10} (\texttt{DELTACHI2})$ and $\texttt{DELTACHI2} > 30$.
Allowing for galaxies passing the $\texttt{DELTACHI2} > 30$
cut and not the $\rm{log}_{10} (\texttt{FOII\_SNR}) > 0.9 - 0.2 \times \rm{log}_{10} (\texttt{DELTACHI2})$
increases the completeness from 93.8\% to 94.8\%, and additionally
allowing for galaxies passing the emission line cut but not the $\texttt{DELTACHI2} > 30$ further increases the completeness to 95.2\%.
Within the VI sample, we do not find any galaxies with only the [OIII] SNR criterion satisfied.
However, this cut modestly increases the completeness on a separate VI sample of 86 quasar targets from the main DESI survey with our fiducial (``high purity'') flux and photometric redshift cuts applied, as described in Section~\ref{sec:qso_selec} below.

The main limitation of the pipeline classification
is that it misses $\sim 20\%$ of the stars, which are classified as bad redshifts. We therefore boost the measured stellar contamination fraction by 28\%, following Table~\ref{tab:vi_validation}.

\begin{table}[]
    \centering
    \begin{tabular}{c|ccc}
      Sample & Completeness & Purity & Good Redshift Purity \\
      \hline
      QSO & 93.9\% & 99.6\% & 100\% \\
      GAL & 94.3\% & 89.6\% & 98.5\% \\
      STAR & 78.1\% & 99.3\% & 99.3\%  \\
    \end{tabular}
    \caption{Purity and completeness of pipeline redshifts,
    measured using the deep VI spectra (with high-quality VI classification) as truth. The quasar pipeline is described in Refs.~\cite{Chaussidon22,Alexander22}. The galaxy pipeline is described in Sec.~\ref{sec:qso_selec} and merges elements of the ELG and LRG redshift success criteria. Stars are spectra with $z < 0.01$ or RedRock classification as stars. The good redshift purity is the fraction of spectra where the redshift is correct to within $\Delta z = 0.1$, regardless of the classification.}
    \label{tab:vi_validation}
\end{table}

For the main target sample, we obtain 60.2\% quasars, 5.4\% stars, 16.4\% galaxies, and 18.0\% unclassified. 
The unclassified fraction is substantially higher than in the deeper VI data. The quasar fraction is consistent given sample variance and the expectation that the contaminated pipeline recovers 95\% of true quasars \citep{Alexander22}.
The consistent object classification between DR1 and VI validates our automated star/galaxy classification.

\subsection{Selecting a cleaner quasar sample}
\label{sec:qso_selec}

The large numbers of stars and unclassified objects are not ideal for this work. Stars imprint complex angular systematics into the sample, since they both directly add to the sample and modulate the density of true quasars (e.g.\ by spuriously modulating the inferred sky background or directly occulting background quasars). Moreover, the stellar fraction of the sample is difficult to characterize, as it is a strong function of sky position (depending most critically on Galactic latitude and position relative to the Sagittarius Stream), raising the issue that the star fraction may be different in the region covered by DESI spectroscopy and the entire imaging sample (Fig.~\ref{fig:imaging_plus_spec}). Unclassified objects pose a problem because they may have a different redshift distribution than the quasars as a whole, biasing predictions for the cross-correlation.

To mitigate these issues, we impose three separate cuts beyond the DESI quasar target selection, which reduce the target density by more than a factor of two but yield a much cleaner selection. We call the full quasar sample the ``main'' sample, and the lower-density sample the ``high purity'' sample.

The high purity sample consists of three separate cuts to maximize the fraction of quasars.
First, we use the Gaussian mixture model of \cite{Duncan22} to separate stars and quasars based on their colors. We reject sources with $P_{\rm star} > 0.01$ as likely contaminants. We find that this reduces the stellar fraction to 1.0\%, increases the quasar fraction to 66.2\%, and keeps the galaxy and unclassified fractions similar at 16.7\% and 16.0\%, respectively.
Second, we increase the purity of the sample by removing roughly the faintest quarter of sources in W2, requiring W2 < 20.8. This cut yields 77.4\% quasars, 1.1\% stars, 9.5\% galaxies, and 12.1\% unclassified objects. Finally, we also remove the faintest quarter in $r$ band, requiring $r < 22.3$. This ultimately yields 90.4\% quasars, 2.0\% stars, 3.6\% galaxies, and 4.0\% unclassified.
While not strictly required, the reduction in galaxies is also helpful, since the galaxies may respond differently than the quasars to imaging systematics, and have a different bias evolution.

We repeat our measurements of the completeness, purity, and good redshift purity using the single-depth deep-field VI data for the high-purity quasar targets (Table~\ref{tab:vi_validation_high_purity}). The sample size is smaller, with (averaging over the five subsets) 660 quasars, 20 galaxies and 12 stars.
The results are largely similar to the full quasar sample;
while the galaxy purity drops to 63\%, the good redshift purity remains high at 97\%. Due to the small sample of galaxies, we additionally visually inspect 86 main DESI spectra of high-purity quasar targets which are not classified as stars or quasars. 44 of them have high-quality redshifts, and 41 are recovered by the galaxy classification of Sec.~\ref{sec:classification}.
This sample of galaxies shows more diversity than the main quasar targets, with only 31 VI galaxies (70.5\%) passing both the [OII] flux and \textsc{DELTACHI2}
cuts, compared to 93.8\% for the main sample.
We additionally obtain 6
good redshifts with \textsc{DELTACHI2} $>30$
only, 2 good redshifts with the [OII] flux cut only, and 2 good redshifts with the [OIII] signal-to-noise cut.Therefore, all three cuts are desirable to maximize the completeness of the pipeline-classified galaxies.

The spectroscopic classifications show some variation across imaging regions, with the stellar fraction ranging from 1.6\% in the North, to 2.1\% in DECaLS N, and 2.7\% in DECaLS S. 
The early DESI spectroscopy does not cover the DES region (Fig.~\ref{fig:imaging_plus_spec}).
The fraction of unclassified sources also rises with the same trend, at 3.4\% (4.0\%, 4.6\%) in North (DECaLS N, DECaLS S). As a result, the quasar fraction drops from north to south: 91.1\% (90.2\%, 88.9\%) in North (DECaLS N, DECaLS S).

Finally, we use the visual-inspection results from the full-depth deep coadds to check our measurement of the sample purity.
Of 721 sources with deep VI data, $94.7 \pm 3.6$\% are quasars, $2.8 \pm 0.62$\% are galaxies, $1.7 \pm 0.48$\% are stars, and $0.9 \pm 0.34$\% are unclassified (errors from Poisson distribution).
This is perfectly consistent with the fractions measured from the first two months of DR1, but with a much smaller fraction of unclassified objects. This implies that most unclassified objects are quasars, and they are not biasing our estimate of the stellar fraction.



\begin{table}[]
    \centering
    \begin{tabular}{c|ccc}
      Sample & Completeness & Purity & Good Redshift Purity \\
      \hline
      QSO & 97.3\% & 99.8\% & 100\% \\
      GAL & 89.0\% & 62.7\% & 96.6\% \\
      STAR & 75.0\% & 97.8\% & 97.8\%  \\
    \end{tabular}
    \caption{Same as Table~\ref{tab:vi_validation}, but for the high-purity sample.}
    \label{tab:vi_validation_high_purity}
\end{table}

\subsection{Redshift distribution}

\begin{figure}
    \centering
    \includegraphics[width=0.45\linewidth]{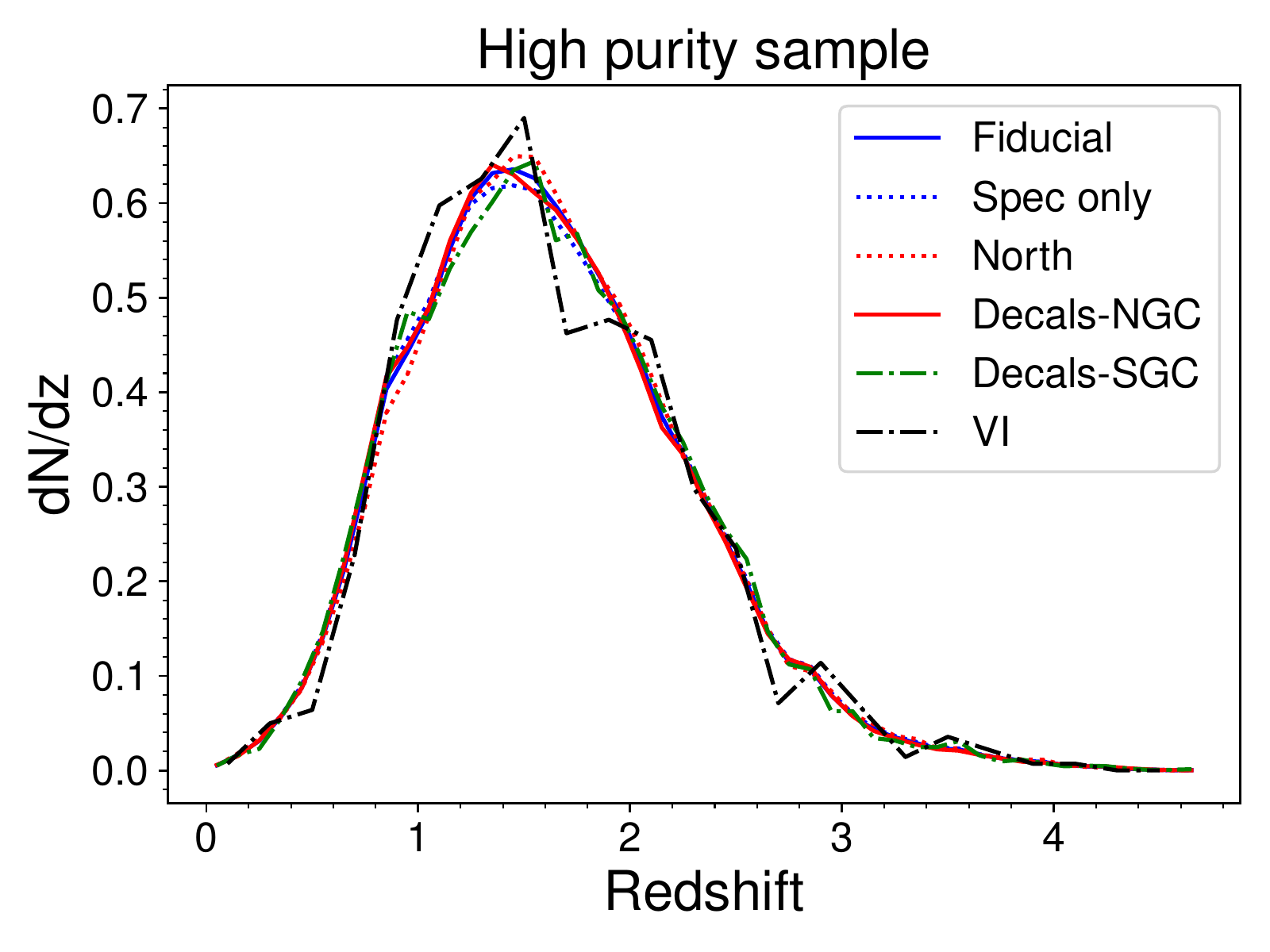}
    \includegraphics[width=0.45\linewidth]{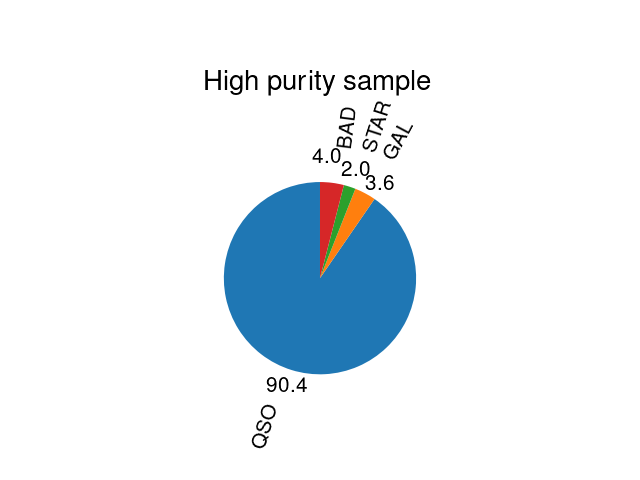}
    \caption{The redshift distribution of ``high purity'' DESI quasar targets.
    \textit{Left:}
    The solid blue line labeled ``fiducial'' uses DESI spectroscopy for targets with good redshifts, and photometric redshifts from \cite{Duncan22} for objects without good redshifts, and not classified as stars. The dashed blue line only uses DESI spectroscopy. The red and green lines show the $dN/dz$ in the separate imaging regions. The dot-dashed black line gives the $dN/dz$ from the VI deep fields. 
    \textit{Right:} Spectroscopic classification of ``high purity'' quasar targets.}
    \label{fig:dndz}
\end{figure}

\begin{figure}
    \centering
    \includegraphics[width=0.45\linewidth]{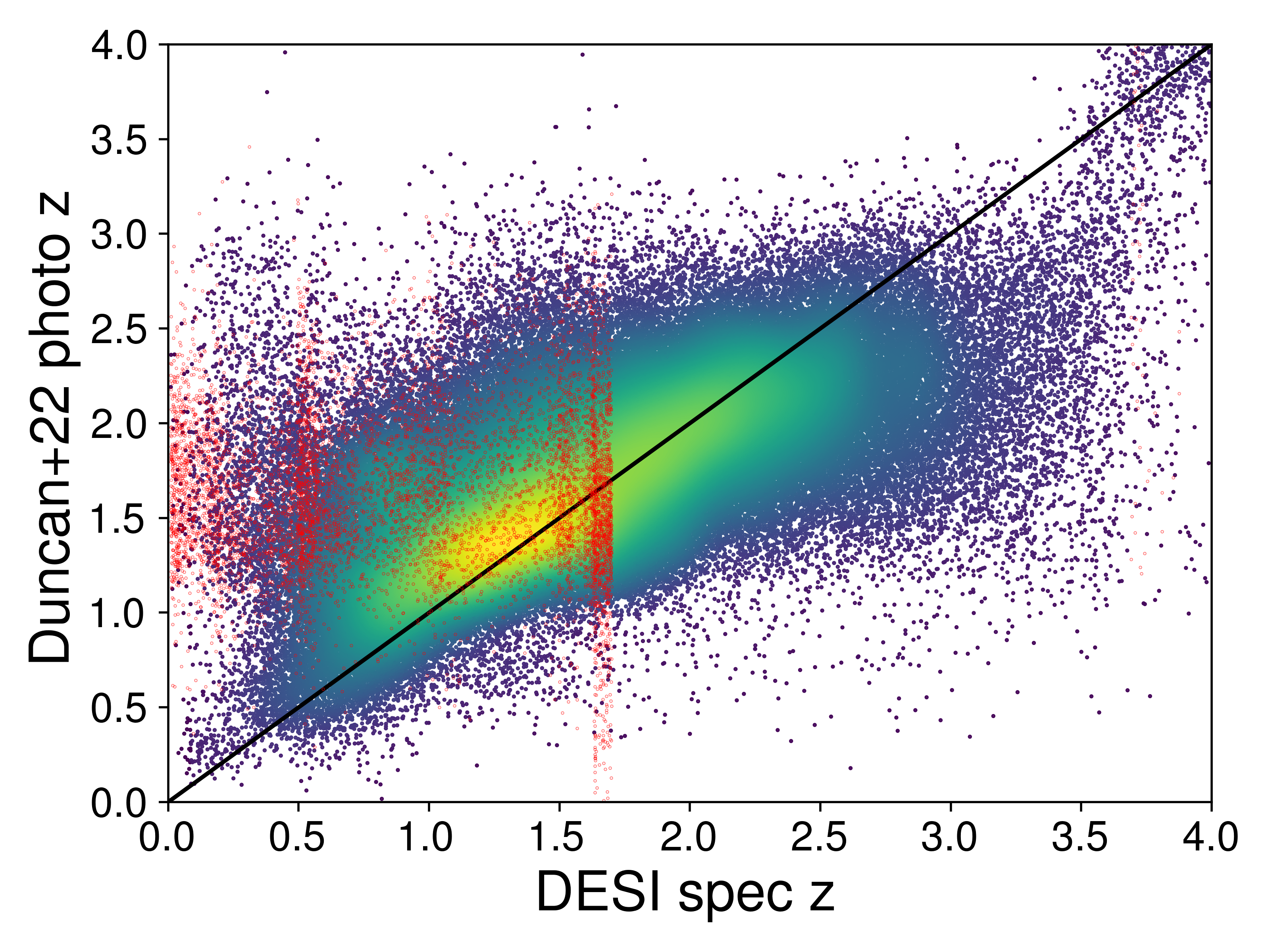}
    \includegraphics[width=0.45\linewidth]{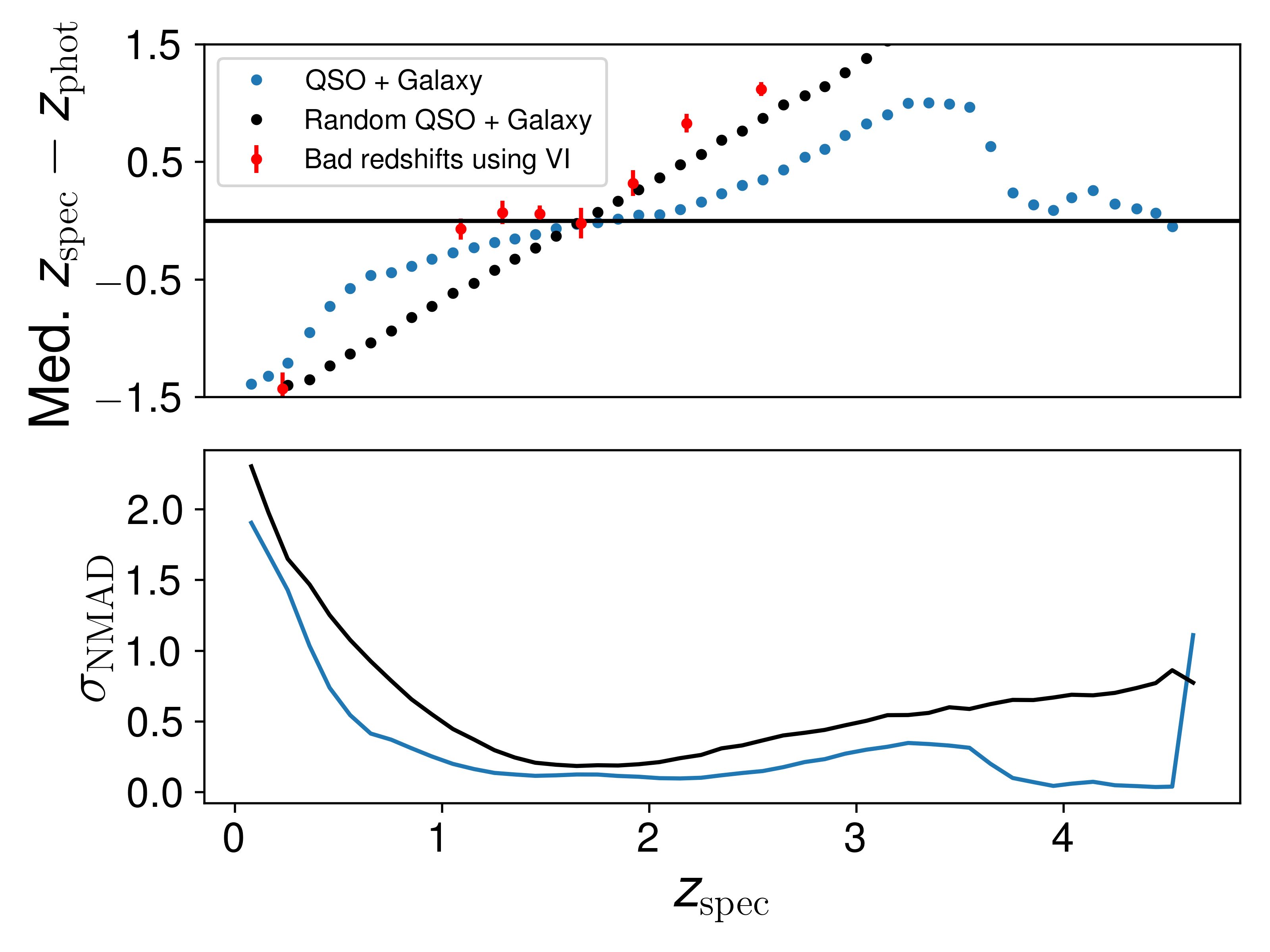}
    \caption{Comparison between DESI redshifts
    and photometric redshifts from Ref.~\cite{Duncan22}.
    \textit{Left:}
    Comparison between photometric and DESI spectroscopic
    redshifts for quasars and galaxies. The overplotted red points are objects without a good DESI redshift,
    which exhibit significant spurious structure in their DESI redshift distribution.
    \textit{Right:} Median bias and normalized median absolute deviation ($\sigma_{\rm NMAD}$) for the quasars and galaxies, and for a randomly shuffled version of the same sample. We also show the median bias for 48 VI targets classified as bad redshifts by our pipeline (not stars, galaxies, or quasars), and not identified as stars by VI, comparing the photometric redshifts to the true spectroscopic redshifts from deep VI (red points).}
    \label{fig:zphot_comparison}
\end{figure}

Fig.~\ref{fig:dndz} shows $dN/dz$ derived from DESI spectroscopy of 194,473 quasar targets passing the ``high purity'' selection cuts.
For the 4.0\% unclassified sources, we use photometric redshifts from \citep{Duncan22}, which are specifically tailored for high-redshift galaxies and quasars. 
Fig.~\ref{fig:zphot_comparison} shows how these photometric redshifts compare to the DESI spectroscopic redshifts of the quasar targets classified as quasars or galaxies, both in the redshift bias (median redshift difference) and normalized median absolute deviation, $\sigma_{\rm NMAD} = 1.48 *  \textrm{median}(|\delta z|/(1 + z_{\rm spec}))$.

The comparison in Fig.~\ref{fig:zphot_comparison} is very similar to Figs.~8 and 10 in Ref.~\cite{Duncan22}; they also find significant biases towards negative $\Delta z$ for faint objects at low redshift, and biases to positive $\Delta z$ for $2.5 < z < 4$ quasars.
We also show the same comparison for DESI redshifts that are classified as bad, showing that the bad DESI redshifts are less correlated with the photometric redshifts and are significantly bunched at known problem redshifts around $z \sim 0.5$ and $z > 1.6$. Finally, we use the deep-field VI to study the relationship between Ref.~\cite{Duncan22} photometric redshift
and true spectroscopic redshift from VI. The sample is quite small, only 64 spectra with good VI and good \textsc{COADD\_FIBERSTATUS}. 16 of these pipeline-classified galaxies are classified as stars: this is expected from the known incompleteness of the stellar classification and the relatively high stellar fraction of the main sample. This fraction will be lower for the high-purity sample, which has a lower fraction of stars to begin. For the remaining galaxies, we plot the performance of the photometric redshifts in the right panel of Fig.~\ref{fig:zphot_comparison}, finding good performance at $1 < z_{\rm spec} < 2$ where 73\% of the bad redshifts lie, and similar degradations at high and low redshift as the overall sample. This justifies our choice
to use the photometric redshifts of Ref.~\cite{Duncan22} for the 4.0\% bad redshifts.


Nevertheless, because of the small fraction of unclassified objects, the $dN/dz$
from spectroscopic DESI redshifts alone is nearly the same as the $dN/dz$ including the photometric redshifts.
We also find that the redshift distribution is very similar across the different imaging regions.
Moreover, we can use the VI data to make a noisy measurement of $dN/dz$ (due to the small number of VI sources), which is also perfectly consistent with the $dN/dz$ from the first two months of DR1.

The redshift distribution can be used to interpret the clustering measurement at an effective redshift
\citep{Modi17,Chen22}
\begin{equation}
    z_{\rm eff} = \int dz \frac{1}{\chi^2} \frac{dN}{dz} W^\kappa(z) z
\end{equation}
slightly different from the mean redshift.
The high-purity sample has an effective redshift of 1.51 using the fiducial redshift distribution.

DESI is a multi-pass survey, with quasars targeted at highest priority in the first pass. Therefore, we expect early (single-pass) DESI data to have a complete sampling of quasar redshifts.
We verify that the quasar redshifts are complete by removing 5\% (25\%) of targets with the lowest
spectroscopic signal-to-noise.
The fraction of unclassified spectra barely changes, from 4.0\% to 3.7\%, and overall the quasar fraction increases from 90.4\% to 90.5\%. Similarly, the mean redshift stays constant at 1.621 when removing the lowest 5\% (25\%) in signal-to-noise.
We conclude that the spectroscopic completeness of the early DESI data is sufficient for this work.

 \begin{figure}
    \centering
    \includegraphics[width=0.45\linewidth]{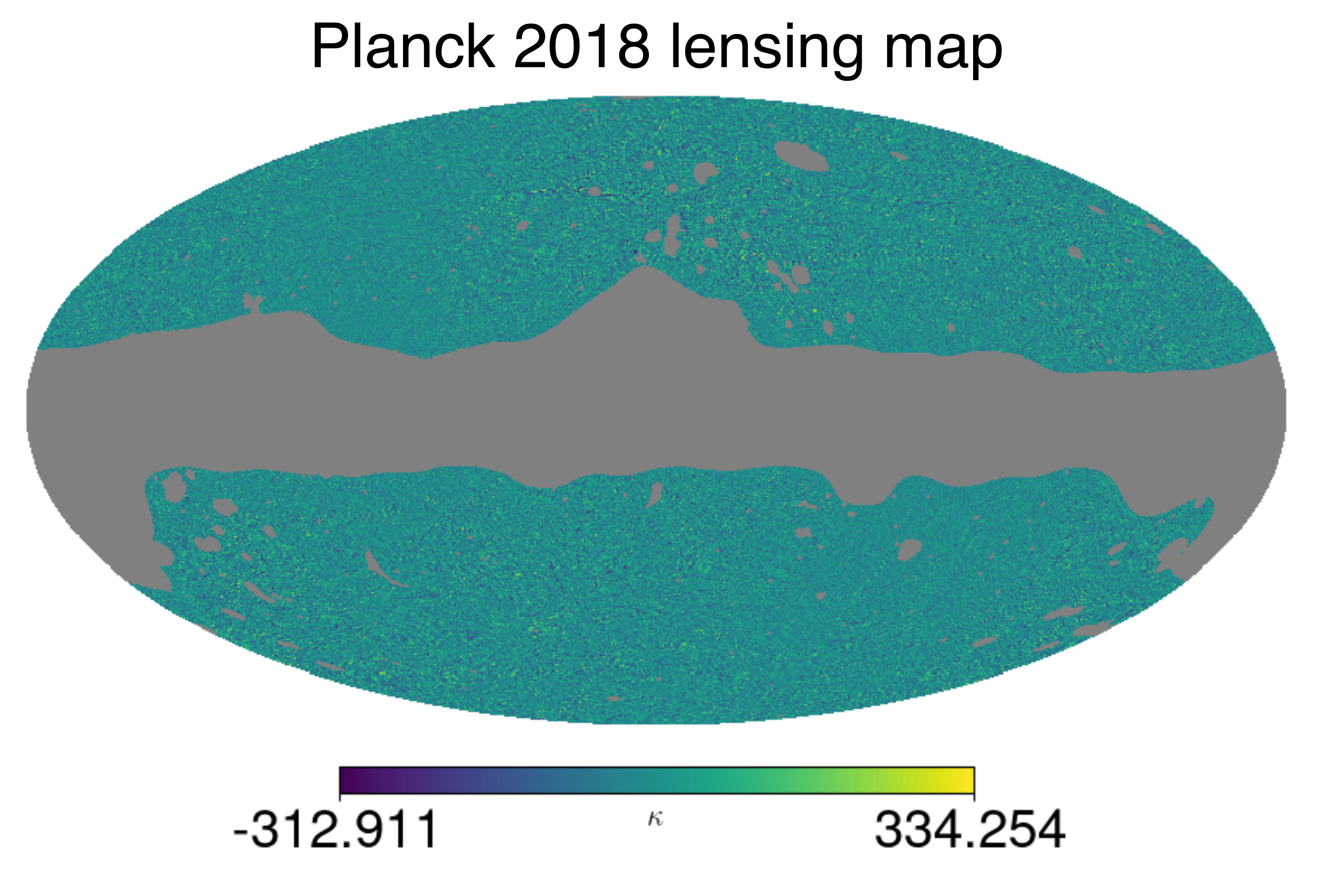}
    \includegraphics[width=0.45\linewidth]{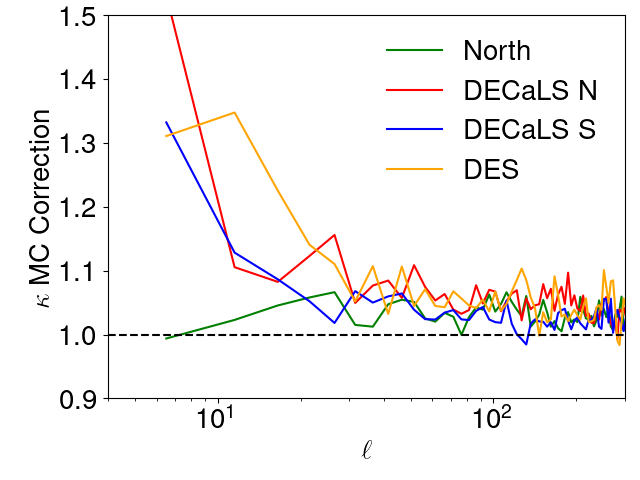}
    \caption{\textit{Left:} Planck 2018 CMB lensing map, with masked regions shown in gray, in Galactic coordinates.
    \textit{Right:} Monte Carlo normalization correction that must be applied to the raw Planck lensing map.
    This is estimated from the 300 Planck lensing mocks, by dividing the autocorrelation of the true lensing map by the cross-correlation between the true map and the reconstructed map. The cross-correlation is measured separately for each of the four regions, and we find that the low-$\ell$ scale-dependence is different in different regions.}
    \label{fig:planck_lensing}
\end{figure}

\subsection{Imaging systematics weights}

To obtain the cleanest measurement of large-scale modes, we also must remove contaminating power from the quasar maps.
We use the linear regression method \citep{Myers06,Ross11ang,Ho12,Ross17,Ross20,Raichoor21,Leistedt16,ElvinPoole18}, as developed in the Regressis software package \citep{Chaussidon22}.\footnote{\url{https://github.com/echaussidon/regressis}}.
In this method, we model the quasar density as a linear function of the various systematics templates, and weight the sample by the inverse of the predicted density, removing linear relationships between quasar density and the imaging systematics.
This is the same method that was used for the full quasar target sample, but we must re-measure the weights since we have substantially modified the selection.
We use the same set of 11 imaging maps as in \cite{Chaussidon22}:
\begin{itemize}

\item{Stellar density, as defined by the density of point sources from Gaia DR2 with $12 < \textrm{PHOT\_G\_MEAN\_MAG} < 17$ \citep{gaia_dr2}}

\item{Extinction E(B-V), from \citep{sfd} and \citep{SchlaflyFinkbeiner11}}

\item{Sagittarius Stream catalog from \citep{Antoja}, with known quasars removed and cut to $r > 18$}

\item{5-$\sigma$ point source  depth in $g$, $r$, $z$, $W1$, and $W2$ magnitudes}

\item{PSF size in $g$, $r$, $z$. Due to the 6'' resolution of WISE, essentially all galaxies are point sources in W1 and W2.}
\end{itemize}

\begin{figure}
    \centering
    \textbf{North}\par\medskip
    \includegraphics[width=1.0\linewidth]{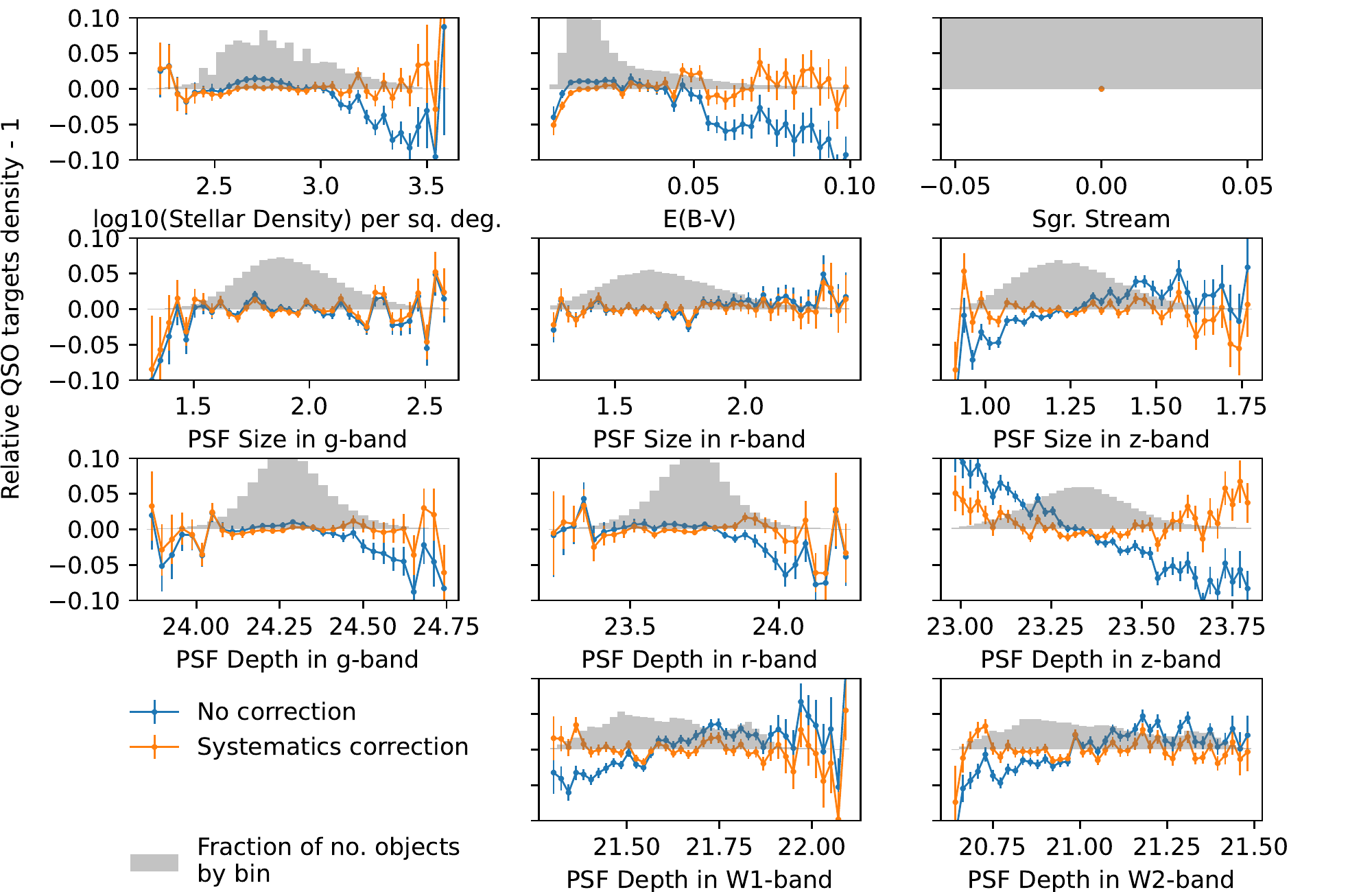}
    \caption{Relationship between quasar density and imaging systematic maps in the North region, before (blue) and after (orange) applying systematics weights.
  Points are the average overdensity in pixels within a specific bin of the systematic template, and errorbars are the standard error of the mean. The Sagittarius Stream template is only defined for the southern imaging regions and hence no points are displayed in the top right panel.
  The gray histograms in the background show the distribution of each systematics property in the quasar catalog.
    }
    \label{fig:north_systematics}
\end{figure}

\begin{figure}
    \centering
    \textbf{DECaLS N}\par\medskip
    \includegraphics[width=1.0\linewidth]{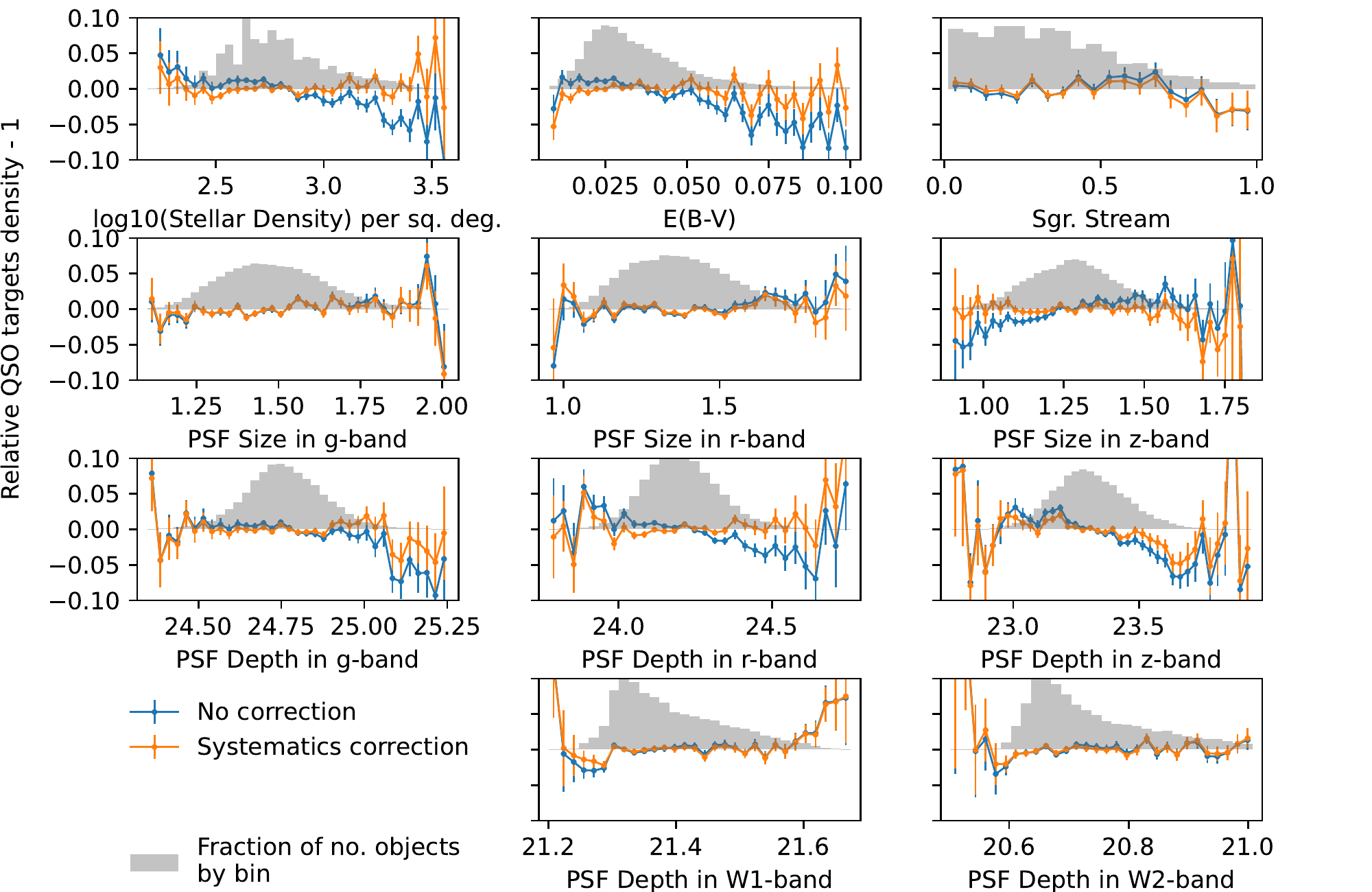}
    \caption{Like Fig.~\ref{fig:north_systematics}, for the DECaLS N region. }
    \label{fig:south_ngc_systematics}
\end{figure}

\begin{figure}
    \centering
    \textbf{DECaLS S}\par\medskip
    \includegraphics[width=1.0\linewidth]{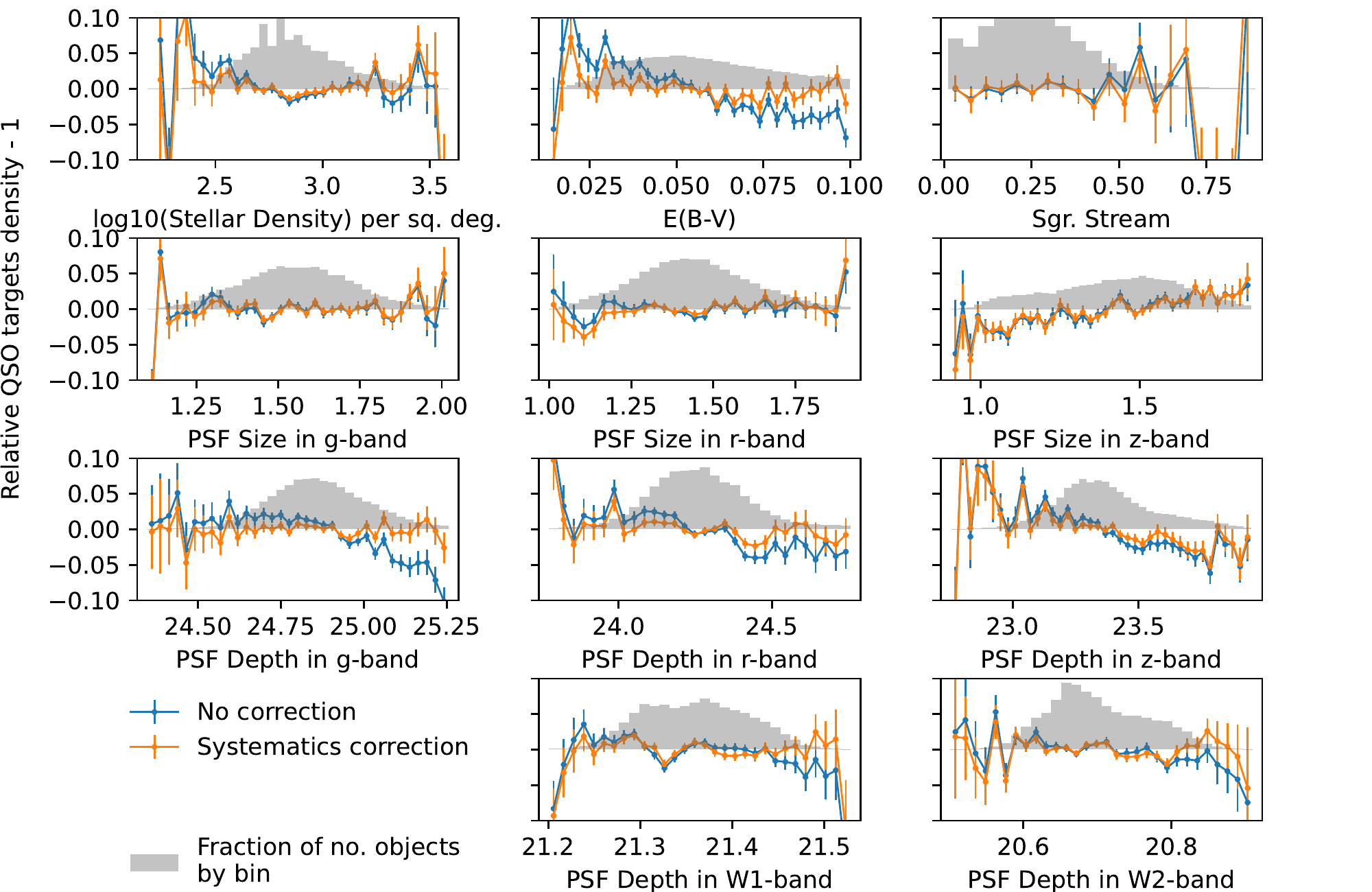}
    \caption{Like Fig.~\ref{fig:north_systematics}, for the DECaLS S region. }
    \label{fig:south_sgc_systematics}
\end{figure}

\begin{figure}
    \centering
    \textbf{DES}\par\medskip
    \includegraphics[width=1.0\linewidth]{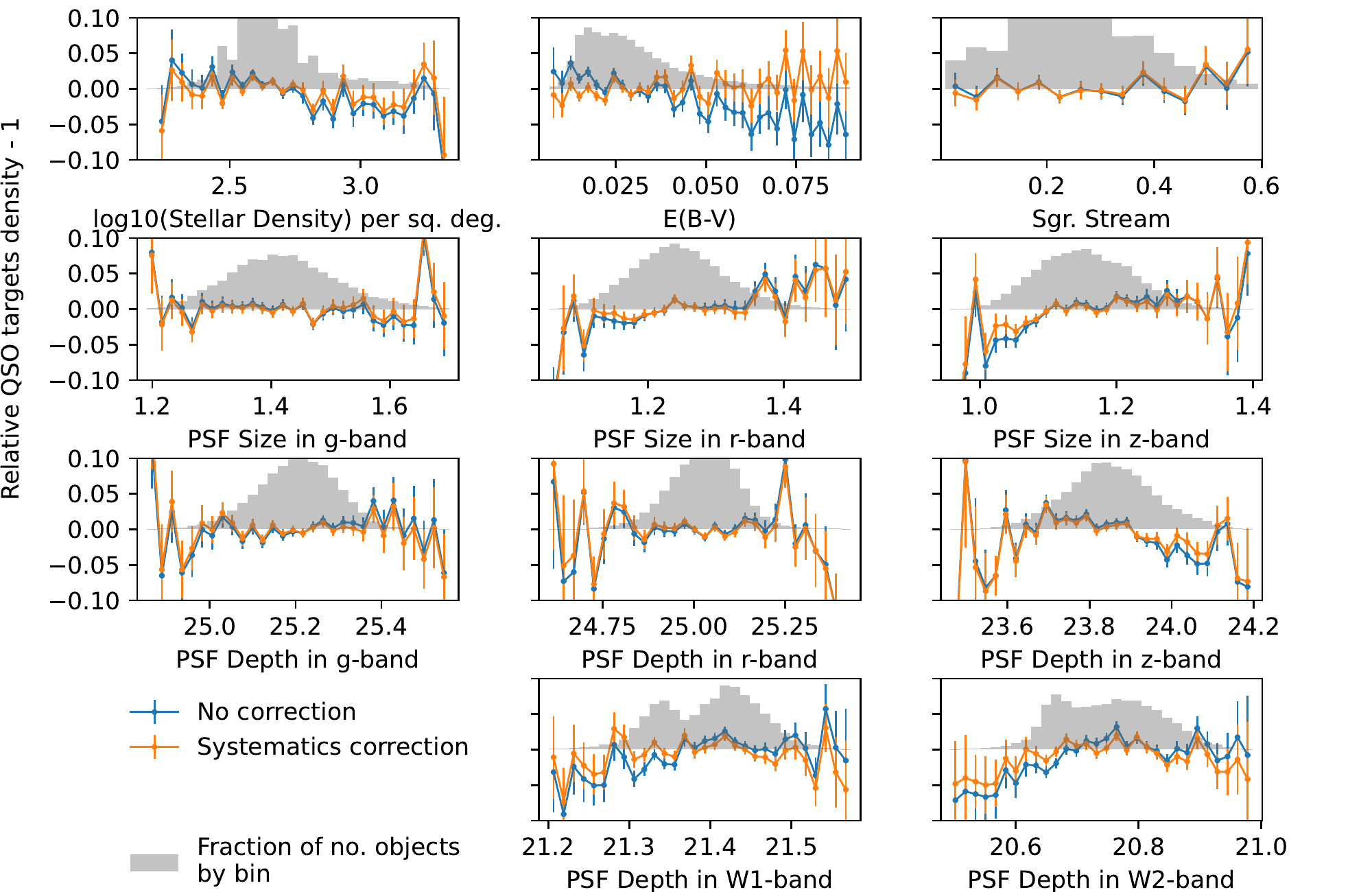}
    \caption{Like Fig.~\ref{fig:north_systematics}, for the DES region. }
    \label{fig:des_systematics}
\end{figure}

We measure correlations between quasar target density and the systematics maps in HEALPixels of NSIDE=256
(\cref{fig:north_systematics,fig:south_ngc_systematics,fig:south_sgc_systematics,fig:des_systematics}).
\textsc{Regressis}
works directly with the counts and template maps in pixel space (rather than binning by systematic property as in \cref{fig:north_systematics,fig:south_ngc_systematics,fig:south_sgc_systematics,fig:des_systematics}), using Poisson errors on the quasar counts in each pixel as an inverse-noise weight.

Overfitting is a major concern for this analysis; if too many templates are used, or a very flexible regression model is employed, the weights may accidentally remove power from the true density field at low $\ell$.
We use mock catalogs to validate the regression procedure and templates used, as described in Section~\ref{sec:mock_tests}.
We use linear regression (as implemented in the regressis package) and treat each of the four regions separately. The templates used in each region are listed in Table~\ref{tab:imaging_templates} and are justified with the contaminated mocks presented in Section~\ref{sec:contaminated_mocks}.
While these weights do a reasonable job of mitigating trends with systematics, some trends still remain---most notably in DECaLS South, where overfitting has the most severe impact and we are limited to using only a single template.


The weights are defined as the inverse of the relationship between the imaging systematics and the quasar target density field. \cref{fig:north_systematics,fig:south_ngc_systematics,fig:south_sgc_systematics,fig:des_systematics} show that the weights reduce the relationship between quasar target density and imaging systematics.

\begin{table}[]
    \centering
    \begin{tabular}{c|p{15cm}}
       Region  & Imaging templates removed \\
       \hline
       North  & STARDENS, EBV, PSFDEPTH\_G, PSFDEPTH\_R, PSFDEPTH\_Z, PSFDEPTH\_W2, PSFSIZE\_Z \\
       DECaLS N & STARDENS, EBV, PSFDEPTH\_R, PSFSIZE\_Z, SGR\_STREAM, PSFSIZE\_R \\
       DECaLS S & EBV \\
       DES & EBV, PSFDEPTH\_W1, PSFDEPTH\_W2
    \end{tabular}
    \caption{Imaging templates removed from each region to minimize extra low-$\ell$ noise from angular systematics. We limit the number of imaging templates removed to prevent overfitting, which will decorrelate the quasar field from the true density field and remove power at low $\ell$. A different set of templates is removed from each region; in particular, only E(B-V) is removed for DECaLS S.
    }
    \label{tab:imaging_templates}
\end{table}

\subsection{Magnification bias slope}

We measure the magnification bias slope $s \equiv \frac{d\log_{10}N}{dm}$ by perturbing the input Legacy Survey photometry by a uniform offset in all bands, re-running quasar target selection, and measuring the change in number density. We try both increasing
and decreasing all magnitudes by 0.05, and find $s = 0.2751$ when we make the input sources fainter, and $s = 0.2777$ when we make the sources brighter.
Given the consistency between these measurements, we use the mean value, $s = 0.2764$.

Since lensing preserves surface brightness,
brightening or dimming is solely caused by a change in the object's size. Our procedure thus assumes that flux measurements capture all of the light from the quasar target. While this is not a good assumption for galaxies \cite{ElvinPoole22,Zhou23}, it is reasonable for the quasars, which are dominated by point sources. The selection requires that the quasar targets are identified as point sources in DR9 imaging, and less than 4\% of the sample are galaxies.
Moreover, quasar target selection is entirely based on total magnitudes rather than fiber magnitudes, which are measured within a fixed angular aperture.
It is therefore a very good approximation to treat lensing magnification as entirely brightening or dimming quasar flux.

This procedure measures
the slope using the observed (post-magnification) magnitude distribution, rather than the true, pre-magnification distribution. Magnification will scatter quasars both above and below the flux limit; given that this is a small effect overall, the change in slope induced by magnification is tiny.

The slope varies slightly in the different imaging regions (which have slightly different selection functions), and we use the value of $s$ appropriate for each region. For the North region, we find $s =0.2880$, for DECaLS-NGC we find $s = 0.2767$, for DECaLS-SGC we find $s = 0.2778$, and for DES we find $s = 0.2548$.

Our value of $s$ is quite consistent with previous measurements
for quasars ($s \sim 0.3$), either from the faint end of the luminosity function \citep{Myers03,Myers05,Irsic16,Wang20}
or from similar methods where the selection is re-run after perturbing
the fluxes by a small amount \citep{Krolewski20}.
This agreement is expected based on the similarity between the quasar
samples and the selection methods used, and provides a good validation
of our method to measure the slope.

\subsection{Stellar contamination}
\label{sec:stars}
 
Approximately 2\% of the DESI quasar targets
 are classified as stars. Provided that the stars
 are uncorrelated with CMB lensing, they add an additional
 source of uncorrelated noise and therefore
 lower the measured power spectrum
 \begin{equation}
     C_{\ell}^{\kappa, g+{\rm stars}} = \frac{C_{\ell}^{\kappa g}}{1+f_\star}
 \end{equation}
 Therefore we write the scale dependent bias as
 \begin{equation}
        \frac{b}{1 + f_\star}  + \frac{2 (b - p) f_{\rm NL}}{1+f_\star} \frac{\delta_c}{\alpha(k)}
 \end{equation}
 In this expression, $b$ is the true bias. The small-scale power spectrum measures $b/(1 + f_\star)$; by fixing $f_\star$ we can therefore measure the true bias $b$. Likewise, $1 + f_\star$ is included in the denominator of the scale-dependent bias. An error on $f_\star$ will change $b$, but will have a smaller effect on $f_{\textrm{NL}}$ since $b-p$ is divided by $1 + f_\star$ in the scale-dependent bias term.
We use a fixed $f_{\star}$ of 2.6\%, which corrects for the redshift pipeline's incompleteness in categorizing stars.

 \subsection{Planck CMB lensing map}
 \label{sec:cmb_lensing}
 
 We use the 2018 Planck CMB lensing map \citep{PlanckLens18} and its associated masks and simulations, downloaded from the Planck Legacy Archive\footnote{\url{https://pla.esac.esa.int/}} (Fig.~\ref{fig:planck_lensing}). We convert the provided spherical harmonic coefficients of convergence, $\kappa_{\ell m}$, into an NSIDE=2048 HEALPix map in Galactic coordinates.
 We use the minimum variance (MV) reconstruction from both temperature and polarization, using the \texttt{SMICA} foreground-cleaned CMB map.
This map does not have the Monte Carlo multiplicative correction applied (Eq.~10 in \cite{PlanckLens18}). Following Ref.~\cite{Farren23}, we apply this correction by measuring
 on the 300 FFP10 simulations the ratio of the true lensing power spectra to the cross-correlation between the input and reconstructed lensing maps, masking
 both in each of our 4 regions (Fig.~\ref{fig:planck_lensing}). We then multiply the measured cross power spectra by this factor (binned with $\Delta \ell = 5$ and then linearly interpolated), which is
 generally small ($\sim$5\%) at high $\ell$, but becomes large on large scales.
 At the precision of this analysis, biases due to CMB foregrounds are expected to be small \citep{Schaan:2018tup,Sailer:2020lal,Ferraro:2017fac}.
 To verify that we aren't affected by foregrounds,
 we also cross-correlate with a lensing reconstruction from the CMB temperature map with the thermal Sunyaev-Zel'dovich (tSZ) effect explicitly nulled (tSZ-free). In addition to testing contamination from the tSZ effect, the tSZ-free map will also be differently sensitive to other foregrounds, such as Galactic dust, which could be correlated with the quasar map.

At very low $\ell$ where $f_{\textrm{NL}}$ sensitivity is maximal, the mean-field correction, which must be obtained from simulations, is 100 to 1000 times larger than the CMB lensing signal (Fig.~B1 in \cite{PlanckLens18}). The Planck lensing cosmology analysis excludes $\ell < 8$ due to concerns about the number and fidelity of the simulations used to derive the mean field.
 However, they note that these very low multipoles do not explicitly look anomalous, except for $\ell = 2$. 
 Unlike the auto-spectrum, where an underestimated mean field would directly bias the recovered power spectrum, in the cross-correlation it will only affect the covariance. Hence our analysis is less sensitive to potential low $\ell$ issues, and we use $\ell_{\rm min} < 8$ in our fiducial analysis.
 
We determine $\ell_{\rm min}$ by cross-correlating the Planck lensing map with the 11 quasar systematics templates.
 We measure the cross-correlation in bins of $\Delta \ell = 10$ starting at $\ell_{\textrm{min}} = 4$ or 6.
  For each template, we subtracted its mean across the union of the four imaging masks. For the stellar density and Sagittarius Stream templates, we turned them into overdensity maps (since these are stellar counts or normalized stellar counts respectively). Errorbars are computing using the Gaussian diagonal approximation, which works well for this case. To avoid instabilities in the mask deconvolution, we measure the convolved power spectrum
  and divide by $f_{\textrm{sky}}$ = 0.45.
  
\begin{figure}
    \centering
    \includegraphics[width=1.0\linewidth]{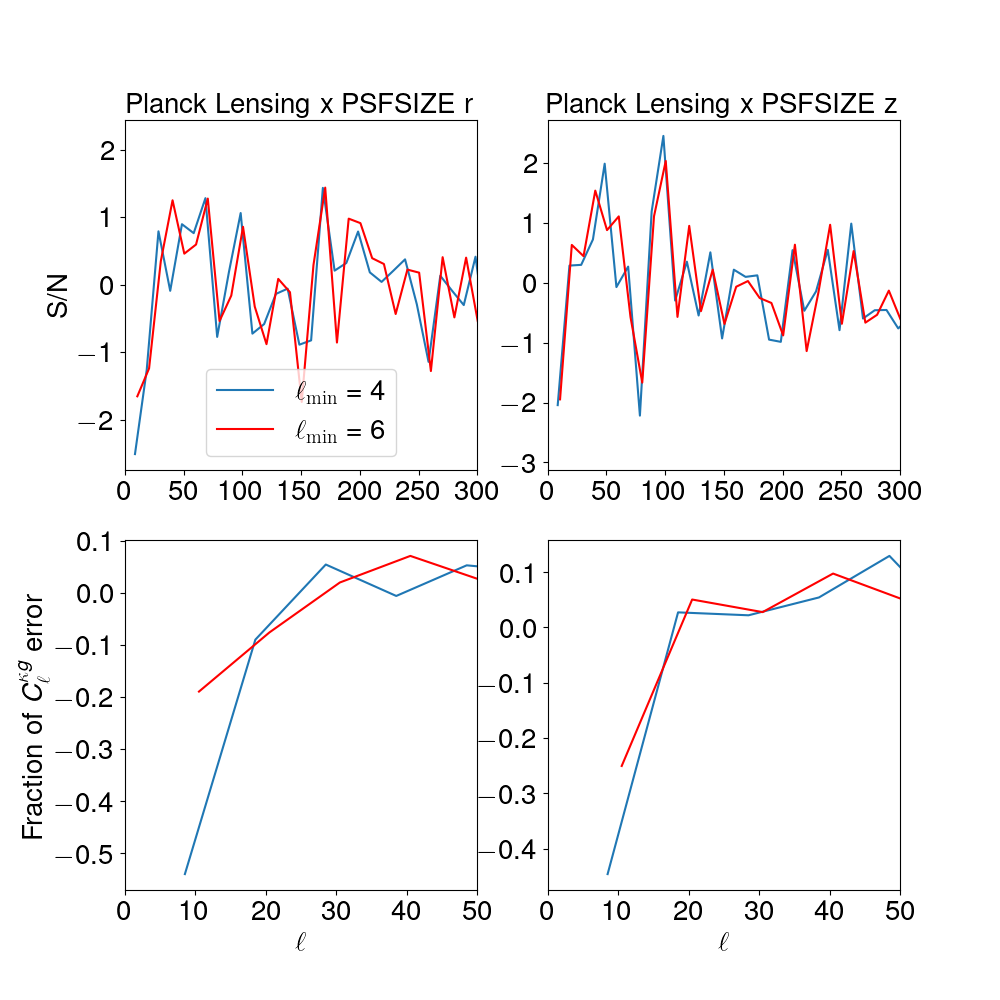}
    \caption{\textit{Top:} Signal to noise ratio of the cross-correlation between the Planck lensing map and $r$-band PSF size (\textit{left}) and $z$-band PSF size (\textit{right}). An anti-correlation is marginally detected in the first $\ell$ bin ($4 \leq \ell < 14$) and is less significant when the bin is changed to $6 \leq \ell < 16$.
    \textit{Bottom:} Estimated impact on $C_{\ell}^{\kappa g}$ from the systematic cross-correlation, using the ``systematics-correction'' slope from Figs.~\ref{fig:north_systematics} to~\ref{fig:des_systematics} to propagate from $r$ or $z$ PSF size to $\delta_g$. We use the slope from DES and these 2 systematics maps because they show the largest impact on the low $\ell$ bin, nearly $0.5 \sigma$. However, the impact on $C_{\ell}^{\kappa g}$ is considerably reduced when we use $\ell_{\rm min} = 6$ instead.}
\label{fig:planck_lensing_cross_systematics}
\end{figure}

While we find no significant correlations between the Planck lensing map and the systematics templates, we do find marginal anti-correlations in the lowest bin ($4 \leq \ell < 13$)
  between the lensing map and PSFSIZE\_G, PSFSIZE\_R, PSFSIZE\_Z, PSFDEPTH\_W1, PSFDEPTH\_W2, and STARDENS. These anti-correlations are generally significant at 2--2.5$\sigma$. We note that some of these maps are very similar (particularly G and R sizes, and W1 and W2 depths), so these are not independent measurements. As an example, we show the signal-to-noise of the cross-correlations with PSFSIZE\_R and PSFSIZE\_Z
  in Fig.~\ref{fig:planck_lensing_cross_systematics}.
  
We translate these measurements of $C_{\ell}^{\kappa, \textrm{syst}}$ into $C_{\ell}^{\kappa g}$ using the slope $\frac{d{\delta_g}}{d{\textrm{syst}}}$ from \cref{fig:north_systematics,fig:south_ngc_systematics,fig:south_sgc_systematics,fig:des_systematics}.
We use either the linear regression slope from \cref{fig:north_systematics,fig:south_ngc_systematics,fig:south_sgc_systematics,fig:des_systematics} (after weights are applied),
or the 95\% upper bound in the case that the slope is not significantly detected.
We find that the estimated bias to the cross-power spectrum can exceed half of the statistical error on $C_{\ell}^{\kappa g}$ at $4 \leq \ell < 13$ in some of the 4 regions. The most problematic templates are PSFSIZE\_R and PSFSIZE\_Z in DES.
Switching to $\ell_\textrm{min} = 6$ reduces the significance of the cross-correlation. In this case, none of the systematics correlate with Planck lensing at $>2\sigma$ in the first bin. Moreover, the impact on $C_{\ell}^{\kappa g}$ is also reduced by 30--50\%. Therefore, with $\ell_\textrm{min} = 6$, the upper bound on biases in $C_{\ell}^{\kappa g}$ is $< 0.3\sigma$.

Because the cross-correlation covariance may be under-estimated at $\ell < 8$, we present our results using both
 $\ell_{\rm min} = 6$ and $\ell_{\rm min} = 8$, which has a 15\% larger $f_{\textrm{NL}}$ error.
 
 



 \begin{figure}
    \centering
    \includegraphics[width=1.0\linewidth]{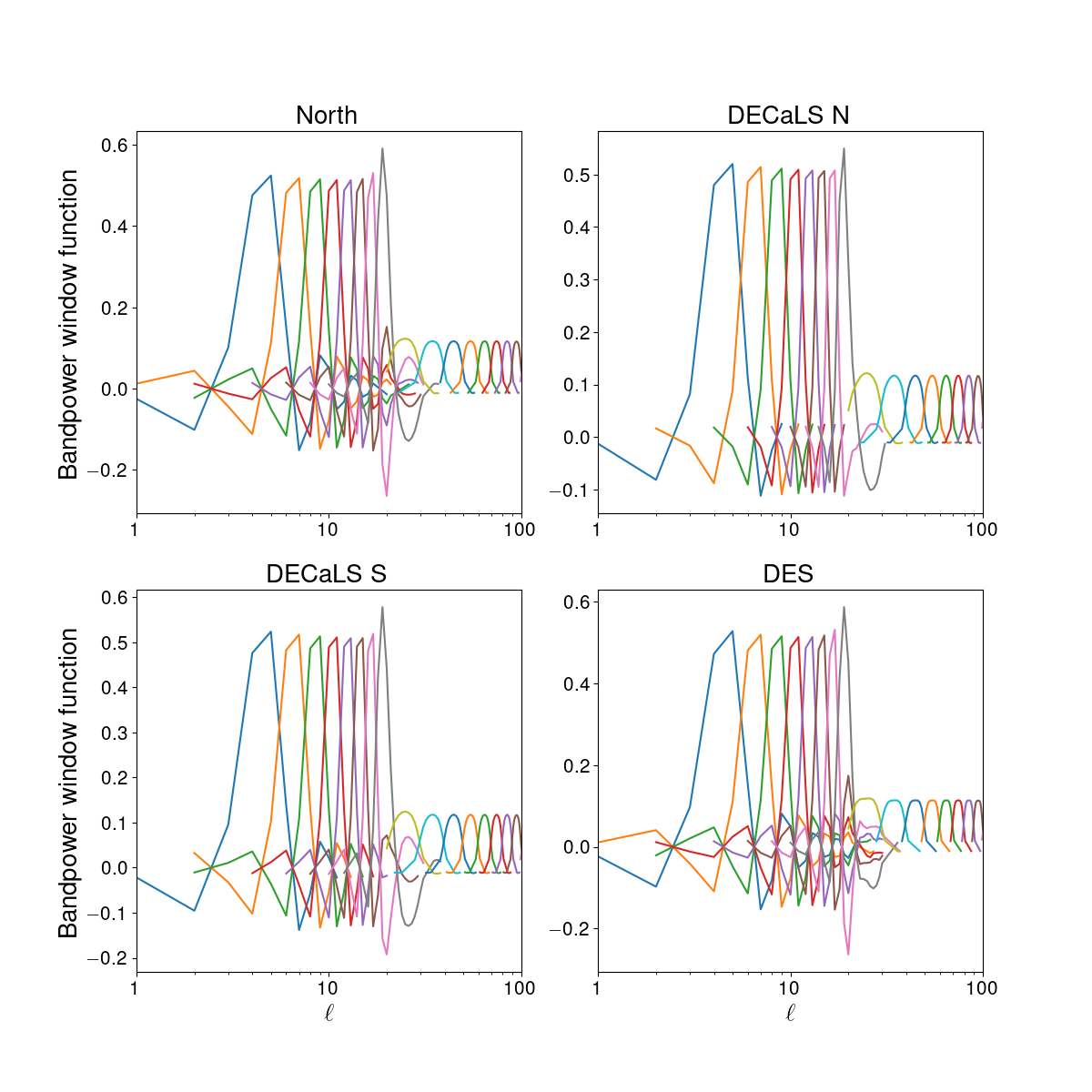}
    \caption{Bandpower window functions that must be multiplied by the theory curve to produce a binned theory prediction. Due to mask convolution and deconvolution, the binning is not a perfect tophat, but rather contains some contributions from multipoles outside the desired range. The drop in the window function at $\ell = 20$ is due to the transition from $\Delta \ell = 2$ bins to $\Delta \ell = 5$ bins.}
    \label{fig:bandpower_window}
\end{figure}

\begin{figure}
    \centering
    \includegraphics[width=1.0\linewidth]{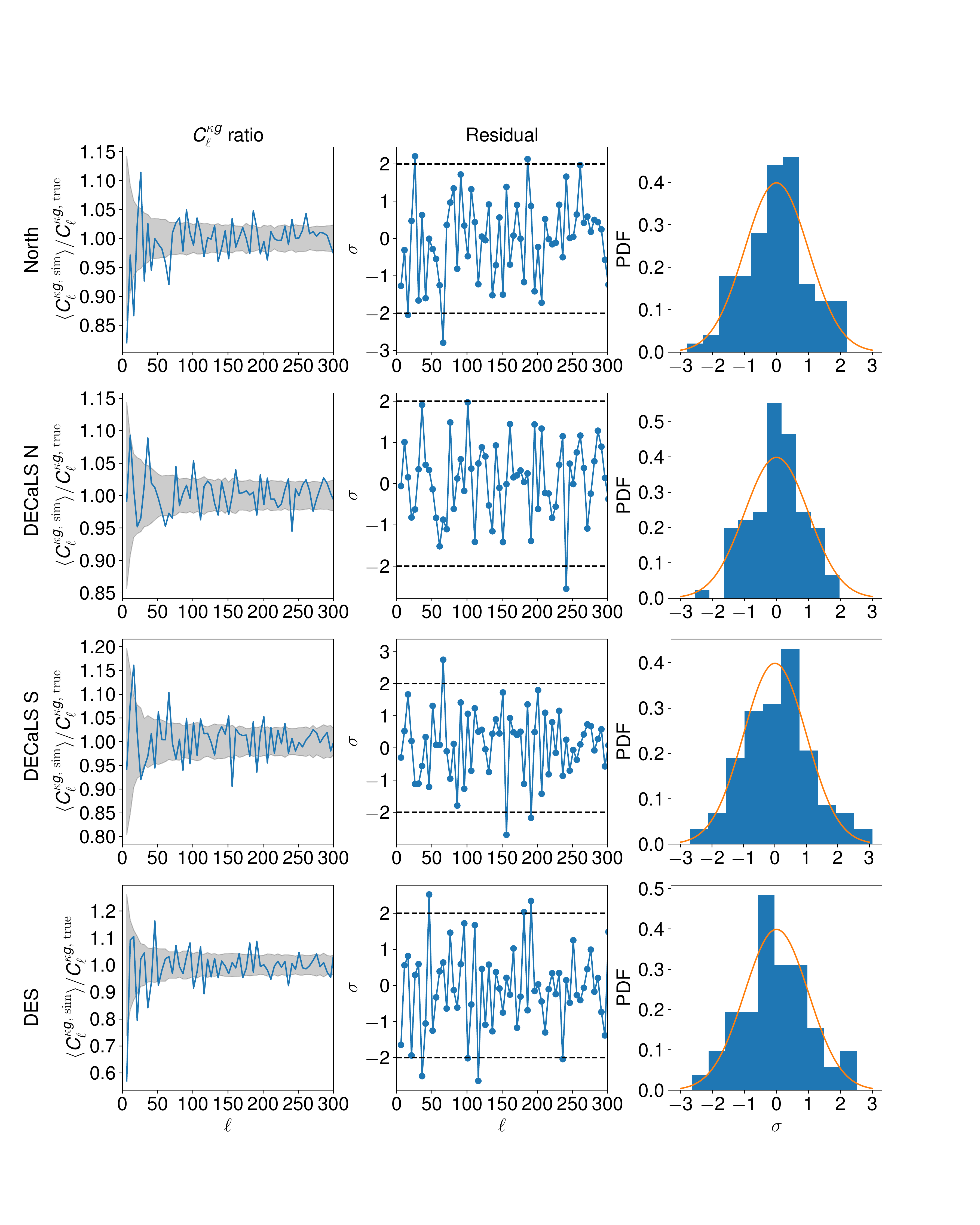}
    \caption{\textit{Left:} Ratio between input $C_{\ell}^{\kappa g}$ and $C_{\ell}^{\kappa g}$ measured from the mean of 100 correlated Gaussian simulations. Galaxies are Poisson-sampled from the expected number within each pixel, corrected for partial coverage using the imaging randoms. Gray band gives 1$\sigma$ error on the ratio. \textit{Center:} Residuals from the left-hand plots in units of the errorbar. \textit{Right:} PDF of the residuals compared to a standard normal distribution. The standard deviations of the residuals are 1.01, 0.88, 1.08, and 1.05.}
    \label{fig:transfer}
\end{figure}

 \begin{figure}
    \centering
    \includegraphics[width=0.48\linewidth]{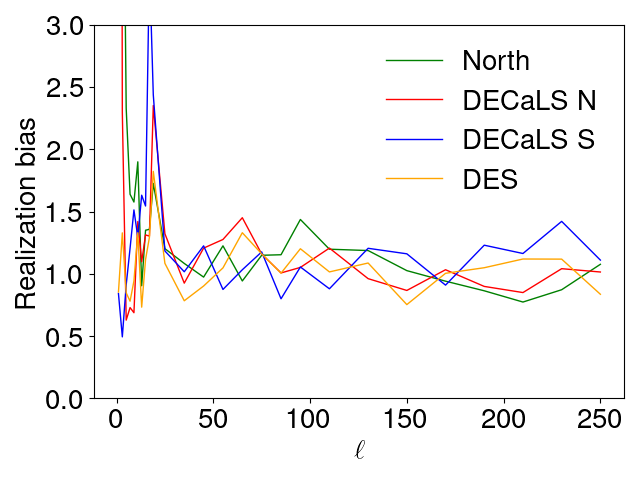}
    \includegraphics[width=0.48\linewidth]{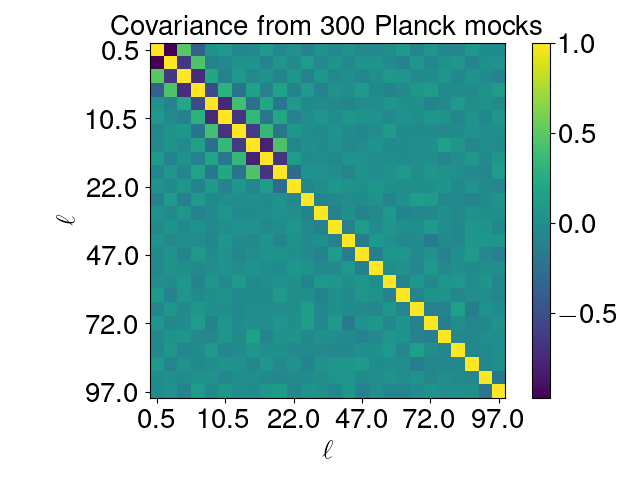}
    \caption{\textit{Left:} Ratio between the diagonal of the covariance matrix between 300 $\kappa$ simulations and a single uncorrelated galaxy simulation (like the situation in data); and the true covariance calculated with 300 correlated $\kappa$ and galaxy simulations.
    \textit{Right:} Correlation matrix for $C_{\ell}^{\kappa g}$ in the North region (without realization bias correction). Covariance for the other 3 imaging regions is very similar.}
    \label{fig:covariance}
\end{figure}

 \begin{figure}
    \centering
    \includegraphics[width=0.9\linewidth]{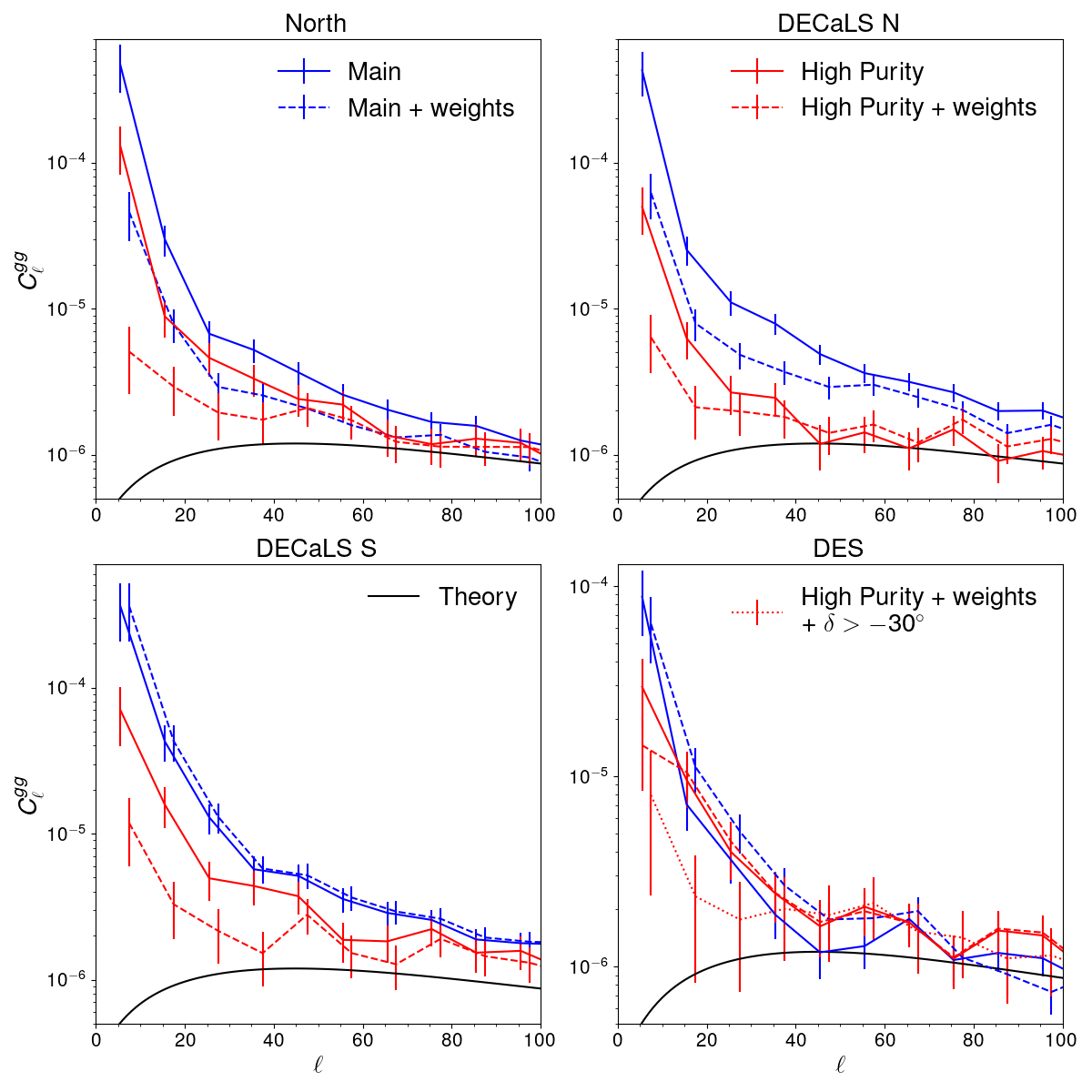}
    \caption{Quasar autocorrelation in the four regions, comparing the main sample (blue) and the high purity sample (red). Linear systematics weights are applied using the templates in Table~\ref{tab:imaging_templates}; dashed lines show weighted power spectra and solid lines show unweighted. 
    The excess systematic power $N_\ell^{\textrm{syst}}$ is defined
    as the difference between the black line theory curve
    and the measured power spectra given by the dashed red curves (or dotted red curve for DES).
    Shot noise equal to $1/\bar{n}$ is subtracted from all spectra to allow comparison between the high purity sample with half the density of the main sample.
    The theory curve is computed using the fiducial cosmology and the 
    bias evolution of Ref.~\cite{Laurent17} (Eq.~\ref{eqn:laurent_b}), with $b_0=1$.
    For DES, we show both the full DES region (4162 deg$^2$; solid and dashed lines); and the fiducial $\delta > -30^{\circ}$ region, which has significantly less excess power on large scales (dotted red line).
    Both high purity and main have excess power at low $\ell$ relative to the theoretical expectation, even with systematics weights applied. However, the excess power is smaller for High Purity than Main, and is further reduced after applying systematics weights.
    The covariance on $C_{\ell}^{\kappa g}$ is directly proportional to $C_{\ell}^{gg}$, including extra noise (Eq.~\ref{eq:knox_cov}). Thus, to achieve the tightest constraints on $f_{\textrm{NL}}$, we must reduce the additional low-$\ell$ contamination power as much as possible.
    }
    \label{fig:clgg}
\end{figure}

\begin{figure}
    \centering
    \includegraphics[width=1.0\linewidth]{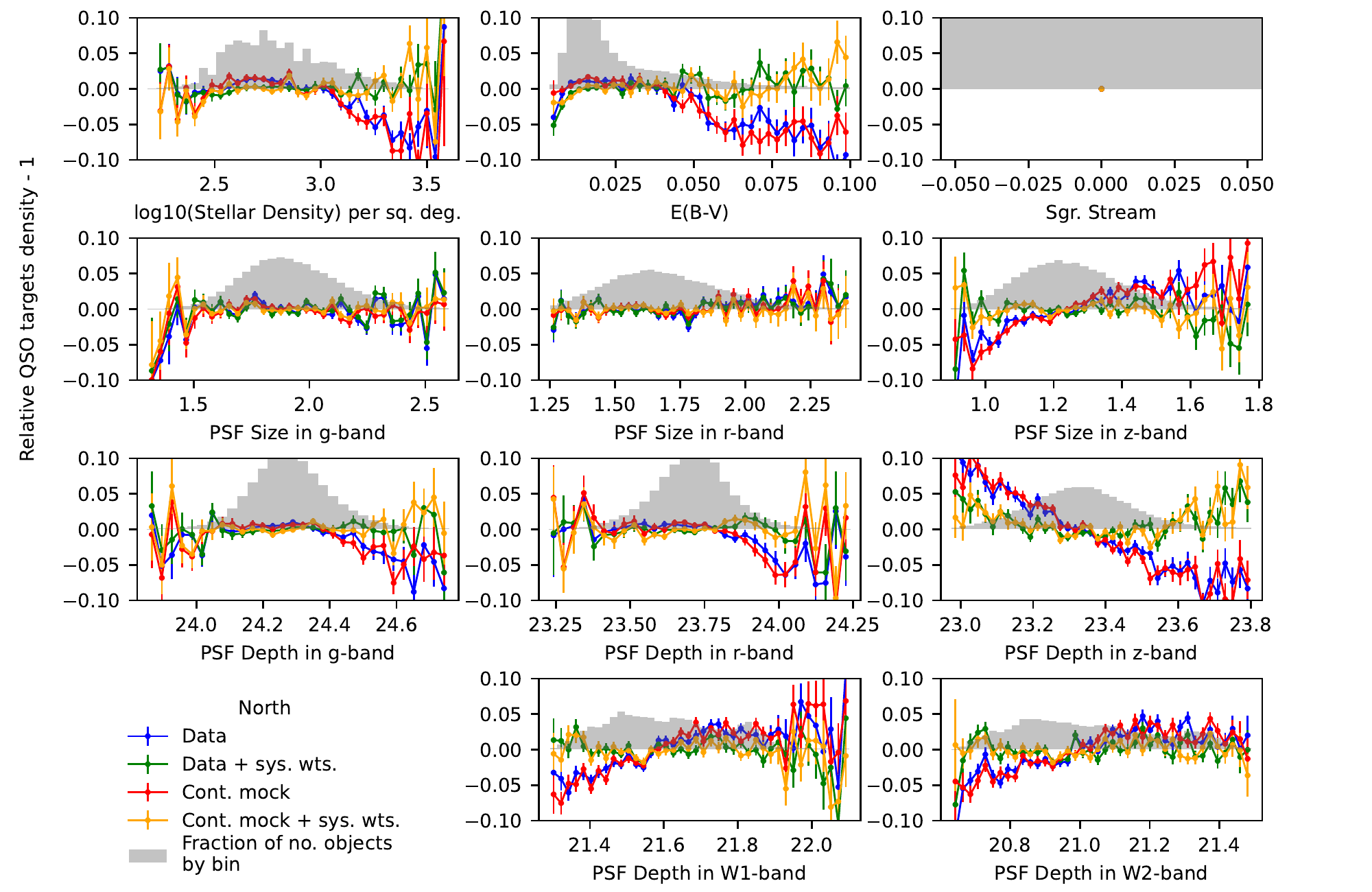}
    \caption{Relationship between imaging systematics and contaminated mock density for the North region.
    Both the data (blue) and contaminated mocks (red) have similar trends with imaging systematics.
    After multiplying by the systematics weights, the trends are similarly mitigated in both data (green) and mocks (orange).}
    \label{fig:contaminated_mocks_imaging_systematic}
\end{figure}

\begin{figure}
    \centering
    \includegraphics[width=0.45\linewidth]{MAIN_QSO_r_W2_Duncan_cuts_4_regions_no_power_loss_targets.png}
    \includegraphics[width=0.45\linewidth]{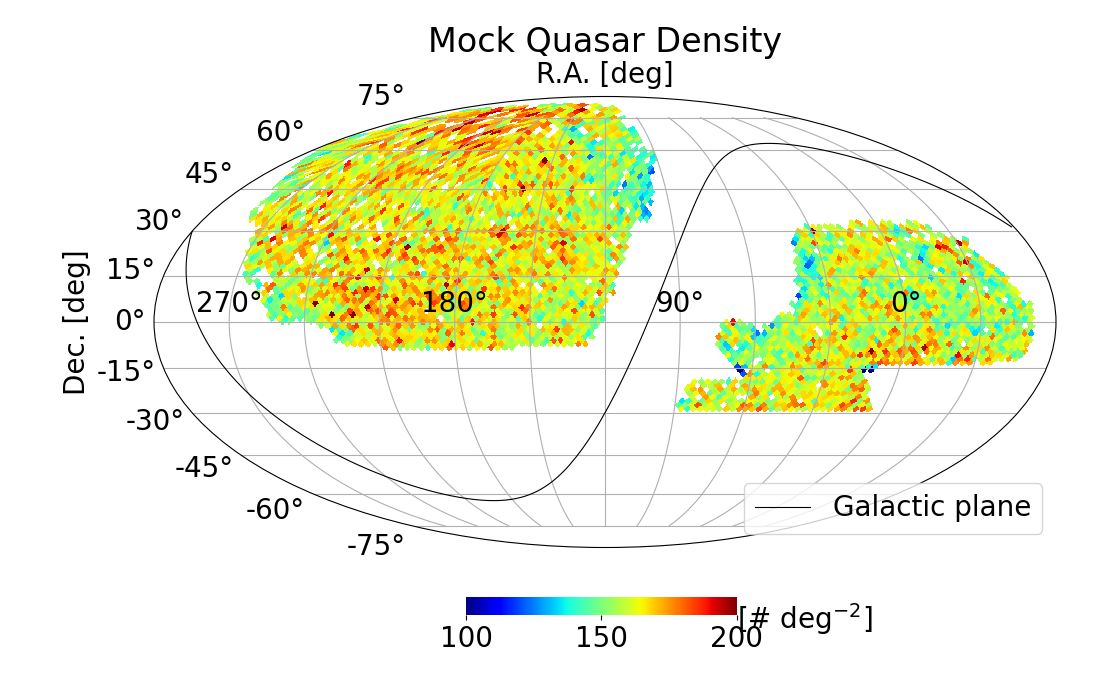}
    \caption{Comparison between the data (\textit{left}) and the contaminated mocks (\textit{right}), before imaging systematic weights are applied.
    The contaminated mocks have a slightly lower density than the data (155 deg$^{-2}$ versus 160 deg$^{-2}$), since the systematic weights used to generate them have a mean of 1.042 (weighted by the fractional coverage of each pixel), not 1. Therefore, for this figure, we multiply the mock quasar density by 1.042 to match the overall density.
    }
    \label{fig:contaminated_mocks_vs_data}
\end{figure}

\begin{figure}
    \centering
    \includegraphics[width=1.0\linewidth]{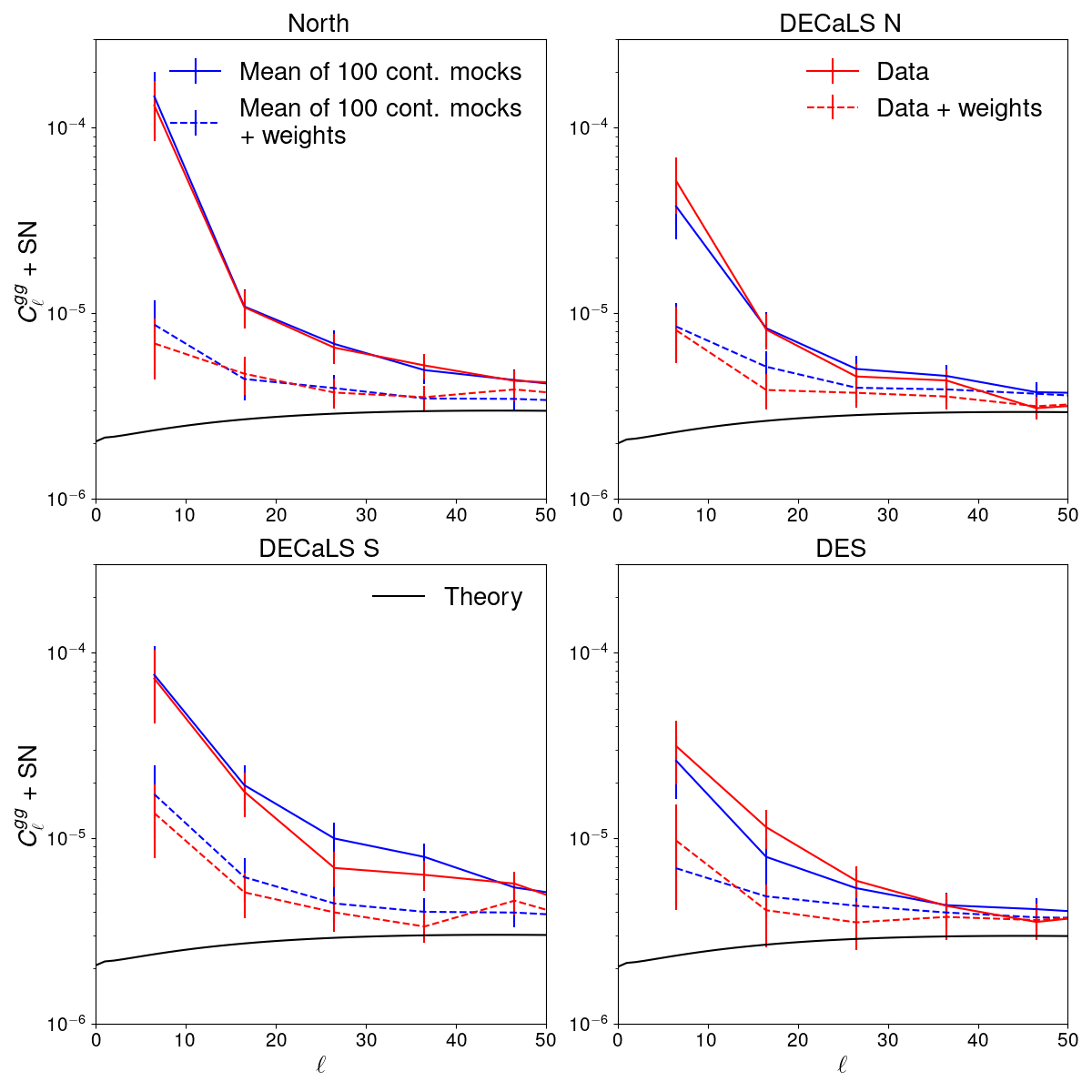}
    \caption{Angular auto-correlation from the mean of 100 contaminated mocks (blue solid line) and after systematic correction (blue dashed line), compared to data (red solid line) and data after systematic correction (red dashed line). A theory curve in the fiducial cosmology with $f_{\textrm{NL}} = 0$ is also shown. Unlike Fig.~\ref{fig:clgg}, we do not subtract shot noise.}
    \label{fig:clgg_contaminated_mocks}
\end{figure}

\begin{figure}
    \centering
    \includegraphics[width=1.0\linewidth]{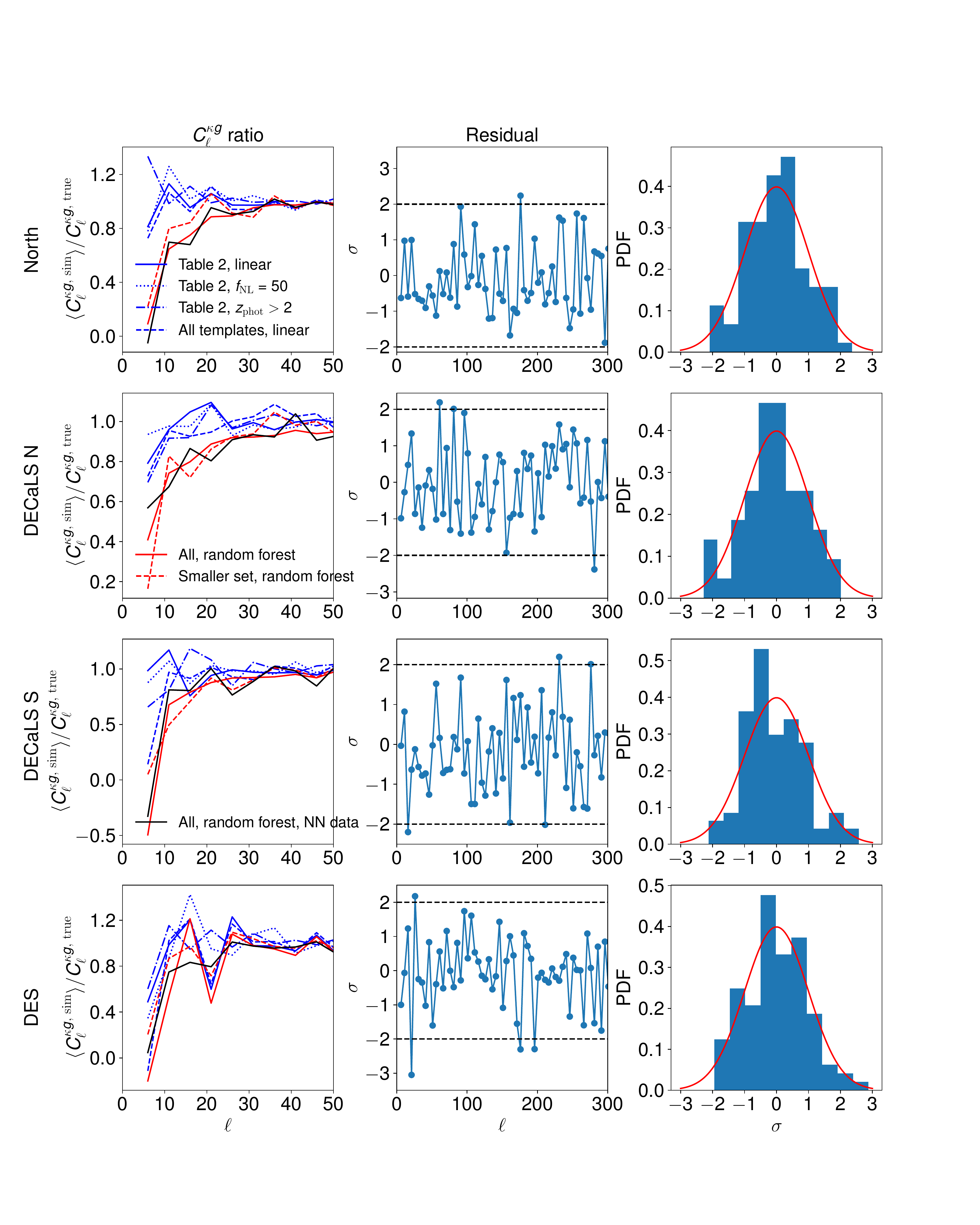}
    \vspace{-50pt}
    \caption{Like Fig.~\ref{fig:transfer}, but using 100 contaminated Poisson-Gaussian mocks instead. A separate regression is run for each contaminated mock to remove imaging systematics. The comparison $C_{\ell}^{\kappa g}$ is the true input cross-power spectrum. The error bars are computed using jackknife resampling. The left panels show the result both using the fiducial linear regression on the restricted set of templates  from Table~\ref{tab:imaging_templates} (solid blue) and the Random Forest regression with all 11 templates (solid red). 
    We also show that the input power is recovered well for a nonzero $f_{\textrm{NL}}$ (dotted blue), and for the high-redshift sample (dot-dash blue), which uses a slightly different template set (Appendix~\ref{sec:highz}).
    For comparison, we also show the result for all 11 templates with linear regression (blue dashed); all 11 templates with Random Forest but with input weights generated with a neural net (black solid); and 7 templates with random forest (red dashed).
    The center and right panels show the residuals from the linear regression, which agree well with a normal distribution (red).}
    \label{fig:transfer_contaminated}
\end{figure}

\section{Angular power spectrum}
\label{sec:ps}
\subsection{Mask deconvolution}

The presence of masked regions of the sky causes the measured power spectrum
to differ from the theoretical expectation. The measured power spectrum is approximately scaled by $f_{\rm sky}$, and neighboring $\ell$ modes are coupled.  On small scales, it is adequate to simply divide the measured power spectrum by $f_{\rm sky}$ to account for the mask. However, at low $\ell$, the mask also changes the shape of the power spectrum, and a more correct treatment is needed.

We use a pseudo-$C_{\ell}$ estimator \citep{Hivon02,NaMaster} with the measured power spectrum
given by
\begin{equation}
    \tilde{C}_{\ell}^{\kappa g} = \frac{1}{2 \ell + 1} \sum_m g_{\ell m} \kappa^*_{\ell m}
\end{equation}
 The expectation value of the pseudo-$C_{\ell}$ is related to the true $C_{\ell}$ through
the mode coupling matrix:
\begin{equation}
    \langle \tilde{C}_{\ell} \rangle = \sum_{\ell'} M_{\ell \ell'} C_{\ell'}
\end{equation}
where the natural generalization of the mode-coupling
matrix to two different input masks is
given in Eqs.~7 and 12 in \cite{Alonso18}.
We use the MASTER algorithm \citep{Hivon02}
as implemented in NaMaster \citep{NaMaster}
to deconvolve the mask from the windowed power spectra.
While the unbinned mode coupling matrix is often singular, MASTER bins the power spectrum to make the (binned) mode coupling matrix invertible. Then the binned, deconvolved bandpowers are estimated by multiplying the measured bandpowers by the inverse of the binned mode-coupling matrix:
\begin{equation}
    C_{b} = \sum_{b'} \mathcal{M}_{bb'}^{-1} \tilde{C}_{b'}
\end{equation}
To account for the binning and unbinning, a bandpower window matrix must be applied to the theory curve
\begin{equation}
    C_{b} = \sum_{\ell'} W_{bl} C_{\ell}
\end{equation}
\begin{equation}
    W_{b\ell} = \sum_{b'} \mathcal{M}^{-1}_{bb'} \sum_{\ell \in b'} M_{\ell \ell'}
\end{equation}

We bin the quasars into NSIDE=2048 HEALPixels in Galactic coordinates \citep{Gorski05}.  We use bins of $\Delta \ell = 2$ until $\ell = 20$, $\Delta \ell = 5$ until $\ell = 100$, and $\Delta \ell = 20$ thereafter.  The resulting bandpower window functions are shown in Fig.~\ref{fig:bandpower_window}.  Since we are concerned with very low $\ell$, we do not correct for the HEALPix window function, which is nearly one over the entire range of scales considered.

We use separate masks for the CMB lensing and galaxy fields. To create the CMB lensing mask, we multiply the binary Planck lensing mask by the binary region masks.
The product of the two masks yields an $f_{\rm sky} = 0.3364$.
We do not apodize either mask, because we find
that apodization slightly increases the low-$\ell$ noise in the cross-correlation.

We verify that our pipeline can correctly recover the angular power spectrum in the presence of a mask using 100 Poisson-sampled Gaussian simulations.
We begin by creating correlated Gaussian quasar and CMB lensing fields.
We then multiply each pixel of the mock quasar map by the imaging completeness mask. Next we Poisson
sample the resulting field (after setting any negative pixels to zero), and then divide by the imaging completeness map to account for the partial coverage of each pixel.
From this map, we compute $\delta$ within the four imaging regions, apply the binary region masks, and measure $C_{\ell}^{\kappa g}$ in the four regions. As in the data, we do not consider any pixel with imaging completeness $< 80\%$.

We find that the average $C_{\ell}^{\kappa g}$ from these 100 simulations is statistically identical to the bandpower-convolved input theory spectrum (Fig.~\ref{fig:transfer}). The scatter in Fig.~\ref{fig:transfer} is consistent with Gaussian noise: we find $\chi^2$ = 69.1, 50.1, 62.6, and 75.3 over 60 bins from $4 < \ell < 304$ for the North, DECaLS N, DECaLS S, and DES regions, and with no significant outliers. The mean ratio is $0.9955 \pm 0.0012$, $1.0029 \pm 0.0013$, $1.0048 \pm 0.0025$, $0.9912 \pm 0.0040$ across these bins. While the deviation in the North is significant at $3.75\sigma$, and the means of the other regions are $\sim 2\sigma$ away from one, these deviations are tiny compared to the statistical errors on the bandpowers, so we do not apply any correction after multiplying by the bandpower window matrix.

\subsection{Covariance matrix}
\label{sec:covariance}

If there is no mask and the underlying field is Gaussian, the covariance matrix for the
binned bandpowers is given by
\begin{equation}
    \textrm{Cov}(C_{b b'}) = \frac{1}{(2 \ell_b + 1)\Delta \ell_b} \left((C_{\ell, b}^{\kappa g})^2 + \left(C_{\ell, b}^{gg} + \frac{1}{\bar{n}}\right) (C_{\ell, b}^{\kappa \kappa} + N_{\ell, b}^{\kappa \kappa})\right) \delta_{bb'}
\end{equation}
In the cut-sky case, if the bandpowers are sufficiently broad that the off-diagonal terms are small,
the covariance can still be well-approximated as diagonal, but with the number of modes in each bin reduced by $f_{\rm sky}$ (``Knox formula'') \citep{Knox95,Hivon02}:
\begin{equation}
    \textrm{Cov}(C_{b b'}) = \frac{1}{(2 \ell_b + 1)\Delta \ell_b f_{\textrm{sky}}} \left((C_{\ell, b}^{\kappa g})^2 + \left(C_{\ell, b}^{gg} + \frac{1}{\bar{n}}\right) (C_{\ell, b}^{\kappa \kappa} + N_{\ell, b}^{\kappa \kappa})\right) \delta_{bb'}
\label{eq:knox_cov}
\end{equation}
However, our measurement is not in this regime, due to the small width of the bins used and the large
correlation between neighboring bins, particularly at low $\ell$.
We therefore primarily rely on a mock-based covariance and check it with a Gaussian
analytic covariance from NaMaster \cite{Garcia-Garcia:2019bku}.
The mock-based approach has the additional advantage of correctly accounting for the excess noise arising from systematic power in $C_{\ell}^{gg}$ at low $\ell$.

\subsection{Excess low-$\ell$ power in quasar autocorrelation}
\label{sec:qso_auto}

The quasar auto-correlation disagrees significantly with theoretical expectations at low $\ell$ (Fig.~\ref{fig:clgg}). 
Due to strong excess power at low $\ell$, mask deconvolution in NaMaster yields negative binned bandpowers at low $\ell$.  To prevent this issue, we instead plot the convolved bandpowers, averaged over bins of width $\Delta \ell = 10$ and starting at $\ell_{\textrm{min}} = 2$,
divided by $f_{\textrm{sky}}$ for each sample.
Modifying the quasar sample to remove likely stars and faint quasar targets significantly reduces this excess power (blue to red line in Fig.~\ref{fig:clgg}). Applying systematics weights also reduces the low-$\ell$ excess power in all samples except DES (dashed lines).
As discussed in Section~\ref{sec:qso_target_selection}, for DES we must also remove the region below $\delta = -30^{\circ}$ to reduce the large-scale power.


Nevertheless, there is still significant excess power at low-$\ell$ even after applying the systematics weights.
This is possibly due to the clustering of the residual stellar contaminants, $f_{\rm star}^2 C_{\ell}^{\star \star}$. Measuring the stellar power spectrum using Gaia stars with $12 < G < 17$ and assuming $f_{\rm star} = 0.026$, $f_{\rm star}^2 C_{\ell}^{\star \star}$ exceeds the $C_{\ell}^{gg}$ theory curve at $\ell < 20$.
If the stars' only impact is extra power, it can be removed by linear regression; however, stars may also reduce the density of nearby quasars by occulting them or raising the effective sky background. Disentangling these competing effects can be difficult with linear regression. We try to mitigate this effect by first subtracting an estimate for the stellar contamination, $f_{\rm star} \bar{n}_{\rm qso, tot}/\bar{n}_{\star} N_{\star}$, before applying the systematics weights (where $\bar{n}_{\rm qso, tot}$ is the total observed density of quasar targets, including stars).
However, this does not reduce the excess power in the unweighted low-$\ell$ power spectrum (similar to the findings of \cite{Chaussidon22}).
This perhaps suggests that a different stellar template is neeeded, since the stellar contaminants occupy a very particular region in color space that may have a different spatial distribution from all stars. Using only samples of spectroscopically confirmed quasars is another option, which will soon be available when early DESI data covers a sufficiently large fraction of the sky.

Given the significant excess power at low $\ell$, we conclude that the DESI quasar targeting data is not sufficiently clean to use the low-$\ell$ autocorrelation.
This is similar to previous work constraining $f_{\textrm{NL}}$ from quasar samples \citep{Pullen13,Giannantonio14a}, who concluded that the quasar samples should only be used in cross-correlation, where correlated systematics are much less likely
(though see \citep{Leistedt14} for an attempt to fully remove clustering systematics in the quasar autocorrelation).

\subsection{Mock-based covariance}
\label{sec:mock_covariance}

To estimate the covariance (shown in Fig.~\ref{fig:covariance}), we follow Ref.~\cite{Ho08} and cross-correlate CMB lensing
maps from 300 realistic Planck mocks (after applying the mean-field correction) with the galaxy maps from the data. Despite lacking
correlations between the CMB lensing field and the galaxies, this approach correctly accounts
for the excess noise arising from systematic power in the quasars at low $\ell$.
It leads to two sources of bias compared to the correct covariance matrix. First,
because the galaxy and CMB lensing fields are not correlated, the measured
covariance does not contain the $(C_{\ell}^{\kappa g})^2$ term, the ``correlation
bias.'' 
However, the covariance is dominated by the product $C_{\ell}^{g g} N_{\ell}^{\kappa \kappa}$ ($>5\times$ larger than $(C_{\ell}^{\kappa g})^2$), so the correlation bias is $<5\%$ of the estimated covariance. 
Second, there is a ``realization bias'' arising
from the fact that we use only one realization of the galaxy field rather than many.
In Ref.~\cite{Ho08}, the realization bias was argued to be small, but this argument relies on the fact that their broad bins average over many modes. This is not the case for our low $\ell$ bins, which are crucial in constraining $f_{\rm NL}$.
We estimate the realization bias from Gaussian mocks and apply the correction
factor to the covariance matrix.

To estimate the realization bias, we begin by generating 300 Gaussian mock skies with correlated $\kappa$ and galaxy fields. We compute the Gaussian covariance $\textbf{C}^{\rm full}$ from these 300 mocks.
We then generate a further single uncorrelated galaxy field, measure its cross-correlation with the 300 $\kappa$ mocks, and compute the single-realization covariance $\textbf{C}^i$. The realization bias
is then the ratio of the matrices
$\textbf{C}^{\rm full}$ to $\textbf{C}^i$.
These matrices are diagonal dominated but still have significant off-diagonal structure.
If the matrices were diagonal, the realization bias would be simply defined as the ratio of their diagonal elements.
However, their off-diagonal elements
may be different as well. 

We enforce that the single-realization covariance matrix derived from the Planck simulations has the same eigenvectors as the covariance matrix from the 300 Gaussian simulations.
We define the eigenvectors $\textbf{Q}^{\textrm{full}}$ and eigenvalues
$\mathbf{\Lambda}
    ^{\textrm{full}}$
\begin{equation}
     (\textbf{Q}^{\textrm{full}})^T \textbf{C}^{\textrm{full}} \textbf{Q}^{\textrm{full}} \equiv
    \mathbf{\Lambda}
    ^{\textrm{full}} 
\label{eqn:4.8}
\end{equation}
We then rotate $\textbf{C}^{\textrm{data}}$ and $\textbf{C}^{i}$
into the basis defined by $\textbf{Q}^{\textrm{full}}$
\begin{equation}
     (\textbf{Q}^{\textrm{full}})^T \textbf{C}^{\textrm{data}} \textbf{Q}^{\textrm{full}} \equiv
    \mathbf{\Lambda}
    ^{\textrm{data,full}}   
\end{equation}
\begin{equation}
    (\textbf{Q}^{\textrm{full}})^T \textbf{C}^{i} \textbf{Q}^{\textrm{full}} \equiv 
    \mathbf{\Lambda}
    ^{i, \textrm{full}}
\end{equation}
We scale $\mathbf{\Lambda}^{\textrm{data, full}}$
by the ratio of $\mathbf{\Lambda}^{\textrm{full}}/\mathbf{\Lambda}^{i,\textrm{full}}$ and neglect its off-diagonal elements
\begin{equation}
   \mathbf{\Lambda}^{\textrm{data, rb}} \equiv  \delta_{ij} \mathbf{\Lambda}^{\textrm{data, full}}_{ij} \frac{\mathbf{\Lambda}^{\textrm{full}}_{ij}}{\mathbf{\Lambda}^{i,\textrm{full}}_{ij}}
\end{equation}
We then rotate the basis back to find $\textbf{C}^{\textrm{data, rb}}$
\begin{equation}
    \textbf{C}^{\textrm{data, rb}}_{ij}
\equiv \textbf{Q}^{\textrm{full}} \mathbf{\Lambda}^{\textrm{data, rb}} (\textbf{Q}^{\textrm{full}})^T
\label{eqn:4.12}
\end{equation}
We test this method on simulations and find that it adds negligible variance ($\leq 5\%$) compared to the statistical error.
We show the realization bias in the right panel of Fig.~\ref{fig:covariance}.

\begin{table}[]
    \centering
    \begin{tabular}{l|ccc|ccc}
       \multirow{2}{*}{Region}  &  Fisher  & Fisher + $N_\ell^{\textrm{syst}}$ & Fisher + $N_\ell^{\textrm{syst}}$ & Mock Cov. & Mock Cov. & Analytic \\
       & (no $N_\ell^{\textrm{syst}}$) & (+ wts.) & (no wts.) & & Transfer & Cov. \\
       \hline
       North & $0 \pm 60.0$ &  $0 \pm 83.1$ & $0 \pm 123.6$ & $17^{+120}_{-84}$   & $17^{+115}_{-81}$  & $17^{+124}_{-87}$ \\
       DECaLS N & $0 \pm 55.8$ & $0 \pm 76.7$ & $0 \pm 100.4$ & $18^{+128}_{-87}$ &  $23^{+130}_{-90}$ & $14^{+112}_{-80}$ \\
       DECaLS S & $0 \pm 71.3$ & $0 \pm 107.9$ & $0 \pm 164.1$  & $32^{+208}_{-135}$  & $39^{+205}_{-135}$ & $29^{+185}_{-119}$ \\
       DES & $0 \pm 95.6$ & $0 \pm 126.7 $ & $0 \pm 165.3$ & $35^{+230}_{-145}$   & $37^{+237}_{-151}$ & $35^{+223}_{-138}$ \\
       \hline
       Total & $0 \pm 33.2$ & $0 \pm 46.5$ & $0 \pm 64.8$ & $-7^{+55}_{-48}$ & $-5^{+56}_{-48}$ & $-7^{+53}_{-46}$

    \end{tabular}
    \caption{
    Fisher forecasts and constraints from noiseless mocks. Fisher forecasts are run for three cases: using the theory curve for $C_{\ell}^{gg}$ in Eq.~\ref{eq:knox_cov}; adding realistic low-$\ell$ noise (denoted by $N_\ell^{\textrm{syst}}$), partially mitigated by systematics weights  (dashed red line in Fig.~\ref{fig:clgg}); 
    and adding low-$\ell$ noise without any systematics mitigation (solid red line in Fig.~\ref{fig:clgg}).
    For the noiseless mocks, we use
    both the fiducial mock-based covariance, requiring a multivariate $t$-distribution likelihood to account for uncertainties in the covariance, and an analytic Gaussian covariance. We also show the impact of applying a correction factor for residual overfitting, as measured from $C_{\ell}^{\kappa g}$ of 100 contaminated mocks (Fig.~\ref{fig:transfer_contaminated}).}
    \label{tab:fisher}
\end{table}

\section{Mock Tests}
\label{sec:mock_tests}

\subsection{Contaminated mocks}
\label{sec:contaminated_mocks}

Given the strong contamination in the quasar catalogs, systematics weights are crucial in reducing low-$\ell$ noise in the CMB lensing cross-correlation and thus tightening $f_{\textrm{NL}}$ constraints (Fig.~\ref{fig:clgg}).
However, over-subtraction of systematics templates can decorrelate the quasar map with the true density field at low $\ell$, removing precisely the cosmological information that we need to constrain $f_{\textrm{NL}}$. We therefore create a set of contaminated mocks with similar systematics trends and contamination power as the data, but with a correlated CMB lensing field.

We generate systematics-correcting weights
and measure the cross-correlation of the systematics-corrected,
contaminated mocks with the mock CMB lensing maps, to ensure
that we can recover an unbiased cross-correlation on all scales.
We use this test to determine
whether to use a linear or nonlinear (Random Forest) systematics correction, and to 
determine
the number of imaging templates that we can remove before incurring a bias in $C_{\ell}^{\kappa g}$
on large scales.
We find that Random Forest
generically removes real power at low $\ell$ even with a very small number of templates, and therefore
use linear regression instead.
These tests justify our choices
of imaging templates to remove in Table~\ref{tab:imaging_templates}.

We create contaminated mocks by generating Poisson-sampled Gaussian mocks with number density 10 times higher than in data. We down-sample the mocks so that the average overdensity in any NSIDE=256 pixel is proportional to the inverse of the imaging systematic weight.
This yields mocks with the same imaging systematics trends as the data,
while also preserving the correlated fluctations from the mock density field (Fig.~\ref{fig:contaminated_mocks_imaging_systematic} and~\ref{fig:contaminated_mocks_vs_data}).

The weights $w_i$ used to contaminate the mocks are produced using the Random Forest (RF) method with all 11 templates, and are thus different from the weights that we use for the $f_{\textrm{NL}}$ analysis on data
(from linear regression with templates given in Table~\ref{tab:imaging_templates}).
The random forest weights are useful for this test because they ensure fully realistic contaminated mocks, and also allow for the existence of extra systematics in the contaminated mocks beyond the templates
that we remove.
This mimics the situation in the real data,
where we don't know the  correct imaging templates to remove.
Fig.~\ref{fig:clgg_contaminated_mocks} validates the contaminated mocks: the auto-power spectrum $C_{\ell}^{gg}$ qualitatively matches the data power spectrum both before and after applying weights.\footnote{In principle, one could use the mean of the contaminated mocks in Fig.~\ref{fig:clgg_contaminated_mocks} as a model, and Fig.~\ref{fig:clgg_contaminated_mocks} shows that this model can adequately fit the auto-correlation. However, one would need to determine how the extra contaminant power in the model ($N_\ell^{gg}$) depends on the input theory spectrum (e.g.\ by interpolating over $N_\ell^{gg}$ determined from many simulations with different $f_{\textrm{NL}})$. Furthermore, one would need to propagate uncertainty in the model for $N_\ell^{gg}$, e.g.\ from uncertainty in the fitted weights used to create the contaminated mocks \citep{Karim23}. By avoiding these complications, our cross-correlation method is considerably simpler and more robust.}

We then process each contaminated mock in exactly the same way as the data: linearly regressing against the contaminant templates in Table~\ref{tab:imaging_templates}, applying the resulting systematic weights, and measuring the cross-spectrum with the correlated CMB lensing field.


In Fig.~\ref{fig:transfer_contaminated}, we show the ratio between the mean of 100 contaminated mocks with systematic weights applied, and the mean of 100 uncontaminated mocks. We compare a variety of methods with linear or Random Forest regression, and a varying number of templates used. We use jackknife resampling to derive the errorbars shown in the figure.
This figure justifies our choice of regression method and imaging templates in Table~\ref{tab:imaging_templates}.

The solid red lines in Fig.~\ref{fig:transfer_contaminated}
show that Random Forest is unsuitable for this work because it leads
to a significant drop in $C_{\ell}^{\kappa g}$ at $\ell < 50$
measured after systematics correction.
To test if this result is sensitive to the input weights used to generate the contaminated mock, we tried using neural net input weights rather than random forest, and found very similar results (black solid line).
To mitigate overfitting, we first tried Random Forest with a subset of the templates, removing PSFSIZE\_G, PSFSIZE\_R, SGR\_STREAM, and PSFDEPTH\_W1. The first three templates have the least-significant correlations with the density field, and we remove PSFDEPTH\_W1 because it is very degenerate with PSFDEPTH\_W2. However, Random Forest with the smaller set of 7 templates still does not recover $C_{\ell}^{\kappa g}$ well on large scales (dashed red lines).

A linear regression with all 11 templates also leads to overfitting, but less significantly (dashed blue lines). 
To determine which templates to use in the regression, we used a \textsc{Regressis} output called \textit{permutation importance}.
Permutation importance is defined for each feature under consideration and measures how significantly that feature contributes to the fit. It randomly permutes the elements of that feature, measuring the change in goodness of fit (i.e.\ $\chi^2$), and averages this over many random realizations.
Therefore, it measures how much the fit is driven by each parameter.

We started by removing PSFSIZE\_G, PSFSIZE\_R, SGR\_STREAM, and PSFDEPTH\_W1,
which have both low correlation with the contaminated density field and low
permutation importance.
With the restricted set of 7 templates and linear regression, we found adequate recovery of $C_{\ell}^{\kappa g}$ in the North region, but overfitting in the other three regions.

In the other three regions, we removed the templates with the lowest
permutation importance. We also needed to add back SGR\_STREAM and PSFSIZE\_R for DECaLS N, and PSFDEPTH\_W1 for DES,
which had relatively high permutation importance.
Indeed, PSFDEPTH\_W1 had the largest permutation importance of any template in the DES region.
We ultimately settled on the template set in Table~\ref{tab:imaging_templates} (solid blue lines in Fig.~\ref{fig:transfer_contaminated}),
which struck the right balance between removing the most important and correlated features (and thus removing low-$\ell$ power in $C_{\ell}^{gg}$); and avoiding overfitting from too many templates or too complicated a regression method.

The linear regression method and reduced template set that we use differs from the Random Forest method with all 11 templates used in \cite{Chaussidon22}. This is due to our requirement of avoiding overfitting on very large scales and in the CMB-lensing cross-correlation; in contrast, \cite{Chaussidon22} is concerned with the angular correlation function on smaller scales.

With this template set, Fig.~\ref{fig:transfer_contaminated} shows we can correctly recover $C_{\ell}^{\kappa g}$ at all multipoles.
Over 60 bins at $\ell < 300$, we find $\chi^2 = 54.8$, 61.7, 63.9, and 63.3 for North, DECaLS N, DECaLS S, and DES, respectively. We find unbiased recovery of $C_{\ell}^{\kappa g}$, with the mean of ratio $0.9959 \pm 0.0024$, $0.9970 \pm 0.0018$, $0.9924 \pm 0.0054$, and $0.9898 \pm 0.0080$.

Fig.~\ref{fig:transfer_contaminated} shows that we can accurately
recover $C_{\ell}^{\kappa g}$ for contaminated mocks with $f_{\textrm{NL}} = 50$ (dotted lines). Indeed, the ratio between the simulated and recovered $C_{\ell}^{\kappa g}$ is very similar between the $f_{\textrm{NL}} = 50$
mocks
and the default $f_{\textrm{NL}} = 0$ mocks.
Fig.~\ref{fig:transfer_contaminated} also shows that we can accurately recover the cross-correlation for the $z_{\rm phot} > 2$ sample,
with a slightly different set of systematics templates
as described
in Appendix~\ref{sec:highz}.

\begin{figure}
    \centering
    \includegraphics[width=1.0\linewidth]{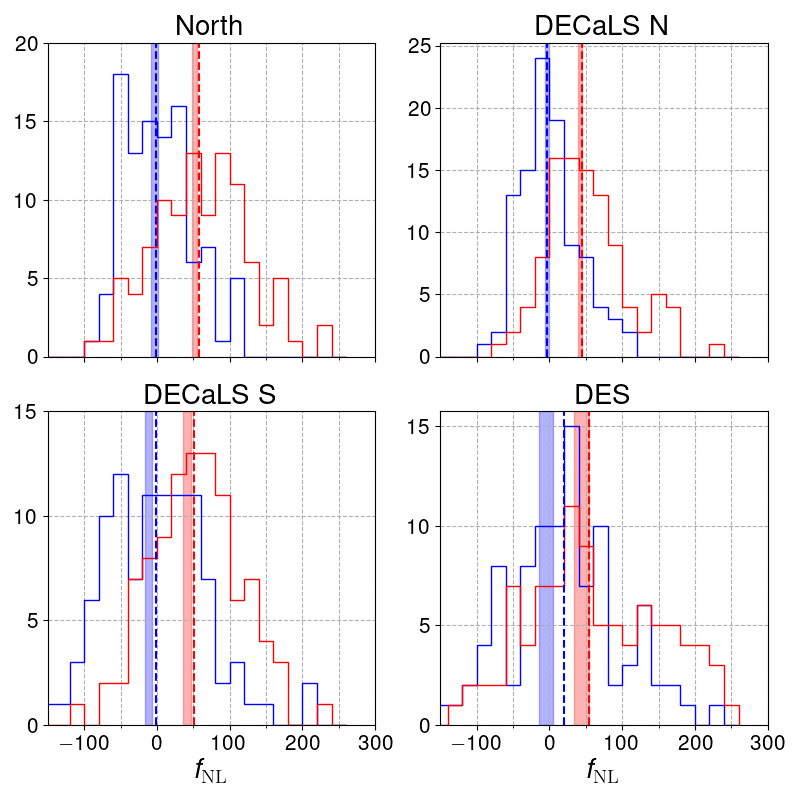}
    \caption{$f_{\textrm{NL}}$ constraints from contaminated mocks, with a true value of $f_{\textrm{NL}} = 0$ (blue) and 50 (red). The histograms show the distribution of the median marginalized $f_{\textrm{NL}}$ for the 100 mocks. The dashed lines are the medians of each of the histograms, and the shaded ranges are the 16th to 84th percentiles of the marginalized distribution for a fit to the mean of the 100 mocks, with the covariance scaled down by 100.}
    \label{fig:fnl_from_mocks}
\end{figure}

We propagate these $C_{\ell}^{\kappa g}$ measurements on contaminated mocks to $f_{\textrm{NL}}$ constraints in Fig.~\ref{fig:fnl_from_mocks}.
For each mock, we measure $f_{\textrm{NL}}$ using the analytic covariance matrix for each region and flat priors on $f_{\textrm{NL}}$ between -500 and 1000 and on $b_0$ between 0.5 and 2.
We measure both the distribution of median marginalized $f_{\textrm{NL}}$ and $b_0$ from the 100 mocks individually, and
$f_\textrm{NL}$ from the mean of the 100 mocks, with the covariance scaled down by a factor of 100.
Fig.~\ref{fig:fnl_from_mocks}
shows the distribution of the marginalized
median $f_{\textrm{NL}}$ from all mocks,
and Table~\ref{tab:fnl_from_mocks} gives key summary statistics.
As the $f_{\textrm{NL}}$ error gets large (as for DES and DeCALS S), we find that some simulations can scatter to very high values of $f_{\textrm{NL}}$, biasing the mean recovered $f_{\textrm{NL}}$ and even the median. However, if we fit $f_{\textrm{NL}}$ to the average power spectrum from the 100 simulations and scale down the covariance, the long tails in the PDF are suppressed and the fitted value is much closer to the input value.

We find that we can successfully recover
the input value of ${f_\textrm{NL}}$ for both $f_{\textrm{NL}} = 0$ and $f_{\textrm{NL}} = 50$ mocks,
with biases $\Delta f_{\textrm{NL}} \lesssim 10$. This is acceptable given our statistical error bars of $\sigma_{f\textrm{NL}} = 40$.
These mocks are created with less noise than the data, to allow us to use fewer mocks to test the $C_{\ell}^{\kappa g}$ pipeline. Therefore, the standard deviations of the median $f_{\textrm{NL}}$ are somewhat smaller than the data errorbars in Table~\ref{tab:data1} or the Fisher forecasts in Table~\ref{tab:fisher}. To check the simulations, we determine the Fisher errorbars for the mock setups: no CMB noise and 10 times higher number density than in the data. These agree reasonably well with the $f_{\textrm{NL}}$ standard deviation in the contaminated mocks for $f_{\textrm{NL}} = 0$. As expected, $f_{\textrm{NL}} = 50$ has a higher standard deviation because the true $C_{\ell}^{gg}$ and $C_{\ell}^{\kappa g}$ are larger, increasing the covariance. Finally, we find that the recovered $b_0$ matches the input extremely well, with deviations of $\lesssim 2\%$.

\begin{table}[]
    \centering
    \small
    \setlength{\tabcolsep}{1.5pt}
    \begin{tabular}{l|cc|cc|c|cc|cc}
   \multirow{3}{*}{Region} &  \multicolumn{4}{c}{$f_{\textrm{NL}}$ fit} & & \multicolumn{4}{c}{$b_0$ fit}  \\
  &  \multicolumn{2}{c}{$f_{\textrm{NL}} = 50$} & \multicolumn{2}{c}{$f_{\textrm{NL}} = 0$} & Fisher & \multicolumn{2}{c}{$f_{\textrm{NL}} = 50$} & \multicolumn{2}{c}{$f_{\textrm{NL}} = 0$}  \\
   & Ind. & Mean & Ind. & Mean & $\sigma_{f\textrm{NL}}$ & Ind. & Mean & Ind. & Mean \\
    \hline
    North & $58 \pm 68$ & $52 \pm 4$ & $-1 \pm 47$ & $-3 \pm 4$ & 48 & $0.98 \pm 0.05$ & $0.99 \pm 0.005$ & $0.99 \pm 0.03$ & $1.00 \pm 0.003$ \\
    DECaLS N & $45 \pm 54$ & $42 \pm 3$ & $-3 \pm 41$ & $-3\pm 4$ & 44 & $0.99 \pm 0.04$ & $0.99 \pm 0.004$ & $0.99 \pm 0.03$ & $0.99 \pm 0.003$ \\
    DECaLS S & $50 \pm 67$ & $42 \pm 5$ & $-2 \pm 68$ & $-12 \pm 5$ & 49 & $0.99 \pm 0.05$ & $1.00 \pm 0.003$ & $0.99 \pm 0.05$ & $1.00 \pm 0.005$ \\
DES & $54 \pm 106$ & $45 \pm 10$ & $20 \pm 89$ & $-4 \pm 10$ & 111 & $0.98 \pm 0.07$ & $0.99 \pm 0.003$ & $0.98 \pm 0.05$ & $1.00 \pm 0.005$ \\
    \end{tabular}
    \caption{ $f_{\textrm{NL}}$ measurements from the contaminated mock $C_{\ell}^{\kappa g}$ in Fig.~\ref{fig:transfer_contaminated}. For each true value of $f_{\textrm{NL}}$, the left columns (labelled ``Ind.'') give the mean and standard deviation of the median marginalized $f_{\textrm{NL}}$ from each  of the 100 mocks. The right columns (labelled ``Mean'') give the median marginalized $f_{\textrm{NL}}$ from a fit to the mean of the 100 mocks}, with the covariance scaled down by a factor of 100. The Fisher column is a Fisher forecast for the $f_{\textrm{NL}}$ error, using the setup of the contaminated mocks (no lensing noise and 10 times higher number density than the data).
    \label{tab:fnl_from_mocks}
\end{table}

\subsection{Noiseless mock tests}
\label{sec:noiseless_mocks}
 

 We estimate an error budget for our measurement by comparing our constraints to  a Fisher forecast. In the Fisher forecast, we use
the Knox formula (equation~\ref{eq:knox_cov}) for the covariance, set $\ell_{\rm min} = 6$, and test
both including and excluding the excess low-$\ell$ power, $N_\ell^{\textrm{syst}}$, in the $C_{\ell}^{gg}$ term.

Table~\ref{tab:fisher} shows constraints on $f_{\textrm{NL}}$ using noiseless mocks with the same covariance as the data. In this case, the ``noiseless mock'' is the $f_{\textrm{NL}} = 0$ theory curve multiplied by the bandpower window functions of Fig.~\ref{fig:bandpower_window}.
The fiducial covariance matrix is measured from 300 Planck mocks, with the realization bias correction as described in Section~\ref{sec:covariance}. To account for uncertainties in the covariance matrix
due to the finite number of mocks, we use the multivariate $t$ distribution rather than a Gaussian distribution \citep{Percival21}. 
Throughout this work, we quote the median of the marginalized posterior and distance to the 16th and 84th percentiles as the upper and lower errorbars.

We find good agreement between the Fisher errorbar (including the additional low-$\ell$ systematic power, i.e.\ $C_{\ell}^{gg}$ from the dashed red line in Fig.~\ref{fig:clgg}) and the errorbar from the noiseless mocks ($\sigma_{f\textrm{NL}} = 47$ vs.\ 52). 
The constraining power is significantly degraded by the extra noise, with $\sigma_{f\textrm{NL}}$ increasing by 50\%.
On the other hand, if we didn't apply any systematics weights at all, $\sigma_{f\textrm{NL}}$ would be 60\% higher ($\sigma_{f\textrm{NL}} = 65$).
To test the covariance matrix, we replace the mock-based covariance with an analytic covariance from NaMaster (switching back to a Gaussian likelihood); we find very similar $f_{\textrm{NL}}$ constraints. 
Furthermore, we find that if we do not apply the matrix rotation of Section~\ref{sec:covariance} and instead use the ``raw'' mock-based covariance, the combined $f_{\textrm{NL}}$ constraint changes from $-7^{+55}_{-48}$ to $-5^{+53}_{-46}$, so this rotation has a very small impact on the constraints.
We also test the simpler diagonal Knox formula for the covariance (Eq.~\ref{eq:knox_cov}), which underestimates the $f_{\textrm{NL}}$ errorbar by 10\%, due to the strong correlations between the narrow low-$\ell$ bins (Fig.~\ref{fig:covariance}). Therefore, we use the Planck mocks for our fiducial covariance matrix.

Finally, we test the impact of potential biases in $C_{\ell}^{\kappa g}$ recovery from linear regression of the systematics templates.
While  we do not detect a significant difference
between the input and measured $C_{\ell}^{\kappa g}$ after applying the systematics weights (Fig.~\ref{fig:transfer_contaminated}), it is possible that
the contaminated $C_{\ell}^{\kappa g}$ does have a bias that is smaller than the errorbars. Therefore, we apply a correction factor to the noiseless mock data vector, defined as the ratio between the input and output $C_{\ell}^{\kappa g}$ in Fig.~\ref{fig:transfer_contaminated}. This leads to almost no difference on the $f_{\textrm{NL}}$ constraint, with only a 1\% increase in the errorbar.
We therefore conclude that any residual overfitting does not affect our measurement.


\section{Measurement on data}
\label{sec:measurement}

\begin{figure}
    \centering
    \includegraphics[width=1.0\linewidth]{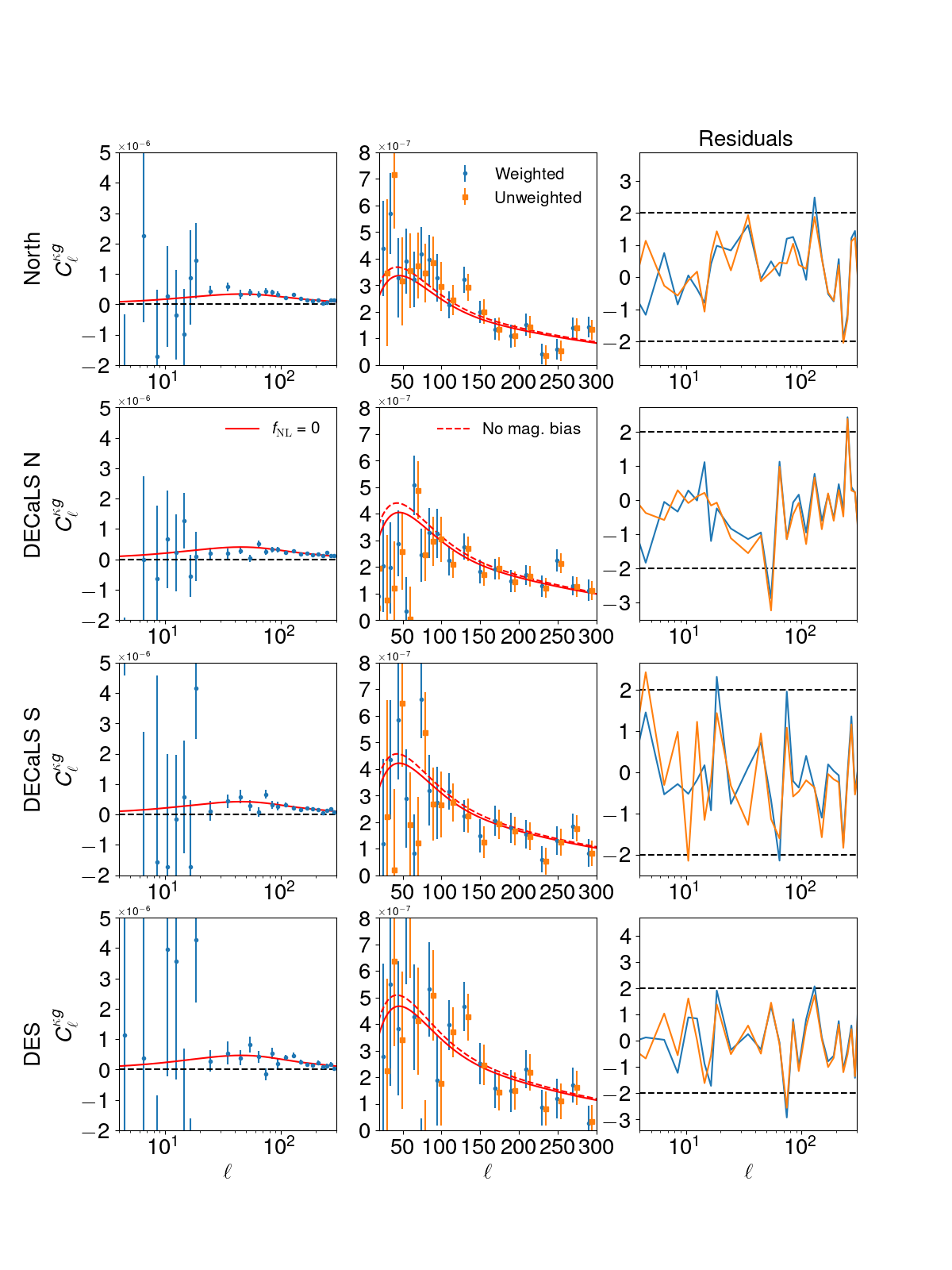}
    \caption{Quasar-CMB lensing cross-correlation, split into the four imaging regions. The central panel is a zoom-in on $C_{\ell}^{\kappa g}$ at $20 < \ell < 300$. The right panel gives the residuals from the theory curve with $f_{\textrm{NL}} = 0$ and the best-fit bias.
    The blue and orange points correspond to the cross-correlation measured with the weighted or unweighted galaxy field. The solid red line is the theory curve with $f_{\textrm{NL}} = 0$ and the best-fit bias for each imaging region from Table~\ref{tab:data1}; the dashed red line is the same theory curve, but excluding magnification bias.}
    \label{fig:data}
\end{figure}

The measured $C_{\ell}^{\kappa g}$ is shown in Fig.~\ref{fig:data} and compared to a theory curve with $f_{\textrm{NL}}$ fixed to zero. We detect the cross-correlation with total signal-to-noise (S/N) 26.2 at $\ell < 300$ combining all 4 regions. 
Signal-to-noise is defined as the difference between the $\chi^2$ with respect to no detection, and the best-fit $\chi^2$:
S/N $\equiv \sqrt{\vec{d}^T C^{-1} \vec{d} - (\vec{d} - \vec{t})^T C^{-1} (\vec{d} - \vec{t})}$
where $\vec{d}$ is the data vector
and $\vec{t}$ is the best-fit theory vector.
The covariance $C$ is corrected for realization
bias following Eqs.~\ref{eqn:4.8} to ~\ref{eqn:4.12}.
In the individual regions, the cross-correlation is detected at S/N 14.3, 15.9, 13.2, and 7.5 for North, DECaLS N, DECaLS S, and DES, respectively.

\begin{figure}
    \centering
    \includegraphics[width=1.0\linewidth]{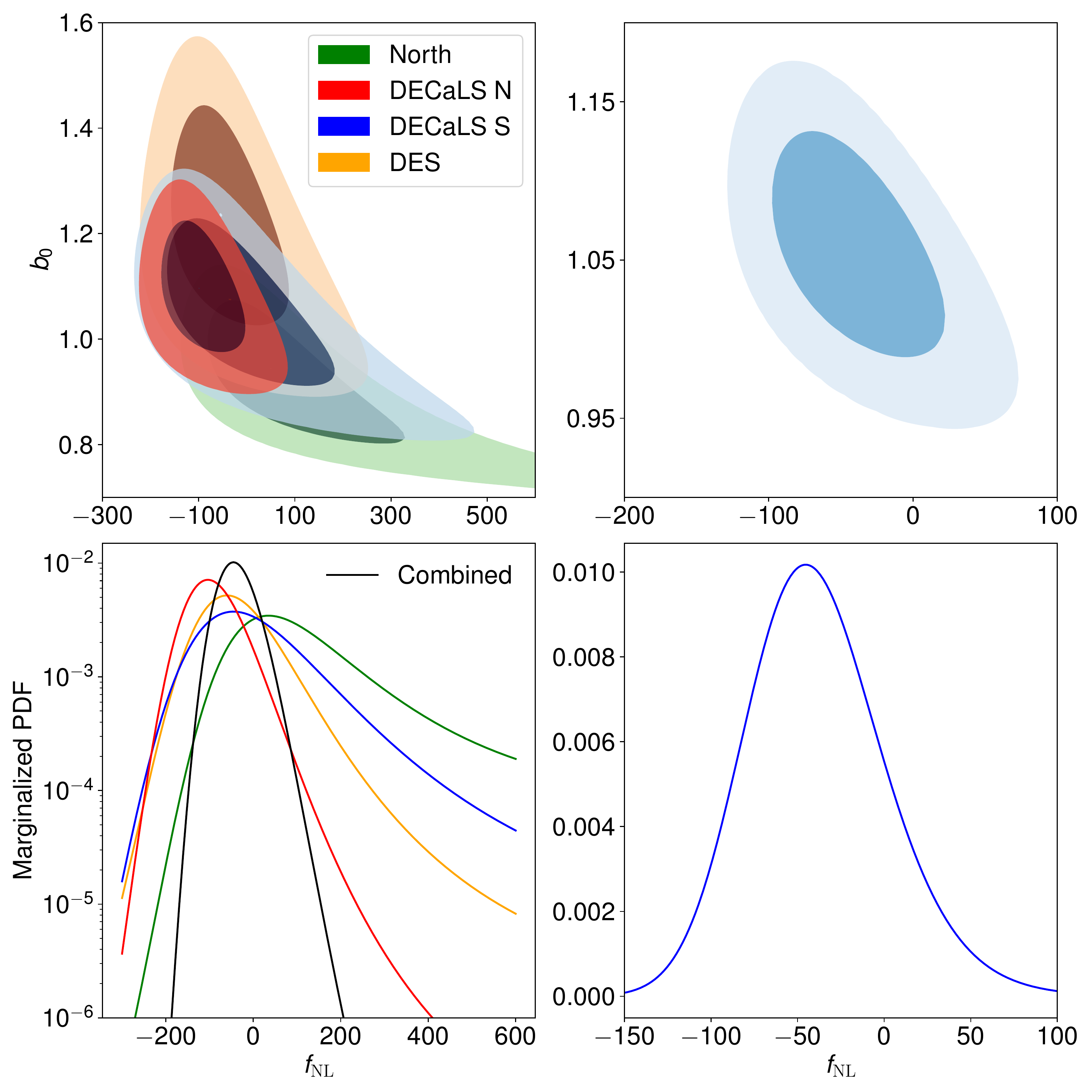}
    \caption{Likelihood for $f_{\textrm{NL}}$ and linear bias, divided into four imaging regions (\textit{left}) and combined (\textit{right}). Bottom panel shows marginalized posterior on $f_{\textrm{NL}}$.}
    \label{fig:likelihood}
\end{figure}

Table~\ref{tab:data1} describes the marginalized $f_{\textrm{NL}}$ and linear bias constraints from data. 
All bias constraints are presented in terms of the scaling amplitude $b_0$ defined in Eq.~\ref{eqn:laurent_b}, and the best-fit measurement of $b_0 = 1.09$ corresponds to $b(z_{\textrm{eff}} = 1.51) = 2.38$.
We find $f_{\textrm{NL}} = -26^{+45}_{-40}$ and a good fit with $\chi^2 = 108.4$ over 90 degrees of freedom. The maximum of the posterior is very similar to the marginalized medians, $f_{\textrm{NL}} = -28$ and $b_0 = 1.09$. The posterior on $f_{\textrm{NL}}$ is notably asymmetric (Fig.~\ref{fig:likelihood}). The asymmetry is worse in the individual regions, where the constraints are poorer.
For the combined constraint, a Gaussian provides a poor fit to the likelihood, and it is best to use the full PDF rather than a summary statistic.

We perform a number of systematics tests on the data, and describe the results in Tables~\ref{tab:data1} and~\ref{tab:data2}.
Table~\ref{tab:data1} first shows tests in which we use the fiducial data vector but change the likelihood or theory, verifying that the results are robust to changing from the mock-based to analytic covariance.\footnote{We also find that our results
change little if we do not perform the realization bias correction of Sec.~\ref{sec:mock_covariance}, dropping $f_{\textrm{NL}}$ to $-42^{+40}_{-37}$ (0.4$\sigma$ shift) and increasing the best-fit
$\chi^2$ to 114.9. We find in tests on simulations
that uncertainty in the realization bias is negligible compared to our statistical errors.} If we use $p=1$ instead of $p=1.6$, the constraints become tighter ($\sigma_{f\textrm{NL}} = 43$ to 28) since the response to $f_{\textrm{NL}}$ is stronger.
Likewise, using $\sigma_8 = 0.77$ \citep[e.g.][]{Hikage19,Heymans21,DES_Y3,Krolewski21,White22,GarciaGarcia21,Philcox_Ivanov,Chen22} raises the linear bias and therefore also tightens constraints on $f_{\textrm{NL}}$ 
to  $-25^{+39}_{-35}$.
On the other hand, using an alternative bias evolution which better fits the high-redshift quasar clustering (Appendix~\ref{sec:highz}) lowers the high-redshift bias and therefore weakens the $f_{\textrm{NL}}$ constraints,  $f_{\textrm{NL}} = -54^{+51}_{-45}$.

Varying the redshift distribution
within the options plotted in Fig.~\ref{fig:dndz} yields tiny changes in the median marginalized $f_{\textrm{NL}}$. 
Changing to the ``spectroscopic only'' $dN/dz$ changes $f_{\textrm{NL}}$ by $\sim 0.1$. Using the $dN/dz$ measured in North or DECaLS S changes $f_{\textrm{NL}}$ by $\sim 4$, whereas using $dN/dz$ from DECaLS N changes $dN/dz$ by $\sim 0.2$. The largest change is from using the VI redshift distribution,
which shifts $f_{\textrm{NL}}$ by $\sim 9$. However, most of this shift is due to the larger noise in the VI $dN/dz$, which has nearly two orders of magnitude fewer quasars. Therefore, uncertainty in the redshift distribution has an impact of $\Delta f_{\textrm{NL}} \lesssim 5$.

The main cosmological parameter
that matters for the $f_{\textrm{NL}}$ inference
is $\sigma_8$, since it controls
the amplitude of the power spectrum
(and hence the bias). 
The other parameter that affects the amplitude is the neutrino mass,
which changes the relationship
between large and small scales by creating a step in the power spectrum
at $k \sim 0.1$ $h$ Mpc$^{-1}$. This affects our $f_{\textrm{NL}}$ measurement, which relies on measuring the bias on $k \sim 0.1$ $h$ Mpc$^{-1}$ scales to calibrate the theory prediction for $f_{\textrm{NL}}$ on large scales. Hence, changing the neutrino mass from its fiducial minimum value of 0.06 eV modestly shifts $f_{\textrm{NL}}$. Fixing $m_{\nu}$ to 0.20 eV (approximately the current
upper limit \cite{PlanckLegacy18}) causes $f_{\textrm{NL}}$ to drop by $\sim 7$. Hence future high-precision $f_{\textrm{NL}}$ measurements
from $C_{\ell}^{\kappa g}$ (with $\sigma_{f\textrm{NL}} < 10$) should likely marginalize over $m_{\nu}$ as well.
The impact of varying the other five parameters within the Planck uncertainties are tiny, and hence
we fix them to their best-fit values.

Table~\ref{tab:data2} shows tests in which we change the data vector: removing the Galactic
$m=0$ mode ($-19^{+47}_{-41}$),
using $\ell_{\rm min} = 8$ ($-8^{+52}_{-45}$),
and using the tSZ-free lensing map ($-25^{+69}_{-57}$).
If we don't apply the systematics weights to the quasars, we obtain a statistically compatible but weaker constraint, $-5^{+64}_{-56}$,
showing that the weights mainly affect the covariance by reducing low-$\ell$ noise (see Fig.~\ref{fig:data} to compare $C_{\ell}^{\kappa g}$ measured with and without the quasar weights).
Interestingly, $f_{\textrm{NL}}$ shifts away from zero for three of the four samples both when turning off systematics weights and when using the tSZ-free lensing map, suggesting that both combinations may be affected
by residual systematics.

There are no significant differences in marginalized $f_{\textrm{NL}}$ constraints between any of the regions.
All except DES have acceptable $\chi^2$; for DES, the $\chi^2$ is slightly high ($p=0.004$).
If we remove DES altogether, we find $f_{\textrm{NL}} = -49^{+46}_{-41}$.
Alternatively, we find acceptable fits for DES when using either the tSZ-free lensing map or when removing the $m=0$ Galactic mode.
If we replace the fiducial DES cross-correlation with the tSZ-free one, we find $f_{\textrm{NL}} = -30^{+46}_{-40}$;
and if we instead use the $m=0$ DES cross-correlation, we find $f_{\textrm{NL}} = -27^{+45}_{-40}$.
Hence, our fiducial results are not driven by the DES region.

The overall errors are similar to the Fisher
forecasts in Table~\ref{tab:fisher}. The individual regions show some differences; this is because the best-fit $b_0$ differs from one, which can substantially change $\Delta b$ and thus sensitivity to $f_{\textrm{NL}}$.

We fix the magnification bias slope $s$ in addition to fixing the
cosmological parameters. The formal statistical error on $s$ is extremely
small due to the large number of sources used to measure the slope.
Hence the uncertainty on $s$ is dominated by systematic errors.
The most rigorous way to determine $s$ is by synthetic source injection
\citep{ElvinPoole22,Everett22}. Ref.~\cite{ElvinPoole22} finds a typical
systematic error of $\Delta s \sim 0.1-0.2$ between source injection
and the simpler flux modification method that we use.
This can propagate to a fairly large change in $f_{\textrm{NL}}$
(though still subdominant to our statistical errors): $\Delta s = 0.1$
yields $\Delta_{f\textrm{NL}} \sim -15$ (anti-correlated with $s$).
This correlation will asymmetrically broaden the $f_{\textrm{NL}}$ constraint,
as it has a much larger impact ($|\Delta_{f\textrm{NL}}| \sim 15$--$20$)
on $f_{\textrm{NL}} \sim 0$ and negative $f_{\textrm{NL}}$ than on $f_{\textrm{NL}} \sim 100$ ($|\Delta_{f\textrm{NL}}| \sim 5$).
However, this systematic error is dominated by size selection effects
(i.e.\ lensing increases size while keeping surface brightness
constant, so galaxy magnitudes measured in a fixed aperture
do not increase by the full amount of magnification).
Because the quasar sample is restricted to point sources, these effects
are much smaller here and thus the typical systematic error
is likely smaller than $\Delta s = 0.1$. 
If we conservatively account for magnification uncertainty by marginalizing over $s$ with a Gaussian
prior with $\sigma = 0.1$, centered on the measured value, our results are nearly unchanged (bottom row of Table~\ref{tab:data1}). However, future measurements with $\sigma_{f\textrm{NL}} < 15$ will require more accurate magnification bias slope measurement, e.g.\ validating the magnification bias slope measurement via source-injection simulations.






\begin{table}[]
\renewcommand{\arraystretch}{1.2}
    \centering
    \begin{tabular}{c|c|ccc}
       Version & Region  &  $f_{\textrm{NL}}$ & $b_0$ & $\chi^2$ \\
       \hline
       \multirow{4}{*}{Fiducial} & North  &  $130^{+186}_{-119}$ & $0.92^{+0.08}_{-0.09}$ & 29.7 \\[0pt]
        & DECaLS N  &  $-67^{+62}_{-58}$ & $1.10^{+0.07}_{-0.09}$ & 18.0 \\[0pt]
        & DECaLS S  &  $-66^{+116}_{-86}$ & $1.14^{+0.10}_{-0.12}$ &  18.3\\[0pt]
        & DES  &  $-34^{+106}_{-81}$ & $1.27^{+0.13}_{-0.15}$ &  42.4 \\[0pt]
        \cline{2-5}
        & Combination & $-26^{+45}_{-40}$ & $1.09^{+0.04}_{-0.05}$ & 108.4 \\[0pt]
        \hline
       \multirow{4}{*}{$p=1$}  
        &  North  &  $60^{+76}_{-59}$ & $0.94^{+0.08}_{-0.09}$ & 29.8 \\
       &  DECaLS N  &  $-45^{+46}_{-41}$ & $1.10^{+0.07}_{-0.09}$ & 18.0 \\
       &  DECaLS S  &  $-50^{+74}_{-61}$ & $1.15^{+0.09}_{-0.11}$ & 18.3 \\
       & DES  &  $-28^{+70}_{-60}$ & $1.28^{+0.14}_{-0.15}$ & 42.4 \\
        \cline{2-5}
       & Combination & $-18^{+29}_{-27}$ & $1.09^{+0.04}_{-0.05}$ & 108.5  \\
       \hline
       \multirow{4}{*}{Analytic cov.}  & North  &  $101^{+160}_{-105}$ & $0.95^{+0.08}_{-0.09}$ & 24.8  \\
        & DECaLS N  &  $-100^{+60}_{-53}$ & $1.10^{+0.07}_{-0.08}$ &  23.2 \\
        & DECaLS S  &  $-16^{+130}_{-89}$ & $1.06^{+0.10}_{-0.11}$ &   22.7 \\
        & DES  &  $-19^{+94}_{-73}$ & $1.27^{+0.12}_{-0.16}$ & 47.9 \\
        \cline{2-5}
        & Combination & $-32^{+41}_{-38}$ & $1.08^{+0.04}_{-0.05}$ & 118.6  \\
           \hline
       \multirow{4}{*}{$\sigma_8 = 0.77$}  
        &  North  &  $102^{+145}_{-97}$ & $0.97^{+0.09}_{-0.09}$ & 29.7 \\
       &  DECaLS N  &  $-61^{+63}_{-53}$ & $1.15^{+0.08}_{-0.09}$ & 18.0 \\
       &  DECaLS S  &  $-62^{+101}_{-79}$ & $1.20^{+0.10}_{-0.12}$ & 18.3 \\
       &  DES  &  $-34^{+93}_{-75}$ & $1.33^{+0.15}_{-0.16}$ & 42.4 \\
        \cline{2-5}
       & Combination & $-25^{+39}_{-35}$ & $1.14^{+0.04}_{-0.06}$ & 108.4 \\
       \hline
       \multirow{4}{*}{Alternate $b(z)$}  
        &  North  &  $105^{+197}_{-128}$ & $1.01^{+0.09}_{-0.09}$ & 29.0 \\
       &  DECaLS N  &  $-95^{+82}_{-69}$ & $1.18^{+0.08}_{-0.09}$ & 19.1 \\
        &  DECaLS S  &  $-104^{+127}_{-95}$ & $1.25^{+0.11}_{-0.13}$ & 18.0 \\
       &  DES  &  $-59^{+117}_{-91}$ & $1.38^{+0.15}_{-0.16}$ & 42.3 \\
        \cline{2-5}
       & Combination & $-54^{+51}_{-45}$ & $1.18^{+0.04}_{-0.05}$ & 108.4  \\
       \hline
       \multirow{4}{*}{Marginalize $s$}  
        &  North  &  $129^{+194}_{-124}$ & $0.91^{+0.09}_{-0.10}$ & 29.6 \\
       &  DECaLS N  &  $-66^{+72}_{-63}$ & $1.10^{+0.08}_{-0.10}$ & 17.8 \\
        &  DECaLS S  &  $-68^{+118}_{-91}$ & $1.13^{+0.11}_{-0.12}$ & 18.3 \\
       &  DES  &  $-35^{+105}_{-83}$ & $1.27^{+0.14}_{-0.16}$ & 42.3 \\
        \cline{2-5}
       & Combination & $-26^{+46}_{-41}$ & $1.08^{+0.05}_{-0.05}$ & 108.0  \\
    \end{tabular}
    \caption{Marginalized constraints on $f_{\textrm{NL}}$ and linear bias scaling amplitude $b_0$ using the fiducial Planck lensing map. In this table, the underlying data is the same, but we change aspects of the likelihood or theory such as the covariance, the assumed value of $p$, the amplitude of the fiducial cosmology, adding marginalization over magnification bias, and the bias evolution. Each region has 21 degrees of freedom.}
    \label{tab:data1}
\end{table}

\begin{table}[]
\renewcommand{\arraystretch}{1.2}
    \centering
    \begin{tabular}{c|c|ccc}
       Version & Region  &  $f_{\textrm{NL}}$ & $b_0$ & $\chi^2$ \\
       \hline
      \multirow{4}{*}{Fiducial} & North  &  $130^{+186}_{-119}$ & $0.92^{+0.08}_{-0.09}$ & 29.7 \\[0pt]
        & DECaLS N  &  $-67^{+62}_{-58}$ & $1.10^{+0.07}_{-0.09}$ & 18.0 \\[0pt]
        & DECaLS S  &  $-66^{+116}_{-86}$ & $1.14^{+0.10}_{-0.12}$ &  18.3\\[0pt]
        & DES  &  $-34^{+106}_{-81}$ & $1.27^{+0.13}_{-0.15}$ &  42.4 \\[0pt]
        \cline{2-5}
        & Combination & $-26^{+45}_{-40}$ & $1.09^{+0.04}_{-0.05}$ & 108.4 \\[0pt]
       \hline
              \multirow{4}{*}{$\ell_{\rm min} = 8$}  & North  &  $139^{+192}_{-126}$ & $0.92^{+0.08}_{-0.09}$ & 29.4 \\
        & DECaLS N  &  $-64^{+78}_{-63}$ & $1.09^{+0.08}_{-0.08}$ & 18.1 \\
        & DECaLS S  &  $-23^{+149}_{-103}$ & $1.09^{+0.10}_{-0.12}$ & 17.4 \\
        & DES  &  $9^{+134}_{-96}$ & $1.22^{+0.15}_{-0.15}$ & 43.5 \\
        \cline{2-5}
       & Combination & $-8^{+52}_{-45}$ & $1.07^{+0.04}_{-0.05}$ & 108.4 \\
       \hline
       \multirow{4}{*}{No sys.\ wts.} & North  &  $265^{+200}_{-177}$ & $0.91^{+0.08}_{-0.07}$ & 25.5 \\
       &  DECaLS N  &  $-179^{+69}_{-59}$ & $1.13^{+0.08}_{-0.08}$ & 21.2 \\
       &  DECaLS S  &  $-57^{+205}_{-129}$ & $1.02^{+0.11}_{-0.13}$ &  22.0\\
       &  DES  &  $130^{+152}_{-104}$ & $1.15^{+0.13}_{-0.13}$ &  44.0 \\
        \cline{2-5}
        & Combination & $-5^{+64}_{-56}$ & $1.04^{+0.05}_{-0.05}$ & 112.7  \\
       \hline
       \multirow{4}{*}{tSZ free} & North  &  $286^{+201}_{-196}$ & $0.82^{+0.07}_{-0.06}$ & 25.6 \\[0pt]
        & DECaLS N  &  $-134^{+67}_{-66}$ & $1.06^{+0.08}_{-0.09}$ & 26.3 \\[0pt]
        & DECaLS S  &  $38^{+267}_{-168}$ & $0.88^{+0.11}_{-0.11}$ &  18.7\\[0pt]
        & DES  &  $-70^{+111}_{-85}$ & $1.23^{+0.15}_{-0.17}$ &  28.1 \\[0pt]
        \cline{2-5}
        & Combination & $-25^{+69}_{-57}$ & $0.98^{+0.04}_{-0.06}$ & 98.7 \\[0pt]
        \hline
       \multirow{4}{*}{Remove $m=0$}  & North  &  $147^{+194}_{-125}$ & $0.91^{+0.09}_{-0.08}$ & 29.4 \\
       &  DECaLS N  &  $-48^{+79}_{-63}$ & $1.09^{+0.07}_{-0.09}$ & 18.2 \\
       &  DECaLS S  &  $-77^{+111}_{-86}$ & $1.15^{+0.10}_{-0.12}$ & 18.4 \\
       & DES  &  $-42^{+104}_{-82}$ & $1.28^{+0.14}_{-0.15}$ & 31.4 \\
        \cline{2-5}
       & Combination & $-19^{+47}_{-41}$ & $1.08^{+0.04}_{-0.05}$ & 97.2  \\
    \end{tabular}
    \caption{ Marginalized constraints on $f_{\textrm{NL}}$ and linear bias scaling amplitude $b_0$. In this table, we show various systematics tests with different data vectors. Each region has 21 degrees of freedom, except for $\ell_{\rm min} = 8$, where each region has 20 degrees of freedom.}
    \label{tab:data2}
\end{table}

\section{Discussion}
\label{sec:discussion}

We use the cross-correlation between Planck 2018 CMB lensing and DESI quasar targets at $z_{\textrm{eff}} = 1.51$ to constrain $f_{\textrm{NL}} = -26^{+45}_{-40}$ with the fiducial $p=1.6$ (or $f_{\textrm{NL}} = -18^{+29}_{-27}$ with $p=1.0$).
Strong systematics are present in the quasar autocorrelation,
exceeding the clustering signal by $\sim 3$---$5\times$ at $\ell = 10$ even after applying systematic weights (Fig.~\ref{fig:clgg_contaminated_mocks}).\footnote{We note that this is nevertheless an improvement over the DR8 quasar autocorrelation measurement from Fig.\ 19 in \cite{Kitanidis20}.} Since this excess noise is uncorrelated with CMB lensing, we obtain an unbiased cross-correlation measurement down to $\ell = 6$ in spite of the strong contamination.
This result highlights the power of cross-correlations to allow cosmological measurements even from galaxy samples with very noisy large-scale clustering.
At low $\ell$ where systematics are unavoidable, cross-correlations thus offer an alternative to precisely modelling and removing the contaminant power \citep{Rezaie21}.

Our $f_{\textrm{NL}}$ constraint is generally weaker than other LSS constraints from the galaxy auto-power spectrum and bispectrum.
The strongest constraints currently come from the 3D power spectrum of eBOSS quasars at $0.8 < z < 2.2$,
with $f_{\textrm{NL}} = -12\pm21$ \citep{Mueller}. A previous result using early eBOSS data found $-81 \leq f_{\textrm{NL}} \leq 26$ at 95\% confidence interval \citep{Castorina} (both results fix $p = 1.6$).
Analyses of the BOSS galaxy power spectrum and bispectrum
using the effective field theory of large scale structure
find $f_{\textrm{NL}} = -33 \pm 28$ \citep{Cabass22} and $f_{\textrm{NL}} = -30 \pm 29$ \citep{DAmico22}.
The constraining power is dominated by the power spectrum, and adding the bispectrum improves the constraint by $\sim 20\%$.
These are significant improvements over early $f_{\textrm{NL}}$ constraints from BOSS DR9 ($\sigma_{f\textrm{NL}} = 60$) \citep{Ross13}, with a larger sample of galaxies and more robust modelling to smaller scales.
While the quasars generally offer slightly better constraints on $f_{\textrm{NL}}$, comparing results is somewhat tricky due to different assumptions on the $b_1-b_\phi$ relationship (or alternatively, assumptions about $p$ in Eq.~\ref{eqn:delta_b}: the quasar papers generally use $p = 1.6$, whereas Ref.~\citep{DAmico22} use $p = 1$ and Ref.~\citep{Cabass22} use $p = 0.55$ for BOSS galaxies).
Further, allowing $b_\phi$ to vary freely can dramatically degrade the constraints \citep{Barreira22}.

Analyses of the angular clustering of photometric samples have generally also found competitive $f_{\textrm{NL}}$ constraints, with $\sigma_{f\textrm{NL}} \sim 20$. \cite{Slosar08} measured $f_{\textrm{NL}} = 28^{+23}_{-24}$ from SDSS photometric samples, and $f_{\textrm{NL}} = 8^{+26}_{-37}$ from photometric quasars specifically.
However, subsequent works found discrepant values of $f_{\textrm{NL}}$ \citep{Xia11,Nikoloudakis12}, and suggested that more robust systematic treatments were needed \citep{Pullen13,Leistedt_Peiris}, degrading $f_{\textrm{NL}}$ constraints.
Later results with more extensive photometric systematic mitigation found $f_{\textrm{NL}} = 12 \pm 21$ from a combination of photometric and spectroscopic SDSS galaxies; photometric SDSS quasars; radio sources; and the X-ray background \citep{Giannantonio14a}; and $-49 < f_{\textrm{NL}} < 31$ at 95\% confidence ($-26 < f_{\textrm{NL}} < 34$ at fixed cosmology) from SDSS photometric quasars alone \citep{Leistedt14} (see also \citep{Ho15}).

Our constraint, with $\sigma_{f\textrm{NL}} \sim 43$, has comparable constraining power to the recent results of Ref.~\cite{McCarthy22}, who find
$\sigma_{fNL} \sim 41$ from CIB-CMB lensing cross-correlations.
We have a 50\% smaller errorbar than LSS-CMB lensing and ISW constraints from Ref.~\cite{Giannantonio14b}.
This is primarily due to the lower noise in the Planck 2018 CMB lensing map that we use, compared to Planck 2013 in \cite{Giannantonio14b}. 

Future cross-correlation measurements with a cleaner quasar sample can significantly improve the constraining power on $f_{\textrm{NL}}$.  If the ``high purity'' DESI quasar target sample had no excess low-$\ell$ power, we would find $\sigma_{f\textrm{NL}} = 26$ (and a further improvement to $\sigma_{f\textrm{NL}} = 24$ if we could use the entire DES footprint).
A fully spectroscopic quasar sample would allow us to remove the residual contamination from stars and unclassified spectra.
Moreover, the cleaner spectroscopic sample may allow us to use the higher-number density ``main'' sample, which would further reduce $\sigma_{f\textrm{NL}} \sim 15$.
A cleaner quasar sample will also allow $f_{\textrm{NL}}$ constraints from the autocorrelation, likely with more statistical power than the cross-correlation (depending on the performance of the low-$\ell$ systematics removal).
However, we emphasize that while systematics add power to the autocorrelation, they only add noise to the cross-correlation. Interpreting the autocorrelation measurement will always require some level of systematics subtraction, and the cross-correlation measurement is more robust to unmodelled systematics.

Redshift binning is another clear area for improvement.
Already, our broad redshift bin combines quasars with positive and negative $f_{\textrm{NL}}$ response ($b(z)-1.6$), partially cancelling its constraining power. In Appendix~\ref{sec:highz}, we explore whether we can improve the $f_{\textrm{NL}}$ constraint by constructing a second redshift bin that excludes the $z < 1$ tail with negative $f_{\textrm{NL}}$ response.
The high-redshift bin is 30\% more constraining on $f_{\textrm{NL}}$ in a Fisher forecast. However, in data we find the bias of the sample is 30\% below expectation, dramatically reducing $b(z)-p$ and thus yielding a weaker $f_{\textrm{NL}}$ constraint than the main sample.
The reason for this discrepancy is unknown, potentially pointing to an issue with the photometric redshift bins, which can be resolved with spectroscopic redshifts for all DESI quasar targets.
Moreover, spectroscopic redshifts will allow for far more granular redshift bins; the resolution of the photometric redshifts is poor enough that the $z_{\rm phot} > 2$ split in Appendix~\ref{sec:highz} is essentially the best we can do. This will allow for further improvements in the $f_{\textrm{NL}}$ constraints from CMB lensing cross-correlations.

Finally, considerably better CMB lensing data is expected in the near future. Already, the Planck PR4 lensing maps have 10-20\% less noise than Planck 2018 \citep{Carron22},
as well as better mean-field estimation and improved systematics control at low $\ell$. Further in the future, the Simons Observatory is forecast to detect CMB lensing at S/N $\sim140$, or 
twice the signal-to-noise of Planck \citep{SO18}.
SO's CMB lensing reconstruction noise is low enough to allow for sample variance cancellation between the galaxy and CMB lensing fields \citep{SchmittfullSeljak}, leading to a forecasted $\sigma_{f\textrm{NL}} \sim 2$.

\section{Conclusions}
\label{sec:conc}

We detect the cross-correlation between Planck 2018 CMB lensing and DESI quasar targets at $26 \sigma$ over the range $6 < \ell < 300$. Rather than using the DESI main quasar target sample (with density 320 deg$^{-2}$),
we remove faint quasar targets and potential stars to create a ``high purity'' quasar sample (density 160 deg$^{-2}$).
This sample has much higher completeness, with only 2\% stars and 4\% redshift failures, compared to 6\% (7\%) stars (redshift failures) in the main sample.
Nevertheless, even after applying  weights to mitigate correlations with imaging systematics templates, the quasar autocorrelation has considerable excess power at low $\ell$, rendering any cosmological interpretation impossible.

Despite the excess noise in the quasar auto-correlation, the quasar-CMB lensing cross-correlation shows no evidence of contamination. The systematics tests pass, and we constrain $f_{\textrm{NL}} = -26^{+45}_{-40}$. CMB lensing cross-correlation therefore allows us to extract cosmological information out of this sample, despite the highly contaminated autocorrelation.

When regressing the imaging systematics templates, overfitting can decorrelate the quasar and CMB lensing fields, leading to a dramatically lower cross-correlation at low $\ell$. Contaminated mocks are critical to verify that we correctly recover the low $\ell$ cross-correlation, and we use them to determine the maximum number of systematics templates that we can regress against.
We find that regression can reduce the cross-correlation while still recovering approximately the correct auto-correlation on all scales.
These results emphasize that large-scale cross-correlations are especially sensitive to overfitting when mitigating imaging systematics.

We present the first cosmological analysis of the clustering of DESI quasar targets on large scales.
As a large scale, Fourier-space analysis, this work is complementary to the small-scale, configuration space clustering measurement of Ref.~\cite{Chaussidon22}. 
By lowering the excess systematic power at low $\ell$ and allowing for finer redshift binning, the full spectroscopic quasar sample will allow for improved constraints on $f_{\textrm{NL}}$ from CMB lensing cross-correlations.

\section*{Data Availability}

All data shown in the paper (including angular spectra, covariances, and a gridded likelihood) is publicly available at this site: 
\url{https://doi.org/10.5281/zenodo.7921905}.


\section*{Acknowledgments}

We would like to thank Gerrit Farren for bringing the Monte Carlo normalization
correction to our attention,
and Niayesh Afshordi for helpful conversations.

This material is based upon work supported by the U.S. Department of Energy (DOE), Office of Science, Office of High-Energy Physics, under Contract No. DE–AC02–05CH11231, and by the National Energy Research Scientific Computing Center, a DOE Office of Science User Facility under the same contract. Additional support for DESI was provided by the U.S. National Science Foundation (NSF), Division of Astronomical Sciences under Contract No. AST-0950945 to the NSF’s National Optical-Infrared Astronomy Research Laboratory; the Science and Technologies Facilities Council of the United Kingdom; the Gordon and Betty Moore Foundation; the Heising-Simons Foundation; the French Alternative Energies and Atomic Energy Commission (CEA); the National Council of Science and Technology of Mexico (CONACYT); the Ministry of Science and Innovation of Spain (MICINN), and by the DESI Member Institutions: \url{https://www.desi.lbl.gov/collaborating-institutions}. Any opinions, findings, and conclusions or recommendations expressed in this material are those of the author(s) and do not necessarily reflect the views of the U. S. National Science Foundation, the U. S. Department of Energy, or any of the listed funding agencies.

The DESI Legacy Imaging Surveys consist of three individual and complementary projects: the Dark Energy Camera Legacy Survey (DECaLS), the Beijing-Arizona Sky Survey (BASS), and the Mayall z-band Legacy Survey (MzLS). DECaLS, BASS and MzLS together include data obtained, respectively, at the Blanco telescope, Cerro Tololo Inter-American Observatory, NSF’s NOIRLab; the Bok telescope, Steward Observatory, University of Arizona; and the Mayall telescope, Kitt Peak National Observatory, NOIRLab. NOIRLab is operated by the Association of Universities for Research in Astronomy (AURA) under a cooperative agreement with the National Science Foundation. Pipeline processing and analyses of the data were supported by NOIRLab and the Lawrence Berkeley National Laboratory. Legacy Surveys also uses data products from the Near-Earth Object Wide-field Infrared Survey Explorer (NEOWISE), a project of the Jet Propulsion Laboratory/California Institute of Technology, funded by the National Aeronautics and Space Administration. Legacy Surveys was supported by: the Director, Office of Science, Office of High Energy Physics of the U.S. Department of Energy; the National Energy Research Scientific Computing Center, a DOE Office of Science User Facility; the U.S. National Science Foundation, Division of Astronomical Sciences; the National Astronomical Observatories of China, the Chinese Academy of Sciences and the Chinese National Natural Science Foundation. LBNL is managed by the Regents of the University of California under contract to the U.S. Department of Energy. The complete acknowledgments can be found at \url{https://www.legacysurvey.org/}.

The authors are honored to be permitted to conduct scientific research on Iolkam Du’ag (Kitt Peak), a mountain with particular significance to the Tohono O’odham Nation.




\clearpage

\bibliographystyle{JHEP}
\bibliography{main}

\appendix

\section{Quasar clustering in redshift bins}
\label{sec:highz}

\begin{figure}
    \centering
    \includegraphics[width=1.0\linewidth]{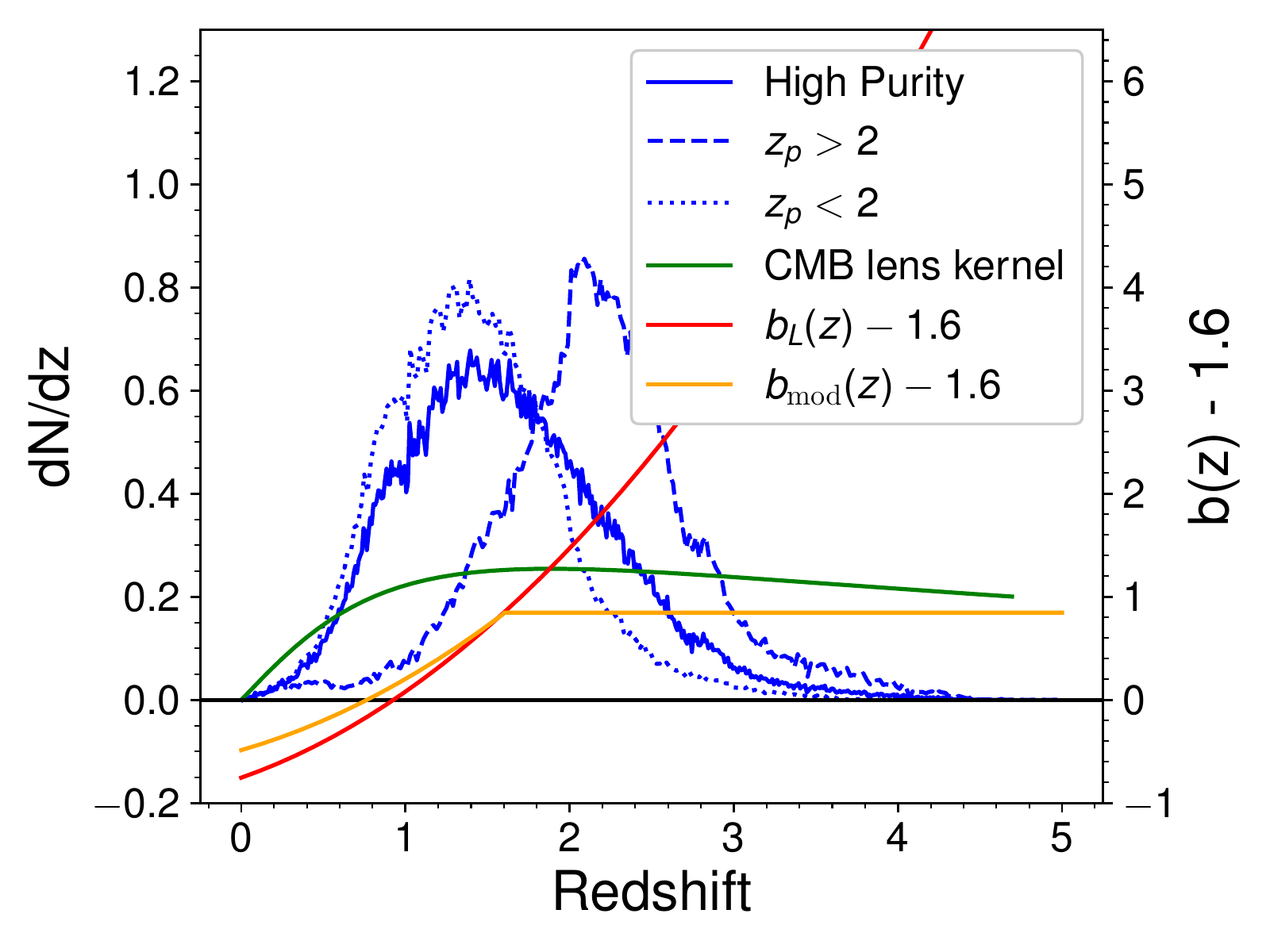}
    \caption{Comparison between the High Purity sample and the high and low redshift samples, split at $z_{\rm phot} = 2$. The CMB lensing kernel is overplotted in green. The $f_{\textrm{NL}}$ response $b_L(z)-1.6$ (using the fiducial bias evolution) is also shown in red; the high redshift sample dramatically reduces the fraction of quasars with negative $f_{\textrm{NL}}$ response.
    We also show the modified bias evolution of Eq.~\ref{eqn:modified_bz}.}
    \label{fig:redshift_bins}
\end{figure}

At $z < 1$, the quasar bias is sufficiently low that the $f_{\textrm{NL}}$ sensitivity, $b(z)-1.6$, becomes negative.
Therefore, a broad redshift bin with a substantial fraction of quasars at $z < 1$ will have reduced $f_{\textrm{NL}}$
sensitivity, as the $z < 1$ tail will cancel the positive $f_{\textrm{NL}}$ response at $z > 1$. By creating tomographic redshift bins, we can potentially improve the constraining power of the data. Using the photometric redshifts of Ref.~\cite{Duncan22}, we split the quasar sample at $z_{\textrm{phot}} = 2$ to create high and low-redshift samples.
These samples have number densities of 117 and 44 deg$^{-2}$ and effective redshifts 1.32 and 2.07, and their redshift distributions are shown in Fig~\ref{fig:redshift_bins}. Due to the low value of $b(z)-1.6$, the forecasted $\sigma_{f \textrm{NL}}$ constraint for the low-redshift sample is 37 without noise, and 88 without systematic weights. In constrast, the higher-redshift sample offers better constraints despite the lower number density, with $\sigma_{f \textrm{NL}}$ of 21 and 41, respectively. The combination of the two samples (correctly accounting for their covariance) offers little improvement over the $z_{\rm phot} > 2$ sample alone, with $\sigma_{f\textrm{NL}} = 18$ and 38 for the noiseless and unweighted cases, respectively. However the $z_{\rm phot} >2$ sample alone offers a 25-35\% increase in constraining power on $f_{\textrm{NL}}$ compared to the single broad redshift bin.

Motivated by this forecast, we also measured the cross-correlation between the high-redshift quasar sample and Planck CMB lensing. Due to the lower number density, we linearly regressed systematics templates at resolution NSIDE=128 rather than 256. We also found that slightly different systematics templates were required for the high-redshift sample. We used E(B-V) alone for North and DECaLS S; E(B-V), stellar density, Sagittarius Stream, and PSFDEPTH in both W1 and W2 for DECaLS N; and E(B-V) and W1 and W2 PSFDEPTH for DES. We verified that this was a sufficiently limited template set to allow us to correctly recover the cross-correlation from mocks.
Using these templates led to no loss of power on large scales in the cross-correlation.
The stellar fraction is also slightly higher, 3.2\% versus 2.0\% for the full sample, and the magnification bias slope $s$ is slightly different, $s = 0.3384$, 0.3455, 0.3481, 0.3162 for North, DECaLS N, DECaLS S, DES, respectively.
We also find an anomalous anti-correlation between quasars and CMB lensing in the lowest-$\ell$ bin in DECaLS N ($\ell = 4$ and 5), significant at $3.3\sigma$. This may be due to the anomalous (and presumably foreground-affected) CMB lensing quadrupole ($\ell = 2$) correlating with systematics in the quasars, and this multipole enters the $\ell = 4$ to 5 bin due to the mask-induced mode coupling. We therefore set $\ell_{\rm min} = 6$ for DECaLS N.

\begin{figure}
    \centering
    \includegraphics[width=1.0\linewidth]{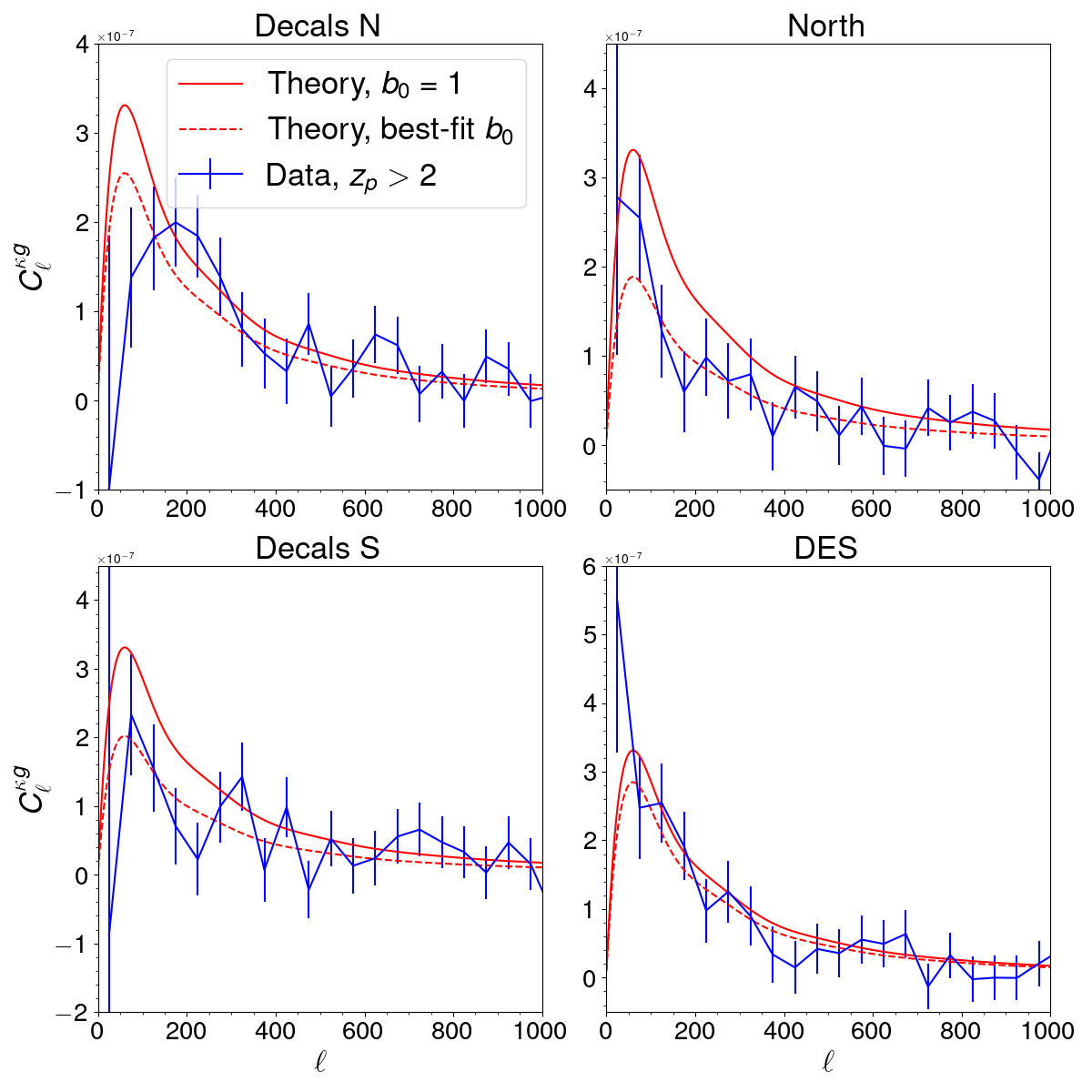}
    \caption{DESI quasar-Planck CMB lensing cross-correlation for the $z_{\rm phot} > 2$ sample. Note that due to the large bin-widths ($\Delta \ell = 50$), this measurement can constrain $b_0$ but not $f_{\textrm{NL}}$, for which we use much smaller bins at low $\ell$. We show both the $C_{\ell}^{\kappa g}$ theory curve for the Ref.~\cite{Laurent17} bias evolution ($b_0 = 1$), and the theory curve with the best-fit amplitude of the bias $b_0$.}
    \label{fig:data_highz}
\end{figure}

The quasar-CMB lensing cross-correlation is detected at 14.4$\sigma$ across all imaging regions and at $0 < \ell < 300$, and the bandpowers are shown in Fig.~\ref{fig:data_highz}.
Despite the optimistic Fisher forecasts, the $f_{\textrm{NL}}$ constraints are much weaker for the high-redshift sample than the full sample,
with $f_{\textrm{NL}} = 124^{+283}_{-129}$. This is due to the fact that the clustering amplitude is much lower than expected: $b_0 = 0.63 \pm 0.07$, yielding a much lower value of $b(z)-1.6$ and thus poor $f_{\textrm{NL}}$ sensitivity.

Defining the effective bias following Refs.~\cite{Modi17,Chen22}
\begin{equation}
    b_{\rm eff} = \int dz \frac{1}{\chi^2} \frac{dN}{dz} W^\kappa(z) b(z)
\end{equation}
the $b_0$ constraint yields an effective bias of $2.26 \pm 0.23$ at an effective redshift of 2.07, compared to the predicted $b(z=2.07) = 3.33$ in Ref.~\cite{Laurent17}.

The poor constraining power on $f_{\textrm{NL}}$ also leads to a highly non-Gaussian and asymmetric posterior.
We repeat the systematics tests in \cref{tab:data1,tab:data2} and do not find any significant deviations. 
We do find $\sim 2\sigma$ discrepancies in the bias measured between different imaging regions, as well as a $\sim 2\sigma$ deficit in power between $\ell < 200$ and $\ell > 200$ in the combination of DECaLS S, DECaLS N, and North. However, given the large number of tests performed, some 2$\sigma$ deviations are expected.
The stellar fraction is slightly higher for the high-redshift sample than the high-purity sample, but fluctuations in stellar fraction are unlikely to explain the $>30\%$ drop in clustering amplitude. We find that the stellar fraction ranges from 2.6\% to 4.3\% between the different imaging regions (following the same trend as the high-purity sample, decreasing from North to DECaLS N to DECaLS S). Moreover, the fitted bias does not significantly vary between the fiducial measurements and the imaging regions restricted to Galactic latitude $|b| > 40^{\circ}$.

We also measure the clustering of the $z_{\rm phot} < 2$ sample. Here we find a clustering amplitude consistent with the evolution of Ref.~\cite{Laurent17}, with $b_0 = 1.04 \pm 0.05$ (fixing $f_{\textrm{NL}} = 0$ as it is poorly constrained anyway). Given these three clustering measurements, we adjust the bias evolution to determine if they are consistent with each other. We try a modified bias evolution with flat bias at $z > 1.6$:
\begin{equation}
        b(z) =
\begin{cases}
    0.23 ((1+z)^2 - 6.565) + 2.393 & \text{if } z< 1.6\\
    2.447,              & \text{if } z\geq 1.6\\
\end{cases}
\label{eqn:modified_bz}
\end{equation}
Using the modified bias evolution, we find (at fixed $f_{\textrm{NL}} = 0$) $b_{0, \textrm{mod}} = 1.08 \pm 0.075$ for the $z_{\rm phot} < 2$ sample,
$1.03 \pm 0.03$ for the entire sample,
and $0.95 \pm 0.07$ for the $z_{\rm phot} > 2$ sample. Hence there is no evidence that the three samples are inconsistent with each other.

\begin{figure}
    \centering
    \includegraphics[width=1.0\linewidth]{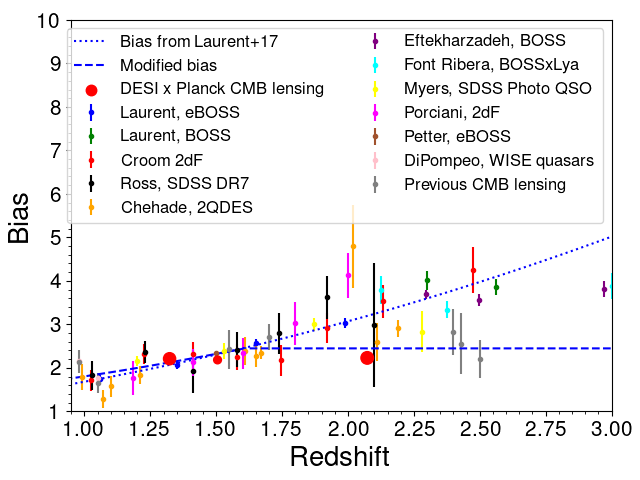}
    \caption{Quasar bias measurements from DESI-Planck CMB lensing cross-correlations (red circles, with the diameter corresponding to the 1-$\sigma$ range). We compare these measurements to previous measurements (colored points) including previous CMB lensing cross-correlations (gray points); the best-fit bias evolution from Ref.~\cite{Laurent17}, and the modified bias evolution (Eq.~\ref{eqn:modified_bz}) that fits our cross-correlation measurements.}
    \label{fig:qso_bias_comparison}
\end{figure}

In Fig.~\ref{fig:qso_bias_comparison},
we compare our cross-correlation bias measurement to a compilation of quasar clustering results (taking the bias measurements at ``face value'' without homogenizing the cosmology
or other aspects of the analysis).
We present both the modified bias evolution and three datapoints showing the effective bias at the effective redshift,
computed using the modified bias evolution.
We note that the effective bias differs by $\sim10\%$ if we use the Ref.~\cite{Laurent17} bias evolution instead.

For comparison to past work, we also show the Ref.~\cite{Laurent17} bias evolution.
This was derived from 3D autocorrelation
measurements of eBOSS \citep{Laurent17}
and BOSS \citep{Laurent16} quasars.
We also show the eBOSS 3D autocorrelation
measurement of Ref.~\cite{Petter22} (averaging over their color-selected bins);
and BOSS 3D autocorrelation measurement from Ref.~\cite{Eftekharzadeh15}.
Ref.~\cite{Chehade} used the 2QDES survey to measure quasar clustering across a wide range in color and luminosity (we do not combine their three luminosity-selected bins per redshift).
They find no relationship between quasar clustering and luminosity. Indeed, they measure the highest bias for the lowest-luminosity bins. This result agrees with several other studies finding no significant relationship between quasar clustering and physical properties \citep{daAngela08,Shen09,Shanks11,Shen13,Krolewski15,Petter22}. This suggests that it is reasonable to compare quasar clustering measurements spanning a wide range in luminosity and selection technique.

We also show older quasar clustering measurements from 2dF \citep{Croom05,Porciani} and SDSS DR7 \citep{Ross09}. We show projected rather than 3D clustering both for SDSS photometric quasars
\citep{Myers06} and for WISE-selected quasars \citep{DiPompeo17}. The quasar bias was also measured by Ref.~\cite{FontRibera} using cross-correlations between BOSS quasars and the Lyman-$\alpha$ forest.

Previous CMB lensing cross-correlation measurements are shown in gray in Fig.~\ref{fig:qso_bias_comparison}. These include cross-correlations with WISE-selected quasars \citep{DiPompeo17}; eBOSS quasars \citep{Petter22,Geach19,Han19}; and BOSS quasars \citep{Alonso18,Doux18,Lin20}.

Our measurements at $z_{\rm eff} = 1.32$ and 1.51 are quite consistent with previous results, but the high-redshift cross-correlation yields a significantly lower bias. Likewise, the modified bias evolution required to fit the high-redshift point agrees well with previous measurements at $z < 1.5$, but is significantly low at higher redshift.
While the scatter of the earlier bias measurements is large, the high-precision BOSS and eBOSS
results are quite close to the Ref.~\cite{Laurent17} bias evolution.
We do note a weak trend for the $z>2$ cross-correlations to be lower than the auto-correlations, with the exception of the lowest-redshift point from \cite{FontRibera}.
These measurements are not independent: the three CMB lensing cross-correlations at $z \sim 2.4$ are all from the same sample (BOSS). Furthermore, the large errorbar of these measurements means that the deviation from Ref.~\cite{Laurent17} is not highly significant.

To investigate this discrepancy further, we measure the 3D quasar correlation function from the Early Data Release (following the large-scale structure catalog creation described in Ref.~\cite{Moon23}). Following Ref.~\cite{Laurent17}, we fit a linear bias and linear (Kaiser effect) RSD model to the monopole $\xi_0$ in the range $20 < r < 80$ $h^{-1}$ Mpc:
\begin{equation}
\xi_0 = \xi_{\textrm{mm}} b^2 (1 + \frac{2}{3}\beta + \frac{1}{5}\beta^2)
\label{eqn:linear_bias_model}
\end{equation}
where $\beta = f/b$ and $f$ is the growth rate $\frac{d\ln{D}}{d\ln{a}}$.
We find that this simple model fits the data well on these scales.
The results are summarized in Table~\ref{tab:b_from_corr_func},
split into a number of spectroscopic redshift bins. 
These  results for the main sample are consistent with the DESI One Percent Survey measurements presented in Ref.~\cite{Prada23}
(measured on a similar scale range,
$10 < r < 85$ $h^{-1}$ Mpc),
and are consistent with the Uchuu SHAM mocks presented in Ref.~\cite{Prada23}.
We find that the 3D correlation function
prefers a similar bias to Ref.~\cite{Laurent17}; the bias is also very similar
between the main sample and the high purity sample.

We therefore conclude that the origin of the low CMB lensing cross-correlation comes from the CMB lensing analysis itself, and not because the DESI quasars have lower bias than previous quasar samples.
One possibility is magnification bias: if $s = 0$ for the $z_{\textrm{phot}} > 2$ sample, this would explain the low clustering measurement. We are unable to re-run the photometric redshift code of Ref.~\cite{Duncan22}
after perturbing the photometry to measure the magnification bias slope, and it is possible (though somewhat unlikely) that this has a massive impact on $s$, lowering it from $\sim 0.34$ to 0. Another possibility is residual
imaging systematics in the $z_{\textrm{phot}} > 2$ sample.
We regress fewer templates for the $z_{\textrm{phot}} > 2$ sample than for the high-purity sample as a whole, and we do find a visually significant trend with Galactic latitude in DECaLS S. However, we do not find any significant shifts in our results if we use the tSZ-deprojected lensing map, or if we turn off the angular systematic weights (or replace them with more-aggressive random forest weights), nor do we find any difference between the cross-correlation above and below Galatic latitude $|b| = 40^{\circ}$. Finally, the largest CMB lensing foreground at these redshifts is CIB contamination, since the CIB peaks at $z\sim2$.
Previous results suggest that the CIB bias to CMB lensing cross-correlated with $z \sim 0.5$ galaxies is similar to the CIB bias on the lensing autocorrelation \citep{Schaan:2018tup}. The Planck lensing team estimates $<1\%$ biases from CIB on the autospectrum (Section 4.5 of Ref.~\cite{PlanckLens18}). On the other hand, the CIB redshift kernel increases from $z \sim 0.5$ to $z \sim 2$, so it is plausible that CIB biases are larger for our sample by a factor of a few. Interestingly, the sign is correct as the CIB bias is negative. However, the average bias shifts by only 3\% when using the tSZ-deprojected lensing map in place of the default SMICA map; these two maps have very different response to CIB. Overall, it seems like it would be difficult to make the CIB contamination large enough to explain the $\sim 30$--$40$\% discrepancy.

\begin{table}[]
    \centering
    \begin{tabular}{c|c|c|c}
   Redshift range & Main $b_{\textrm{qso}}$ & High Purity $b_{\textrm{qso}}$ & Laurent+17 $b_{\textrm{qso}}$ \\
   \hline
   $0.8 < z < 1.1$ & $1.85 \pm 0.13$ & $1.92 \pm 0.14$ & 1.63 \\
   $1.1 < z < 1.6$ & $2.13 \pm 0.09$ & $2.22 \pm 0.11$ & 2.12 \\
   $1.6 < z < 2.1$ & $2.67 \pm 0.10$ & $2.79 \pm 0.13$ & 2.81 \\
   $2.1 < z < 3.5$ & $3.67 \pm 0.15$ & $4.69 \pm 0.21$ & 3.99 \\
    \end{tabular}
    \caption{ Bias measured from the monopole of the 3D quasar correlation function in four spectroscopic redshift bins, using DESI EDR data. The monopole is fit using a linear theory model (Eq.~\ref{eqn:linear_bias_model})
    over $20 < r < 80$ $h^{-1}$ Mpc.}
    \label{tab:b_from_corr_func}
\end{table}

\end{document}